\documentclass{ipol}
\ipolSetTitle{An Analysis and Implementation of the HDR+ Burst Denoising Method}
\ipolSetAuthors{
	Antoine Monod\ipolAuthorMark{1,}\ipolAuthorMark{2}, Julie Delon\ipolAuthorMark{1}, Thomas Veit\ipolAuthorMark{2}
}
\ipolSetAffiliations{
	\ipolAuthorMark{1} MAP5, Universit\'e de Paris, France (\texttt{\{antoine.monod, julie.delon\}@u-paris.fr})\\
	\ipolAuthorMark{2} GoPro Technology, France (\texttt{tveit@gopro.com})
}

\ipolSetID{336}
\ipolSetVolume{11}

\ipolSetSubmissionYear{2021}
\ipolSetSubmissionMonth{03}
\ipolSetSubmissionDay{01}

\ipolSetAcceptedYear{2021}
\ipolSetAcceptedMonth{05}
\ipolSetAcceptedDay{05}

\ipolSetPublicationYear{2021}
\ipolSetPublicationMonth{05}
\ipolSetPublicationDay{25}

\usepackage[font={sf,footnotesize}]{caption}
\usepackage{hyperref,graphicx,amsmath,amssymb,amsthm, subcaption, url}
\usepackage[vlined,ruled]{algorithm2e}

\SetKwInOut{Input}{input}
\SetKwInOut{Output}{output}
\SetKwComment{Comment}{}{}

\SetCommentSty{mycmtsty}

\DeclareMathOperator*{\argmin}{argmin}

\newtheorem*{remark}{Remark}

\begin{document}
\begin{ipolAbstract}
	HDR+ is an image processing pipeline presented by Google in 2016. At its core lies a denoising algorithm that uses a burst of raw images to produce a single higher quality image. Since it is designed as a versatile solution for smartphone cameras, it does not necessarily aim for the maximization of standard denoising metrics, but rather for the production of natural, visually pleasing images. In this article, we specifically discuss and analyze the HDR+ burst denoising algorithm architecture and the impact of its various parameters. With this publication, we provide an open source Python implementation of the algorithm, along with an interactive demo.
\end{ipolAbstract}

\begin{ipolCode}
	This Python implementation of HDR+ has been peer-reviewed and accepted by Image Processing On Line\@. The source code, its documentation, and the online demo are available from \href{\ipolLink}{the associated article web page}. The code is also available on GitHub\footnote{\url{https://github.com/amonod/hdrplus-python}}.
\end{ipolCode}

\ipolKeywords{burst denoising; computational photography; motion estimation; temporal filtering; high dynamic range; image signal processor}

\section{Introduction}

Image noise is a common issue in digital photography, and that issue is even more prominent in smaller sensors used in devices like smartphones or action cameras. Continuous improvement in the processing power of small devices has enabled many advances in computational photography, where algorithms are used to compensate for the limited physical properties of the imaging system. Algorithms that aim at removing digital noise are a very recurrent topic of the computational photography literature. Most of them only use the information stored in the input image (they are usually referred to as ``single-frame'' algorithms), while others suggest the use of information stored in additional images (usually videos or bursts). These multi-frame ``denoisers'' usually make use of similarity measurements in order to combine the information of multiple pixels or patches~\cite{buades2009note}.

A multi-frame denoising algorithm is at the heart of the HDR+ system. It was first officially introduced as a feature within the Google Camera App in the newly released Nexus 5 and Nexus 6 smartphones in fall 2014\footnote{HDR+: Low Light and High Dynamic Range photography in the Google Camera App, \url{https://ai.googleblog.com/2014/10/hdr-low-light-and-high-dynamic-range.html}, 2014}. In 2016, a full raw image processing pipeline with the same HDR+ moniker was presented in~\cite{hasinoff2016burst}. This method is the main subject of this article. The pipeline was embedded in smartphones such as the Google Nexus 6, 5X and 6P and first generation Google Pixel. Further algorithm refinements and optimizations were included in the Google Pixel 2 which was released in fall 2017. This device was the first to embed HDR+ in a dedicated Google-designed Image Processing Unit\footnote{Pixel Visual Core: image processing and machine learning on Pixel 2, \url{https://www.blog.google/products/pixel/pixel-visual-core-image-processing-and-machine-learning-pixel-2/}}. The same raw burst alignment and merging technique was included in a new ``Night Sight'' mode that was launched alongside the Pixel 3A. The changes and optimizations of this mode were presented in a new publication in 2019~\cite{liba2019handheld}. As such, the complex HDR+ system is not only part of a detailed, frequently compared and quoted scientific publication, but also embedded into multiple recent mass-produced consumer electronics devices (which have received praise for their camera performance\footnote{\url{https://www.dxomark.com/google-pixel-2-reviewed-sets-new-record-smartphone-camera-quality/}}), making it a relevant subject of study.

In this article, we will describe the core part of the system, a raw burst denoising algorithm (although it will not be our main focus, we will also mention the rest of the pipeline). This publication also comes with a publicly available HDR+ Python implementation (with a simplified finishing pipeline), along with an interactive demo on IPOL.

\section{Characteristics of the HDR+ Pipeline}

HDR+ was designed as a consumer-centric smartphone photography pipeline. Its goal is to produce individual pictures with good contrast and dynamic range, little noise and motion blur, and pleasing colors in most shooting scenarios, all while looking natural and requiring little to no user input. In order to achieve that goal, the authors of~\cite{hasinoff2016burst} conceived a system with the following constraints:
\begin{itemize}
	\item The images that are captured and processed are raw images, as they typically have higher bit depth and a better known noise model than their post-processed and compressed 8-bit JPEG counterparts.
	\item Images are underexposed at capture time to avoid clipped highlights and saturated pixels, which effectively allows the capture of more dynamic range. Thanks to a custom auto-exposure algorithm, a gain is also memorized to compensate later in the pipeline for that under-exposure.
	\item Images also have high shutter speed (or short exposure time) to avoid motion blur (from the scene or the camera).
	\item For a fixed gain, images that have a shorter exposure time tend to have a lower signal-to-noise ratio; to compensate for that, multiple images are actually captured (usually between 2 and 8 images depending on scene brightness measurements, also selected through the auto-exposure algorithm), and the temporal information of the burst is combined through an alignment and merging procedure. All images have identical shutter speed (meaning that unlike some HDR fusion methods, exposure is not bracketed) and the interval between each frame is constant. This allows both a similar noise profile across the burst and easier, more accurate motion estimation and compensation.
	\item The resulting higher bit depth, high dynamic range, and less noisy image is then tone mapped to compensate for underexposition and to produce a more natural look, even in high contrast scenarios (in that case, shadows are boosted while as few highlights as possible are clipped).
	\item Additional processing is performed to produce a more visually pleasing final image with a distinctive ``HDR+ look'' (c.f. Section~\ref{googlepipeline}).
	\item Total burst processing time on a mobile device is only a few seconds, and the mode is transparent to the user (they do not know that several pictures are being taken or what processing needs to be performed).
	\item The whole process is automatic and parameter-free for the photographer.
\end{itemize}

\begin{figure}[!htbp]
	\begin{center}
		\includegraphics[width=\linewidth]{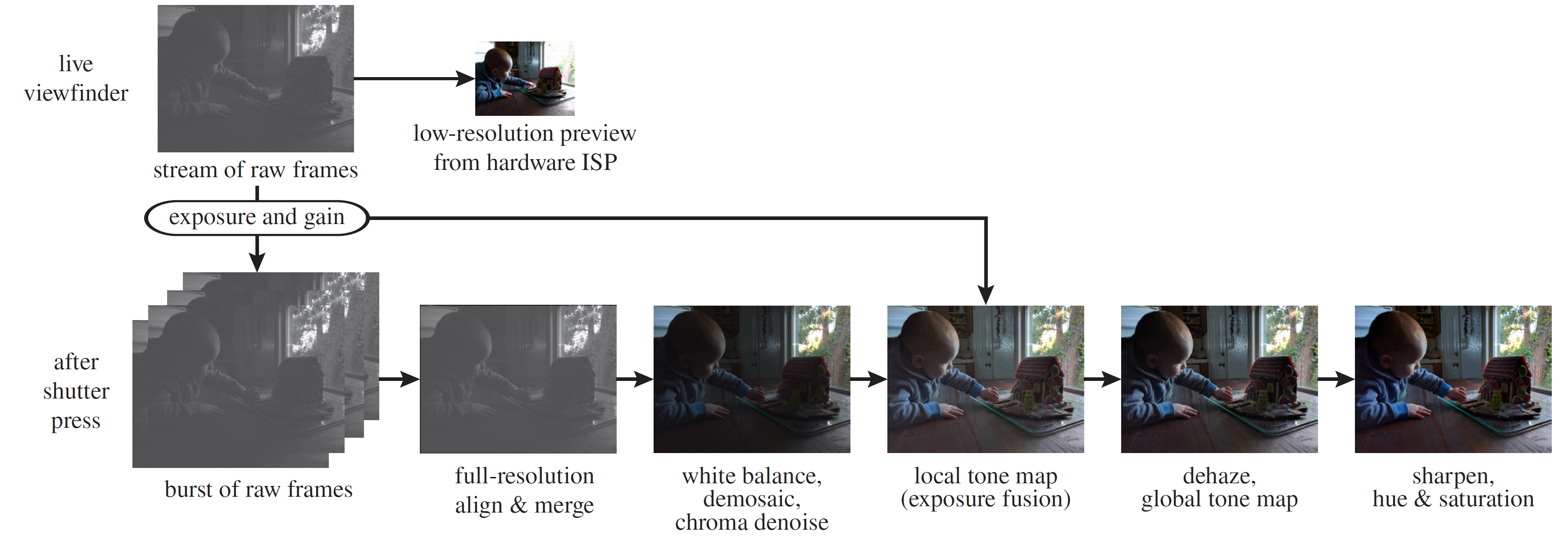}
		\caption{The HDR+ pipeline (extracted from~\cite{hasinoff2016burst}, lens shading and chromatic aberration correction have been omitted for brevity)}
		\label{fig:pipeline}
	\end{center}
\end{figure}

A representation of the whole image processing pipeline can be seen in Figure~\ref{fig:pipeline}. For more information on the capture process, including the functioning of the example-based auto-exposure algorithm, please refer to the original article~\cite{hasinoff2016burst}, and its supplemental material\footnote{\url{https://static.googleusercontent.com/media/hdrplusdata.org/en//hdrplus_supp.pdf}}. In the next sections, we will explain the whole processing pipeline after capture. 

\section{Bayer Tile-based Alignment} \label{alignment}

Before combining the information of multiple noisy images to produce a single, clean image, one must make sure that all images feature similar content. Since they share the same exposure time, scene brightness can be considered constant for the duration of the burst. Even though that exposure time is short, there can be apparent motion in the burst, mainly due to camera shake (the pipeline being designed for smartphones, most pictures will be taken handheld) and objects moving in the scene. That motion has to be compensated.

The strategy to compensate motion across the burst is to cleverly select one image as a reference, and then estimate motion between that image and each one of the other burst images (sometimes referred to as alternate frames). To that end, the reference image is divided into a regular array of square tiles and, for each tile, the corresponding tile in the alternate image is the one of minimal distance within a defined search radius around the reference tile's original location. Estimating the motion of image tiles instead of individual pixels is sufficient for the merging phase, and as we will explain in Section~\ref{merging}, the rest of the algorithm is robust to alignment errors.

\subsection{Reference Image Selection}

The reference image is selected among the first three images of the burst in order to minimize perceived shutter lag (the content of subsequent images of the burst can be quite different from what it was when the user pressed the shutter). Among these, the sharpest image is picked, the reasoning being that images that are less sharp are more likely to feature motion blur, and will not merge well with images that are sharp in the same regions. Sharpness is measured by the magnitude of gradients within a single (green) channel. We did not re-implement this part of the algorithm in our own implementation: for bursts provided by Google (which we'll discuss in Section~\ref{hdrplusdata}), we simply reuse their selected reference image. In the case of new bursts, we let the user select the reference or default to the first image otherwise.

For the remainder of the alignment step, the full resolution Bayer images are converted to single channel, lower resolution grayscale images using a simple $2\times 2$ box filter: each R, G, G, B square Bayer pattern is averaged in order to produce a single intensity value. Since the motion estimation of the grayscale images produces final results at pixel level, motion can only be estimated in multiples of 2 pixels in the original full-size raw images. Although this could be considered as a lack of precision, it is actually convenient because the arrangement of color planes in the Bayer image will be preserved after alignment (the value of a blue pixel will never be replaced by the value of a red pixel).

\subsection{Multi-scale Pyramid Alignment}
 
In order to quickly cover a large search area to find the tiles with minimal distance, the following multi-scale coarse-to-fine alignment strategy is adopted:
\begin{itemize}
	\item Gaussian pyramids of the single channel burst images are constructed (e.g.\ in the HDR+ article supplement, it is typically 4-level, with successive downsampling factors of 2, 4 and 4 from the grayscale image to the coarsest level of the pyramid).
	\item At each pyramid level starting from the coarsest (i.e.\ lowest resolution) level, the reference image is divided in equally spaced square tiles (e.g.\ in the HDR+ supplement tiles of size $8\times 8$ at the coarsest level and $16\times 16$ at other levels). For each reference tile, we compute the L2 or L1 distance between it and each possible tile of the alternate frame within a specified search radius (e.g.\ + or - 4 pixels) around some initial guess (that guess being the location of the reference tile at the coarsest level)
	\begin{equation}
		\label{eq:p_distance}
		D_{p}(u, v)=\sum_{i=0}^{n-1} \sum_{j=0}^{n-1}\left|T(i, j)-I\left(i+u+u_{0}, j+v+v_{0}\right)\right|^{p},
	\end{equation}
	where $T$ is the reference tile of size $n \times n$, $(u,v)$ is the possible location of the alternate image tile within the larger search area $I$ currently evaluated, $p$ is the power of the norm used (1 or 2) and ($u_0, v_0$) is the initial guess which indicates the location of the search area ($u_0=v_0=0$ at the coarsest scale).\\
	As suggested in the HDR+ supplement, the L2 distance (written in the case where $u_0 = v_0 = 0$ for brevity) can be rewritten as
	\begin{equation}
		\label{eq:p2_distance}
		D_{2}(u, v) =\sum_{x=0}^{n-1} \sum_{y=0}^{n-1} T(i, j)^{2}
		+\sum_{i=0}^{n-1} \sum_{j=0}^{n-1} I(i+u, j+v)^{2}
		-2 \sum_{i=0}^{n-1} \sum_{j=0}^{n-1} T(i, j) I(i+u, j+v),
	\end{equation}
	where the second term can be computed by filtering the squared elements of $I$ with a $n\times n$ box filter, and the third is proportional to the cross-correlation of I and T, which can be computed quickly using fast Fourier transforms.\\
	The displacement from the location of the reference tile to the location of the alternate tile with minimal distance gives us the motion vector attributed to the reference tile.
	\item Alignments are upsampled, meaning that motion vectors are scaled and propagated to the correct number of tiles in the finer level. For example, if tiles are $16\times 16$ at both levels and the upsampling factor is 4, a tile with a (2, 1) motion vector at the coarser level implies 16 tiles with a (8, 4) motion vector at the finer level.
	\item An additional step is added in order to mitigate some potential upsampling problems, particularly when coarse-scale tiles straddle over the boundaries of moving objects. For each tile at a new (finer) level, instead of directly using the upsampled motion vector of the corresponding coarse-scale tile as the initial alignment guess ($u_0, v_0$), 3 candidates are evaluated: the alignment of the coarse-scale tile, plus those of the nearest coarse-scale tiles in each spatial dimension. The candidate alignment that minimizes the L1 distance between the reference tile and the corresponding alternate tile at the new pyramid level becomes the new initial alignment guess. This step is illustrated in Figure~\ref{fig:candidates}.
	\begin{figure}[!htbp]
		\begin{center}
			\includegraphics[width=0.5\linewidth]{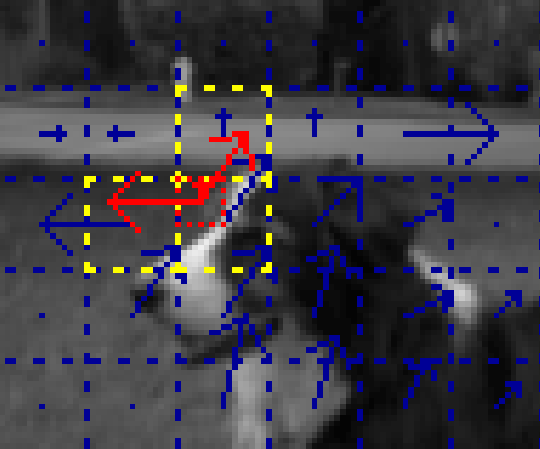}
			\caption{\textbf{Blue}: coarse-scale motion field (arrows: motion vectors, dotted lines: tiles). \textbf{Red}: finer-scale tile. \textbf{Yellow}: its 3 nearest coarse-scale tiles. The 3 motion vectors extracted from those neighboring tiles will be the candidates evaluated at the finer scale (shown in \textbf{red}).}
			\label{fig:candidates}
		\end{center}
	\end{figure}
	\item The selected upsampled motion vectors are then used as the location of the search area for the motion estimation at the subsequent pyramid level. For example, if $(u_m, v_m)$ is the upsampled motion vector resulting from the previous steps, the initial guess $(u_0, v_0)$ will be updated to $(u_0+u_m, v_0+v_m)$.
\end{itemize}

\begin{figure}[!htbp]
	\begin{center}
		\begin{subfigure}[t]{0.33\linewidth}
			\begin{center}
				\begin{subfigure}[t]{.98\linewidth}
					\begin{center}
						\includegraphics[width=\linewidth]{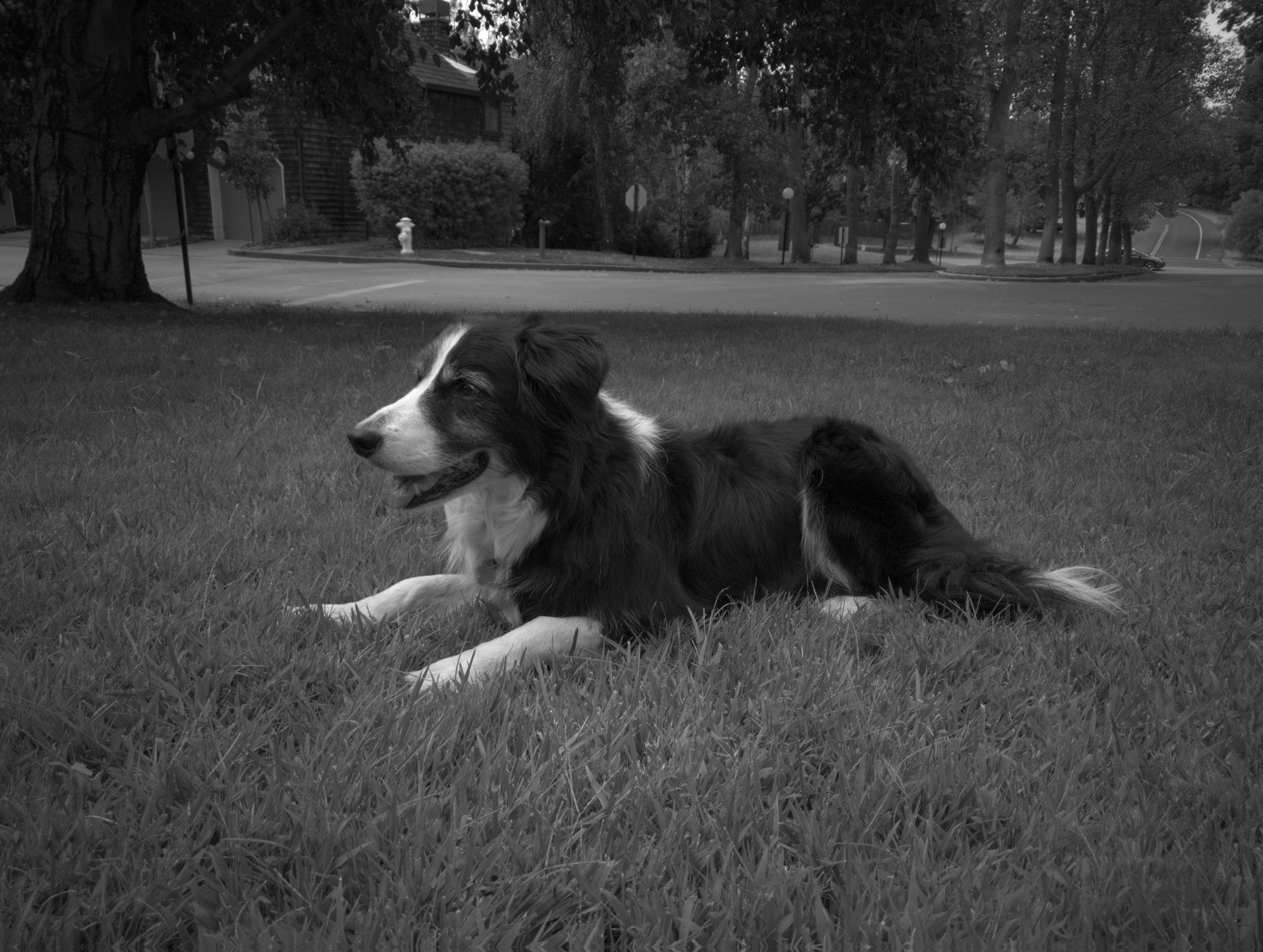}
					\end{center}
				\end{subfigure}
				\begin{subfigure}[t]{.98\linewidth}
					\begin{center}
						\includegraphics[width=\linewidth]{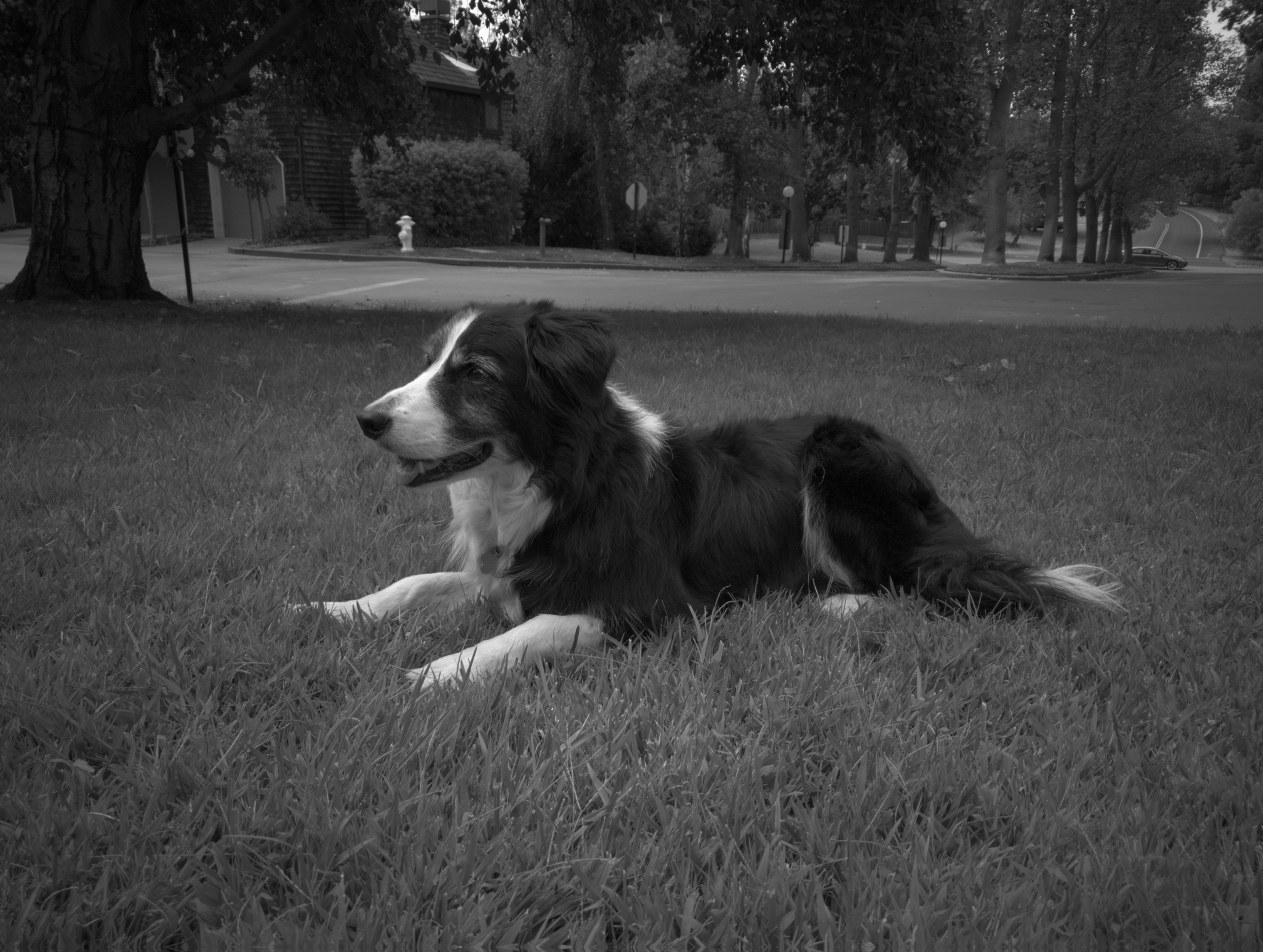}
					\end{center}
				\end{subfigure}
				\caption{Image pair (ref=top)}
			\end{center}
		\end{subfigure}%
		\begin{subfigure}[t]{0.66\linewidth}
			\begin{center}
				\begin{subfigure}[t]{.49\linewidth}
					\begin{center}
						\includegraphics[width=\linewidth]{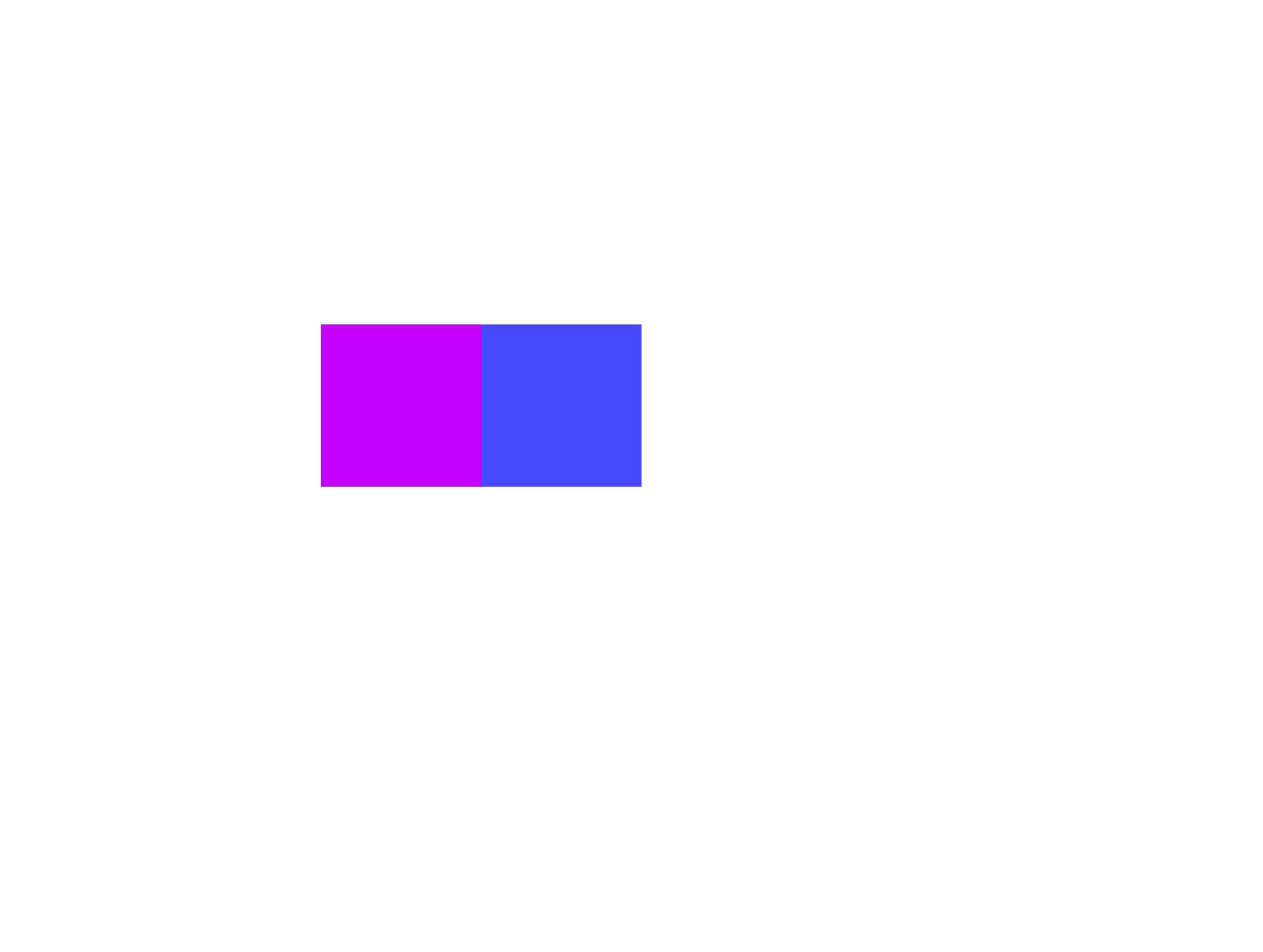}
					\end{center}
				\end{subfigure}
				\begin{subfigure}[t]{.49\linewidth}
					\begin{center}
						\includegraphics[width=\linewidth]{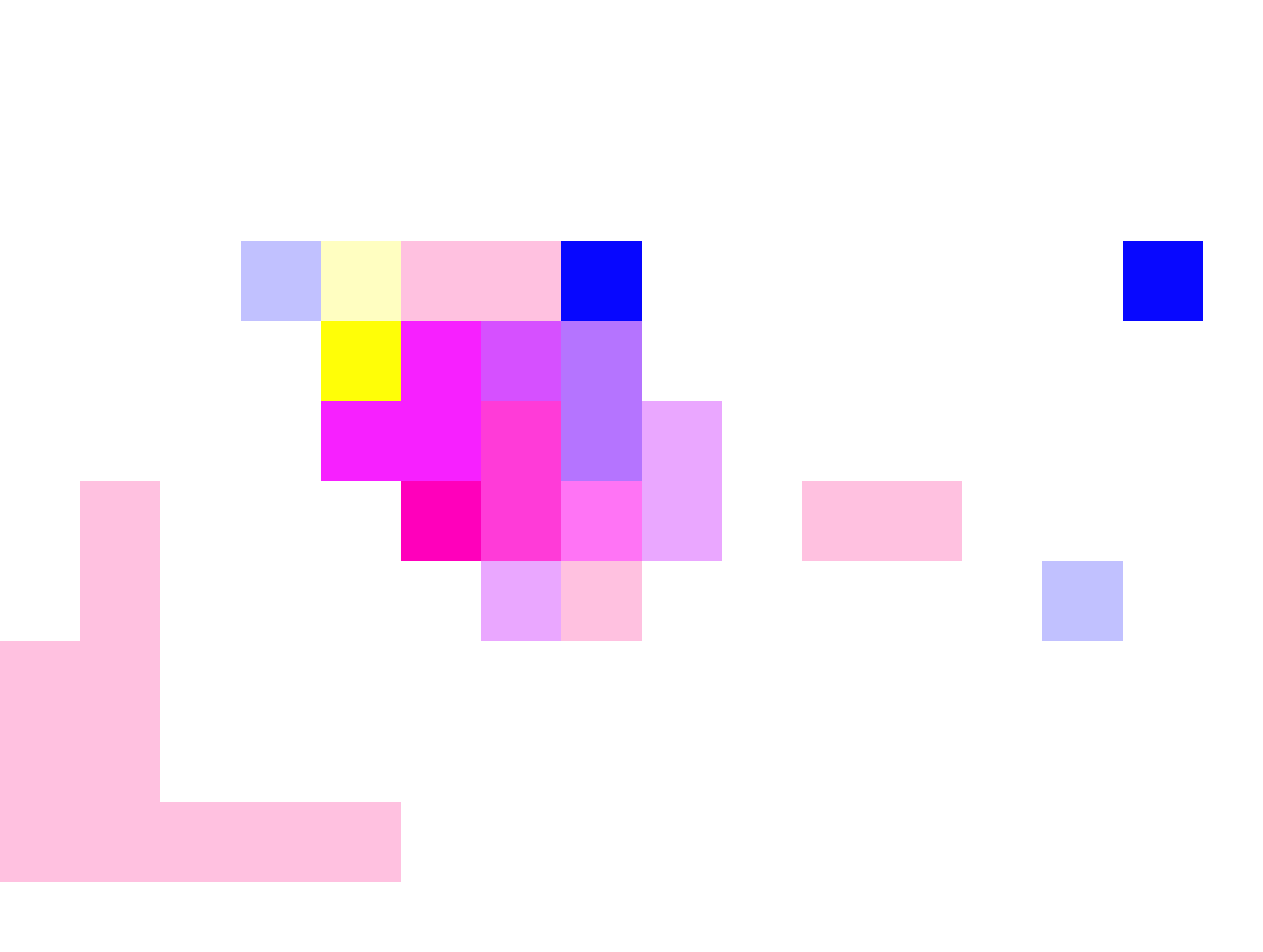}
					\end{center}
				\end{subfigure}
				\begin{subfigure}[t]{.49\linewidth}
					\begin{center}
						\includegraphics[width=\linewidth]{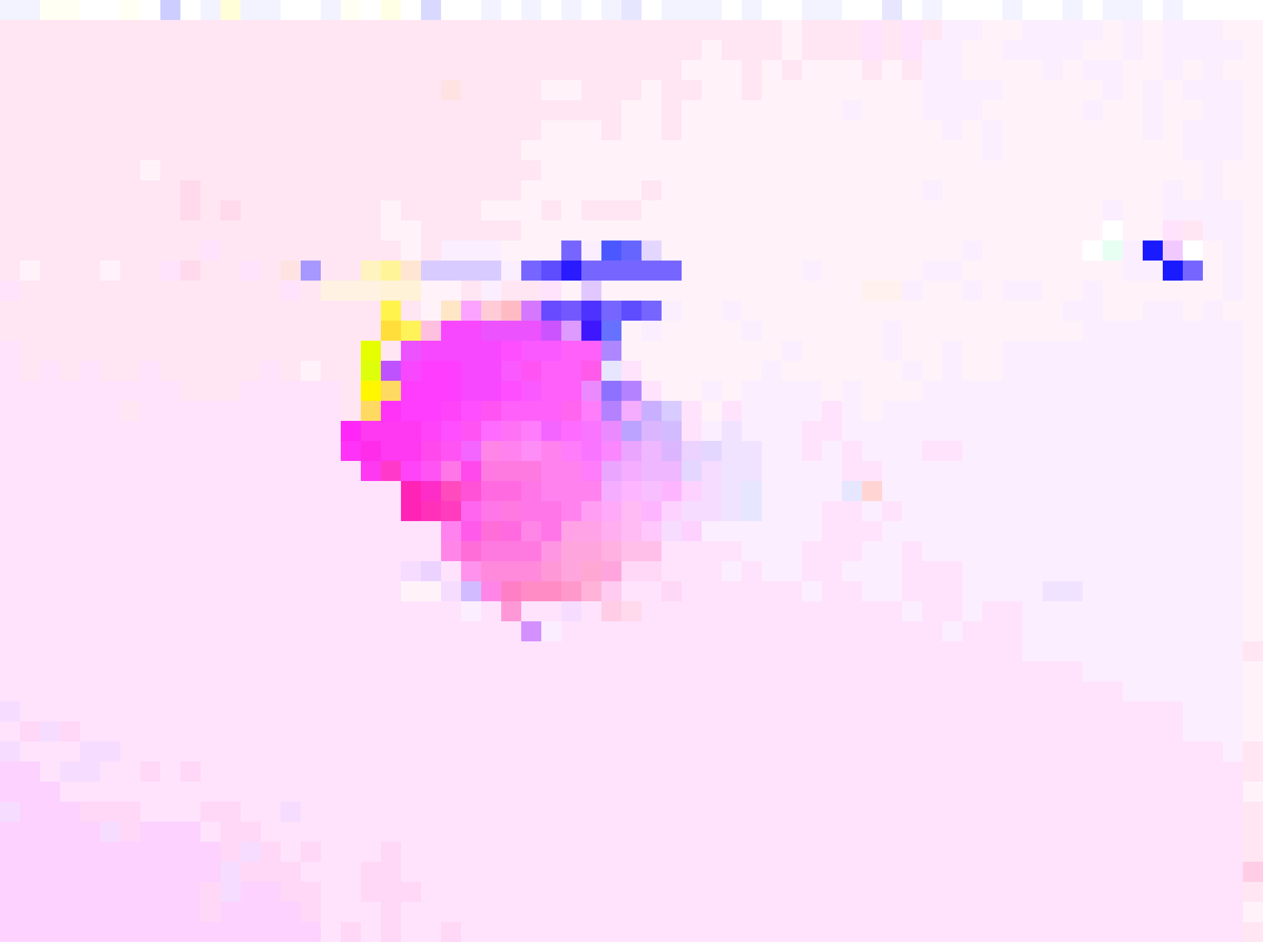}
					\end{center}
				\end{subfigure}
				\begin{subfigure}[t]{.49\linewidth}
					\begin{center}
						\includegraphics[width=\linewidth]{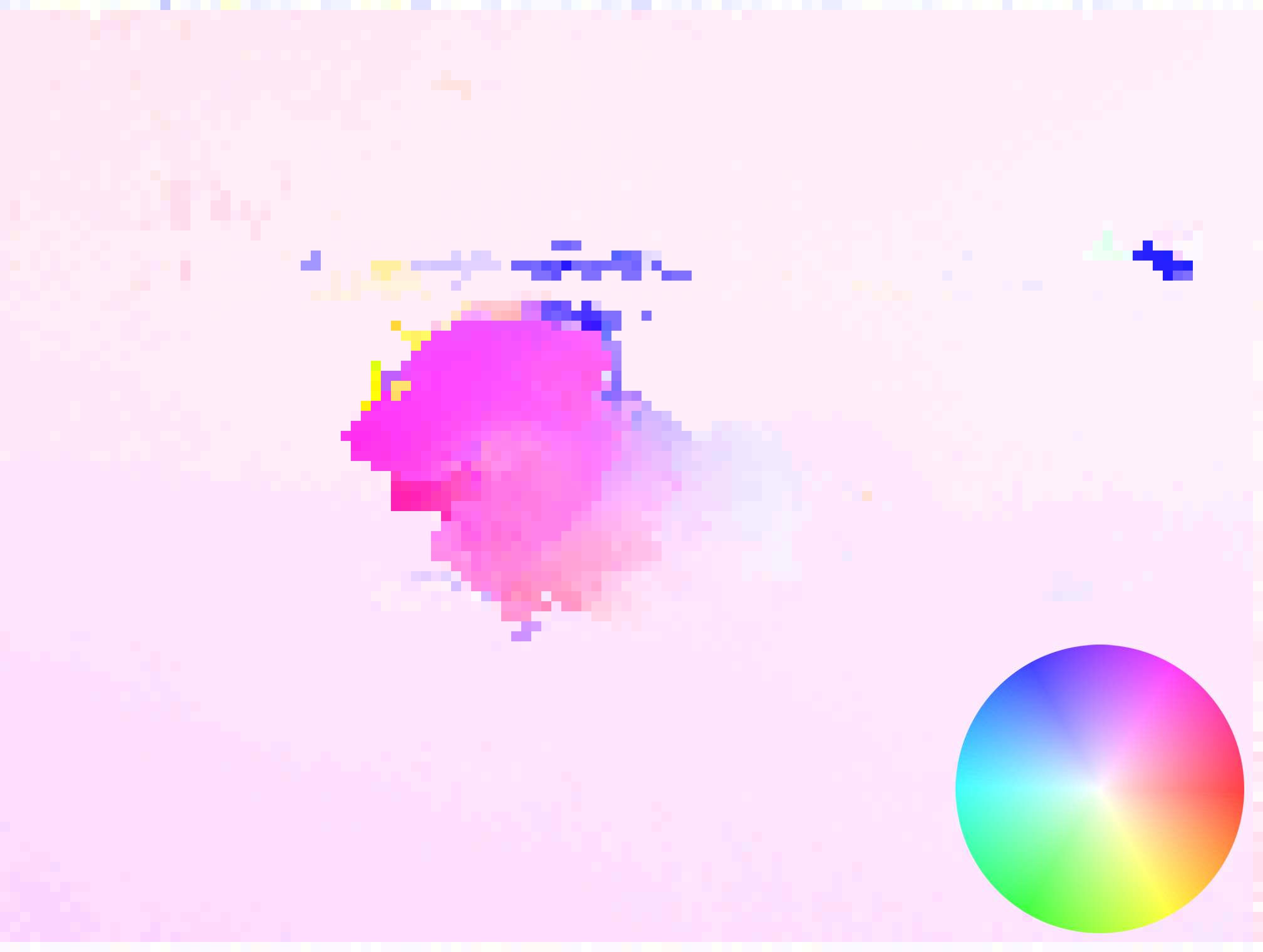}
					\end{center}
				\end{subfigure}
			\caption{Coarse-to-fine motion fields (hue: direction, saturation: magnitude)}
			\end{center}	
		\end{subfigure}
	\caption{Multi-scale tile-based alignment using Gaussian pyramids}
	\label{fig:multiscalealignment}
	\end{center}
\end{figure}

An example of coarse-to-fine motion fields can be seen in Figure~\ref{fig:multiscalealignment}. Given that the alignments are computed for downsampled grayscale versions of the original images, motion vectors must be multiplied by a factor of 2 to go back to the raw Bayer array size.

\subsection{Subpixel Alignment}

We implemented subpixel alignment as defined in the article: between two pyramid scales, for each motion estimation result, we fit a bivariate quadratic polynomial to the $3\times 3$ window surrounding the L2 distance minimum, and find the minimum of the polynomial.

As described in the article supplement\footnote{\url{https://static.googleusercontent.com/media/hdrplusdata.org/en//hdrplus_supp.pdf}}, $D_2$ can be approximated  as
\begin{equation}
	D_2(u,v) \approx \frac{1}{2} \left[u; v\right]^\top \mathbf{A} \left[u; v\right] +\mathbf{b}^\top\left[u; v\right] + c,
\end{equation}
where $\mathbf{A}$ is a $2\times 2$ positive semi-definite matrix (because the shape of $D_2$ is assumed to be an upward-facing quadratic surface near the minimum), $\mathbf{b}$ is a $2 \times 1$ vector and $c$ is a scalar.

If $(\hat{u}, \hat{v})$ is the estimated distance minimum, the technique consists in constructing a weighted least-squares problem fitting a quadratic polynomial to the $3 \times 3$ window of $D_2$ centered around $(\hat{u}, \hat{v})$, which we call $D_{2}^{sub}$. Without loss of generality, we can solve the problem assuming $(\hat{u}, \hat{v}) = (0,0)$ and then shift the estimated subpixel position $\mu$ by $(\hat{u}, \hat{v})$.

The free parameters of the quadratic can be estimated by computing the inner product of $D_{2}^{sub}$ with a set of six $3 \times 3$ filters, each corresponding to an unknown parameter in $(\mathbf{A}, \mathbf{b}, c)$
\begin{equation}
	\begin{aligned}
		\mathbf{A} &=\left[\begin{array}{lll}
			F_{A_{1,1}} \cdot D_{2}^{s u b} & F_{A_{1,2}} \cdot D_{2}^{s u b} \\
			F_{A_{1,2}} \cdot D_{2}^{s u b} & F_{A_{2,2}} \cdot D_{2}^{s u b}
		\end{array}\right], \\
		\mathbf{b} &=\left[\begin{array}{ll}
			F_{b_{1}} \cdot D_{2}^{s u b} \\
			F_{b_{2}} \cdot D_{2}^{s u b}
		\end{array}\right], & \\
		c &=F_{c} \cdot D_{2}^{s u b}.
	\end{aligned}
\end{equation}
In some cases, additional operations might be required so that $\mathbf{A}$ is guaranteed positive semi-definite. Please refer to the supplement for demonstrations, including the derivation of the filters.

Once the quadratic approximation parameters are estimated, we can compute the minimum of the fitted surface
\begin{equation}
	\mathbf{\mu}=-\mathbf{A}^{-1} \mathbf{b},
\end{equation}
which, in the case of a bivariate polynomial is equivalent to
\begin{equation}
	\mu=-\frac{\left[\mathbf{A}_{2,2} \mathbf{b}_{1}-\mathbf{A}_{1,2} \mathbf{b}_{2}, \quad \mathbf{A}_{1,1} \mathbf{b}_{2}-\mathbf{A}_{1,2} \mathbf{b}_{1}\right]^{\mathrm{T}}}{\mathbf{A}_{1,1} \mathbf{A}_{2,2}-\mathbf{A}_{1,2}^{2}}.
\end{equation}

The location of the minimum of the quadratic $\mu$ yields a subpixel displacement vector (because we assumed $(\hat{u}, \hat{v}) = (0,0)$ to solve the problem, when in reality $(\hat{u}, \hat{v})$ can be at any position). We simply need to add this displacement to the original location of the pixel level minimum. This procedure is quite vulnerable to image noise and flat regions, so we only take that displacement into account if it is less than a pixel away from the original distance minimum location: if $||\mathbf{\mu}|| \leq 1$, we update the estimated motion vector $(\hat{u}, \hat{v})$ to $(\hat{u}, \hat{v}) + \mathbf{\mu}$.

Even though this procedure is supposed to help with the upsampling of motion vectors from one level of the pyramid to the next, we have found very few visual improvements to the overall result in our experiments. Subpixel alignment is not applied at the end of the finest pyramid level: because we're dealing with downsampled Bayer to grayscale images, having subpixel motion vectors at the last scale could imply having pixels aligned to pixels from different Bayer CFA channels, which would induce significant color shifts (which we verified in our experiments).

The pseudo-code of the overall alignment strategy of the HDR+ pipeline is described in Algorithm~\ref{alg:alignment}.

\begin{algorithm}[!htb]
	\caption{HDR+ tile-based alignment}
	\DontPrintSemicolon
	\Input{$\left\{I_0,\ldots,I_{N-1}\right\}$ burst of $N$ downsampled grayscale images \Comment*{$I_0$=reference image}}
	\Input{$\left\{p_0,\ldots,p_{3}\right\} \in \left\{1, 2\right\}^{4}$ norm power at each pyramid level \Comment*{Typically $\left\{2,2,2,1\right\}$}}
	\Output{Sets of aligned tiles at reference image location $(i,j)$ $\left\{T_0(i,j), \ldots, T_{N-1}(i,j)\right\}$}
	\;
	\ForEach{$k\in \left\{0,\ldots,N-1\right\}$}{
		Compute $G_k^l$, $l\in \left\{0,\ldots,3\right\}$ \Comment*{4-level coarse-to-fine Gaussian pyramid of $I_k$, $G_k^3 = I_k$.}
	}
	\ForEach{$k\in \left\{1,\ldots,N-1\right\}$}{
		$(u_{-1}, v_{-1})(i,j) \leftarrow (0,0)$ $\forall i,j$ \Comment*{Initial guess at coarse scale: no offset.}
		\ForEach{pyramid level $l \in \left\{0,\ldots,3\right\}$}{
			Divide $G_0^l$ in equally spaced tiles $T_0^l$ of size $n_l\times n_l$\;
			\If{$l > 0$}{
				Upsample the previous level alignments $(u_{l-1}, v_{l-1})(i,j)$, $\forall i,j$
			}
			\ForEach{reference tile $T_0^l$ at location $(i,j)$}{
				\If{$l > 0$}{
					Update the upsampled $(u_{l-1}, v_{l-1})(i,j)$ by keeping the alignment that minimizes $D_1(u_{l-1}, v_{l-1})$ among those of the 3 nearest coarse-scale tiles
				}
				Get all possible $(u,v)$ locations in $G_k^l$ of tiles of size $n_l\times n_l$ in a search area of size $(n_l+2r_l)\times (n_l+2r_l)$ centered around $(i+u_{l-1}, j+v_{l-1})$ \Comment*{$r_l$: search radius.}
				Compute all possible distances $D_{p_l}(u,v)$, $p_l \in \left\{1,2\right\}$ \Comment*{Equation (\ref{eq:p_distance}).}
				$(u_{l}, v_{l})(i,j) \leftarrow (u_{l-1}, v_{l-1})(i,j) + \argmin_{u,v}D_{p_l}(u,v)$ \Comment*{Add computed displacement to initial guess.}
				\If{$l < 3$}{
					Compute subpixel displacement vector $\mathbf{\mu}(u_l,v_l)$\;
					\If{$||\mathbf{\mu}|| < 1$}{	$(u_{l}, v_{l})(i,j) \leftarrow (u_{l}, v_{l})(i,j) + \mathbf{\mu}$}
				}
			}
		}
		Associate to the reference tile at finest scale $T_0^3$ at location $(i,j)$ the tile $T_k^3$ of $I_k$ at location $(i+u_3, j+v_3)$
	}
	\label{alg:alignment}
\end{algorithm}

\section{Fourier Tile-based Merging} \label{merging}

For a given reference tile, we now have a set of corresponding tiles (one per alternate frame according to the results of the alignment step). If these tiles feature the same image content, i.e.\ if motion (or lack thereof) is correctly estimated, intensity differences between them can only be attributed to noise, and we can merge these tiles to obtain a single, temporally denoised tile.\\
This is of course an ideal case, and in practice, tiles can also differ because of errors in the motion estimation. The method presented in Section~\ref{alignment} has several limitations that can lead to alignment errors:
\begin{itemize}
	\item The alignment is tile-based, so pixels that belong to different objects can be part of the same tile and therefore be attributed the same motion.
	\item Motion is estimated locally as the most likely translation of a tile of fixed size in the image plane, according to a minimal L1 or L2 norm. This means that any object motion that is not parallel to the image plane is not guaranteed to be correctly approximated.
	\item Depending on the selected parameters (successive downsampling factors, tile sizes, search radii and norm types) versus the actual motion, motion estimation could be stuck in a local mininum. That said, since most real world scenes will feature motion that is compatible with the short exposure time of the burst, a good choice of parameters will mitigate that problem.
	\item The motion estimation is sensitive to noise: the higher the noise level with respect to the image content, the higher the likelihood of having an incorrect tile minimizing the distance.
	\item As with many motion estimation techniques, it does not bode well with occlusions: areas that are not occluded in the reference frame but are in the alternate frame will probably yield incorrect motion vectors.
\end{itemize}
Such errors are expected given the relatively simple nature of the alignment method. Keep in mind that it is designed to run quickly on systems like smartphone systems on a chip (SoCs). Moreover, this fast, imperfect motion estimation can be sufficient for subsequent processing steps, provided the merging step takes those potential alignment errors into account. In this section, we will discuss the strategy employed by the authors of~\cite{hasinoff2016burst} to perform temporal denoising while being robust to alignment errors.

\subsection{Noise Level Estimation} \label{noiseestimation}

For a set of tiles, the first prerequisite of the merging step is an estimation of the noise level. Since the noise profile is considered identical for all the images of the burst, we perform that estimation on the reference image.

In the HDR+ pipeline, raw image noise follows the Poissonian-Gaussian model described in~\cite{foi2008practical}, where it is composed of two mutually independent parts: a Poissonian signal-dependent component and a Gaussian signal-independent component.
Given the properties of Poissonian and Gaussian distributions, the noise variance $\sigma^2$ can simply be expressed as an affine function of the signal level $x$:
\begin{equation}
	\label{eq:noisemodel}
	\sigma^2(x) = \lambda_s x + \lambda_r
\end{equation}
where the parameter $\lambda_s$ can be tied to shot noise, and $\lambda_r$ can be tied to read noise~\cite{mildenhall2018burst}.

There are multiple ways to obtain these two noise curve parameters:
\begin{itemize}
	\item One can take one or several pictures of a scene that features multiple uniform (typically gray) areas and perform mean and variance measurements of intensities~\cite{foi2008practical, healey1994radiometric}. $\lambda_s$ and $\lambda_r$ can then be computed from simple linear regression.
	\item In~\cite{hasinoff2016burst}, the $(\lambda_s, \lambda_r)$ tuple is claimed to only depend on the analog and digital gain settings selected at capture time. Since the authors implemented their pipeline on a specific set of smartphones and sensors, they could likely use a look-up-table that gives them the appropriate per-camera $(\lambda_s, \lambda_r)$ tuple as a function of the applied gain settings. In some raw DNG files, $\lambda_s$ and $\lambda_r$ can be directly found in the \texttt{NoiseProfile} DNG tag\footnote{\url{https://wwwimages2.adobe.com/content/dam/acom/en/products/photoshop/pdfs/dng_spec_1.4.0.0.pdf}}.
	\item The \texttt{NoiseProfile} tag is not present in all DNG files, and separate analog and digital gain values are not directly accessible in image metadata. A single value is usually specified: ISO, which combines analog and digital gain differently from one manufacturer to another. This makes it difficult to extract noise curve parameters \textit{a posteriori}, especially given that different sensors can have different shot versus read noise curves~\cite{mildenhall2018burst}. That said, even though they might not correspond to the actual values of $\lambda_s$ and $\lambda_r$ of the image, we found that computing a $(\lambda_s, \lambda_r)$ tuple from the image ISO and baseline values of $\lambda_s$ and $\lambda_r$ for an image at ISO=100 produces equally pleasing results for the rest of the merging step. We use the following formula:
	\begin{equation} \label{eq:noiseparams}
		\begin{split}
			\sigma^2\left(\alpha_{\mathrm{ISO}}x\right)	& = \left( \alpha_{\mathrm{ISO}} \right)^2 \sigma^2(x) \\
			& = \left( \alpha_{\mathrm{ISO}} \right)^2 \left(\lambda_{s_{\mathrm{ISO}_{100}}} x + \lambda_{r_{\mathrm{ISO}_{100}}}\right) \\
			& = \alpha_{\mathrm{ISO}} \lambda_{s_{\mathrm{ISO}_{100}}} \alpha_{\mathrm{ISO}} x + \left(\alpha_{\mathrm{ISO}}\right)^2 \lambda_{r_{\mathrm{ISO}_{100}}}
			= \lambda^{'}_{s} x^{'} + \lambda^{'}_{r},
		\end{split}
	\end{equation}
	where $\alpha_{\mathrm{ISO}}=\frac{\mathrm{ISO}}{\mathrm{ISO}_{100}}$ is the ratio between the ISO setting used at capture time and $\mathrm{ISO}_{100}$ (considering that $\mathrm{ISO}_{100} = 100$ is the baseline ISO where no gain is applied), $x^{'} = \alpha_{\mathrm{ISO}} x$ is the image we're actually observing, and $(\lambda_{s_{\mathrm{ISO}_{100}}}, \lambda_{r_{\mathrm{ISO}_{100}}})$ are the baseline noise curve parameters at $\mathrm{ISO}_{100}$ (these could be known for a given camera, averaged from a certain amount of ISO-normalized images, or arbitrary).
	\item Another solution is to use an off-the-shelf noise curve estimation algorithm directly on the input image. These algorithms usually revolve around finding homogeneous regions within the image and computing the variance and means of said regions~\cite{colom2013analysis, ivhc2016}. Using a noise curve estimation algorithm would effectively allow the HDR+ burst denoising algorithm to run blind, as all necessary information is stored in the raw Bayer data. That said, extra care should be taken to verify the robustness of these algorithms and the consistency of their outputs. In our experiments, we found that sticking to image metadata or computing $(\lambda_s, \lambda_r)$ from ISO and baseline values produced more consistent results, all while requiring less computational time and fewer external dependencies.
\end{itemize}

% from KPN paper, A is in the range [0.0001: 0.01], B is in [1e-6, 1e-3]
In order to avoid potential problems due to the undersampled nature of the red and blue channels of a Bayer pattern, the noise estimation and the remainder of the merging method are performed separately for each color plane (the two green channels are also treated independently). Since we computed pixel-level alignment for tiles of downsampled grayscale images, we can actually assign the same tile size and motion vectors to each individual color plane.

For computational efficiency, even though the noise is claimed to follow a signal-dependent model across the reference image, the noise variance is actually considered signal-independent within each reference image tile. This means that for a given tile $T$, a single intensity value $\rho$ is used to evaluate the noise model. Instead of taking the average of all intensities within the tile, the authors of~\cite{hasinoff2016burst} decided to consider the root-mean-square (RMS) of the tile
\begin{equation}
	\label{eq:RMS}
	\rho(T) = \mathrm{RMS}(T) = \left( \frac{1}{n^{2}}\sum_{i=0}^{n-1} \sum_{j=0}^{n-1} T(i, j)^{2}\right)^{\frac{1}{2}}.
\end{equation}
The reasoning being that given two tiles of the same average, one having a higher contrast than the other, the higher contrast tile will have a higher root-mean-square (and thus a higher estimated noise variance), which will allow a more aggressive temporal denoising.

\subsection{Pairwise Wiener Temporal Denoising} \label{sub:pairwisedenoising}

Given a set of tiles (the reference tile and the alternate tiles selected by motion estimation), each alternate tile $T_z$ is compared to the reference $T_0$. The difference between the two tiles $D = T_0 - T_z$ is computed, and that difference is compared to the estimated noise variance $\sigma^2(\rho(T_0))$. That comparison is then used to compute a weighted average of $T_z$ and $T_0$ (if the difference between the two tiles is far greater than the estimated noise level, the alternate tile will have a very small weight and we fall back to the reference).

The final temporally denoised result is the average of all pairwise merges. This method is actually performed in the frequency space (i.e.\ via the 2D DFT of tiles), which allows each spatial frequency bin to be treated individually. This can be useful in cases where alignment failure is partial (i.e.\ some frequencies match well while others don't) as not all the content of the tiles will be discarded.

\begin{figure}[!htbp]
\captionsetup[subfigure]{justification=centering}
	\begin{center}
		\begin{subfigure}[t]{.24\linewidth}
			\begin{center}
				\includegraphics[width=\linewidth]{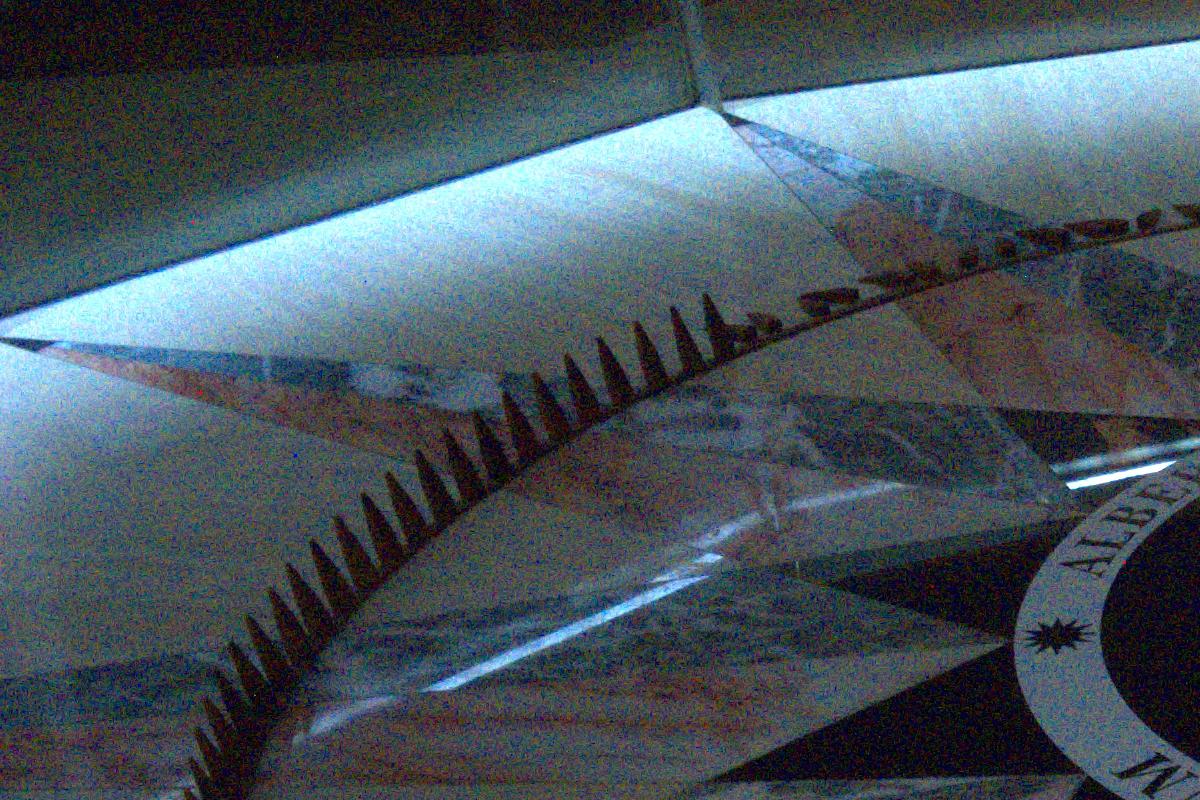}
				\caption{Reference image crop \\($\tau = 0$) }
				\label{fig:c0ref}
			\end{center}
		\end{subfigure}
		\begin{subfigure}[t]{.24\linewidth}
			\begin{center}
				\includegraphics[width=\linewidth]{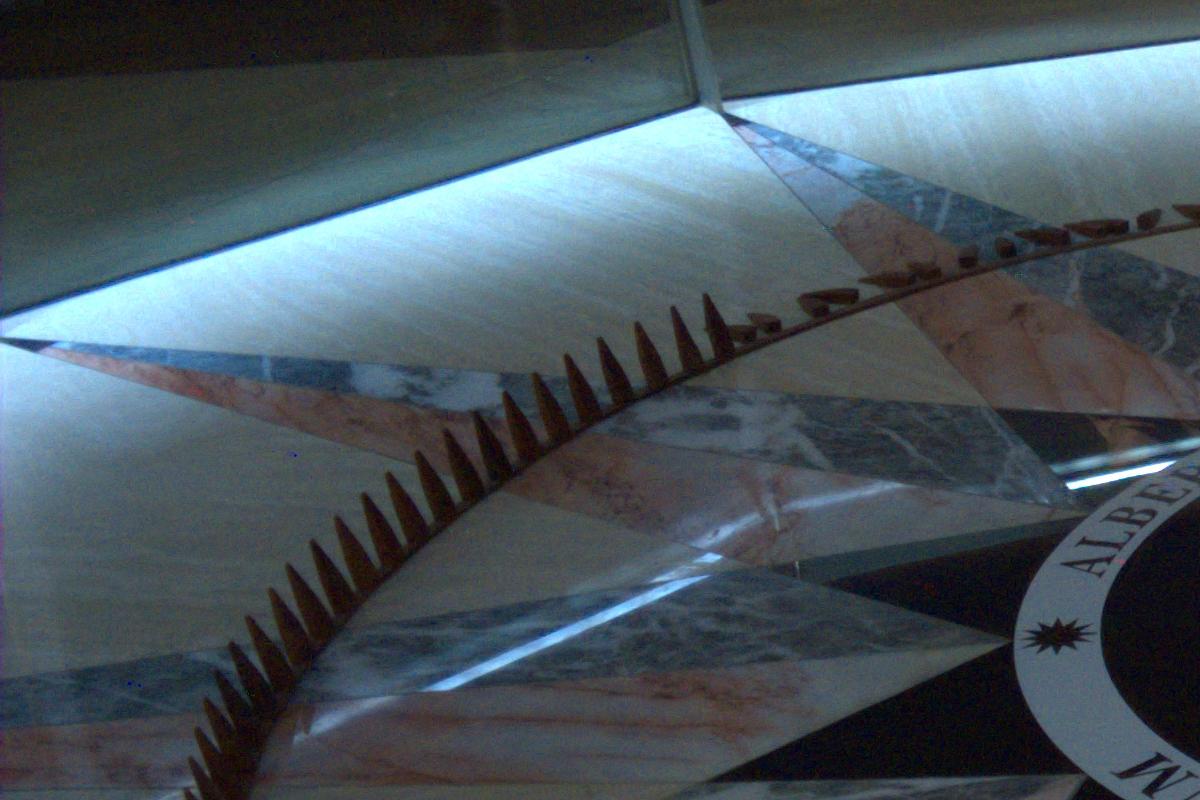}
				\caption{Temporal denoise \\($\tau=75$)}
				\label{fig:coptimal}
			\end{center}
		\end{subfigure}
		\begin{subfigure}[t]{.24\linewidth}
			\begin{center}
				\includegraphics[width=\linewidth]{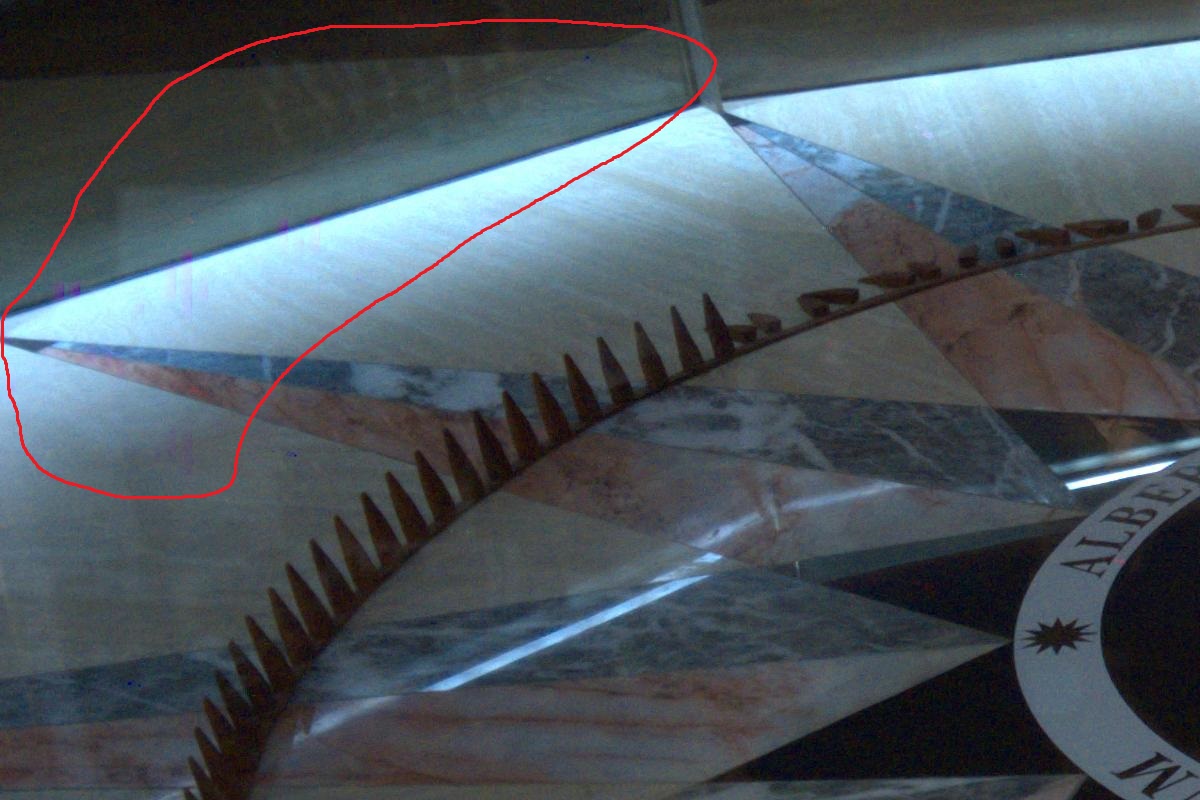}
				\caption{Temporal denoise \\($\tau=+\infty$, ghosting)}
				\label{fig:cinf}
			\end{center}
		\end{subfigure}	
		\begin{subfigure}[t]{.24\linewidth}
			\begin{center}
				\includegraphics[width=\linewidth]{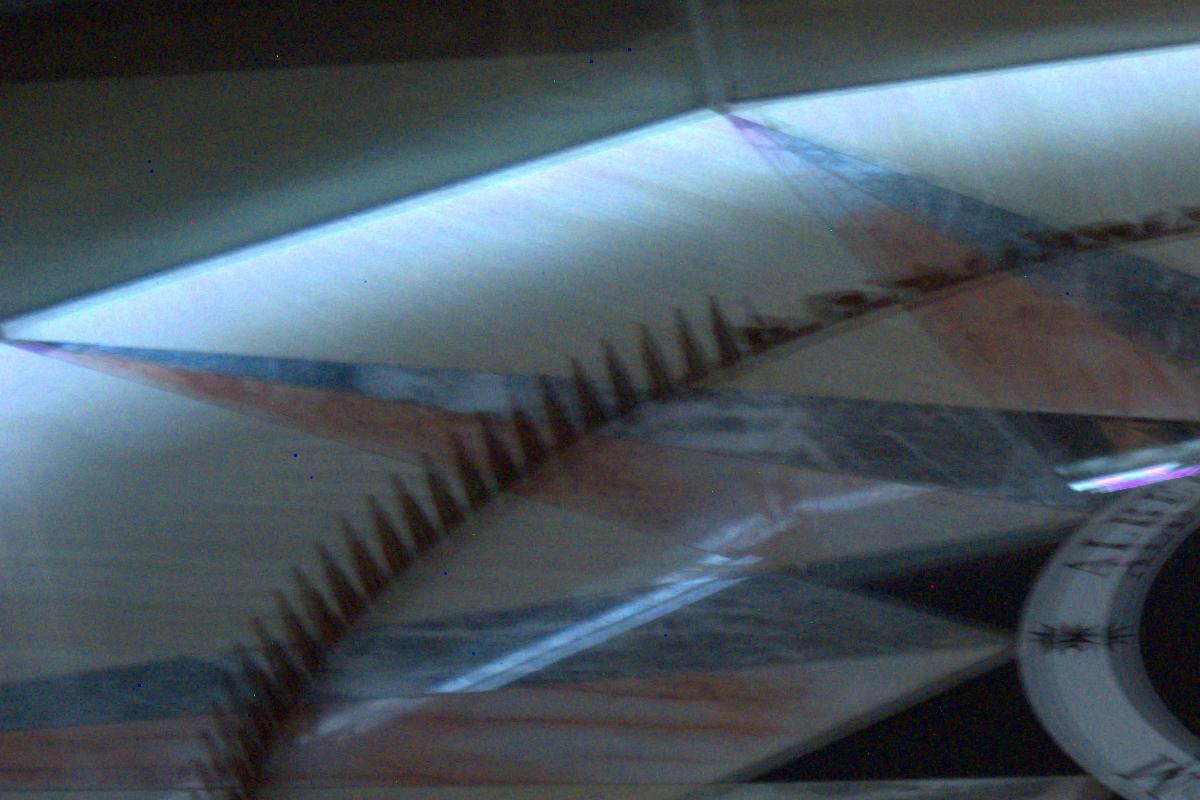}
				\caption{Burst average \\(no alignment)}
				\label{fig:burstaveragenoalignment}
			\end{center}
		\end{subfigure}
		\caption{Impact of the tuning factor $\tau$ on the result of temporal denoising.}
		\label{fig:temporalfactor}
	\end{center}
\end{figure}

Mathematically,
\begin{equation}
	\label{eq:temporaldenoise}
	\tilde{T}_{0}(\mathbf{\omega}) = \frac{1}{N} \sum_{z=0}^{N-1} (1 - A_{z}) T_{z}(\mathbf{\mathbf{\omega}}) + A_{z}(\mathbf{\mathbf{\omega}})T_{0}(\mathbf{\mathbf{\omega}}),
\end{equation}
where the shrinkage operator $A_{z}(\mathbf{\mathbf{\omega}})$ is similar to a Wiener filter
\begin{equation}
	A_{z}(\mathbf{\mathbf{\omega}})=\frac{\left|D_{z}(\mathbf{\omega})\right|^{2}}{\left|D_{z}(\mathbf{\omega})\right|^{2}+ c \sigma^2(\rho(T_0))}, D_z(\mathbf{\omega}) = T_0(w) -T_z(\mathbf{\omega}),
	\label{eq:wienerdenoise}
\end{equation}
where $c$ is defined in the original article as a scaling and tuning factor. For the sake of simplicity in this publication, we will set $c=k \tau$ where $k$ is the scaling factor and $\tau$ is the tuning factor that effectively controls the strength of temporal denoising.
In~\cite{hasinoff2016burst}, $k$ is fixed to $n^{2} \times 1/4^{2} \times 2$, where $n$ is the tile length. However, the justification for this in the article is questionable, because they are trying to scale a single variance value computed in the spatial domain to a per-frequency-bin Wiener filter of the squared difference of two images. Still, we leave the value of $k$ as is, since its influence can be superseded by simply changing the values of $\tau$. The pseudo-code of temporal denoising is described in Algorithm~\ref{alg:merging}.
\begin{remark}
	This 2D frequency-based denoising technique bears similarities with the burst deblurring method of Fourier Burst Accumulation~\cite{fbaIEEE,fbaIPOL}. However, the comparison of 2D frequency bins in HDR+ serves a different purpose: if $T_0$ is sharp, and it is still matched to a blurry tile $T_z$ after alignment, the frequency content corresponding to the blurry elements ot $T_z$ will be discarded, but only because it does not match that of $T_0$. On the contrary, if we have a blurry $T_0$, but a sharp $T_z$, there is also a strong frequency mismatch and we will not merge that content either. In effect, the Wiener filter does not introduce additional blur from other images, but it does not remove blur either.
\end{remark}
If $\tau\rightarrow0$, $\tilde{T}_{0}(\mathbf{\omega}) \rightarrow T_{0}(\mathbf{\omega})$: we stick to the reference frame and the result is not temporally denoised. If $\tau\rightarrow+\infty$, $\tilde{T}_{0}(\mathbf{\omega}) \rightarrow \frac{1}{N} \sum_{z=0}^{N-1} T_{z}(\mathbf{\omega})$: the result is equivalent to an average of all aligned frames in the spatial domain, which can showcase alignment errors. These edges cases can be observed in Figure~\ref{fig:temporalfactor}.

\begin{algorithm}[!htb]
	\caption{HDR+ per-channel, tile-based, pairwise temporal denoising}
	\DontPrintSemicolon
	\Input{Set of aligned tiles $\left\{T_0,\ldots, T_{N-1}\right\}$ of size $n \times n$ at reference image location $(i,j)$ (all tiles correspond to a single channel of the Bayer image)}
	\Input{Noise curve parameters $\lambda_s, \lambda_r$, temporal denoising tuning factor $c$}
	\Output{Single denoised tile $\tilde{T}_{0}$ at reference image location $(i,j)$}
	$\rho \leftarrow \mathrm{RMS}(T_0) = \left( \frac{1}{n^{2}}\sum_{i=0}^{n-1} \sum_{j=0}^{n-1} T_0(i, j)^{2}\right)^{\frac{1}{2}}$ \Comment*{Compute the root-mean-square of $T_0$.}
	$\sigma^2 \leftarrow \lambda_s \rho + \lambda_r$ \Comment*{Compute the noise variance assigned to $T_0$.}
	
	\ForEach{$z\in \left\{0,\ldots,N-1\right\}$}{
		$T_z(\mathbf{\omega}) \leftarrow \mathrm{FFT}(T_z)(\mathbf{\omega})$ \Comment*{DFT of $T_z(x,y)$.}
	}
	$\tilde{T}_{0}(\mathbf{\omega}) \leftarrow T_0(\mathbf{\omega})$ $\forall \mathbf{\omega}$ \;
	\ForEach{$z\in \left\{1,\ldots,N-1\right\}$}{
		\ForEach{frequency bin $\mathbf{\omega} \in \left\{0,\ldots,n-1\right\}^2$}{
		$D_z(\mathbf{\omega}) \leftarrow T_0(\omega) -T_z(\mathbf{\omega})$\;
		$A_{z}(\mathbf{\mathbf{\omega}}) \leftarrow \frac{\left|D_{z}(\mathbf{\omega})\right|^{2}}{\left|D_{z}(\mathbf{\omega})\right|^{2}+c \sigma^{2}}$\;
		$\tilde{T}_{0}(\mathbf{\omega}) \leftarrow \tilde{T}_{0}(\mathbf{\omega}) + (1 - A_{z}) T_{z}(\mathbf{\mathbf{\omega}}) + A_{z}(\mathbf{\mathbf{\omega}})T_{0}(\mathbf{\mathbf{\omega}})$ \Comment*{Sum pairwise temporal denoisings.}
		}
	}
	$\tilde{T}_{0}(\mathbf{\omega}) \leftarrow \frac{1}{N}\tilde{T}_{0}(\mathbf{\omega})$ \Comment*{Average pairwise temporal denoising.}
	$\tilde{T}_{0}(i,j) \leftarrow \mathrm{FFT}^{-1}(\tilde{T}_{0})(i,j)$ \Comment*{Inverse DFT of $\tilde{T}_{0}(\mathbf{\omega})$.}
	\label{alg:merging}
\end{algorithm}

\subsection{Wiener Spatial Denoising} \label{sub:spatialdenoising}

In addition to temporal filtering, spatial filtering is performed on the 2D DFT of tiles in order to remove some of the residual noise. Since we expect the noise level to be significantly reduced after temporal denoising, the estimated noise variance is updated to $\sigma^2(\rho(T_0))/N$ (this implies a perfect averaging of all N frames, which is not the case in practice but allows for a more conservative spatial denoising).
We compute a shrinkage operator similar to the one used for temporal denoising and apply it to the spatial frequency coefficients
\begin{equation}
	\label{eq:spatialdenoising}
	\hat{T}_{0}(\mathbf{\omega}) = \frac{\left|\tilde{T}_{0}(\mathbf{\omega})\right|^{2}}{\left|\tilde{T}_{0}(\mathbf{\omega})\right|^{2}+ f(\mathbf{\omega})\frac{\sigma^2(\rho(T_0))}{N}} \tilde{T}_{0}(\mathbf{\omega}),
\end{equation}
where $f(\mathbf{\omega})$ is a noise shaping function; in~\cite{hasinoff2016burst}, it is a piecewise linear function that increases the ``effective noise level'' for higher spatial frequencies, tuned to subjectively maximize image quality. As we do not know the specifics of said shaping function, we simply replaced it with $f(\mathbf{\omega})=\gamma\left|\mathbf{\omega}\right|$, which also effectively increases the noise level for higher frequencies, and features a scaling and tuning factor $\gamma$. We set $\gamma = \frac{k}{2}s$, where $k$ is the scaling factor defined in Section~\ref{sub:pairwisedenoising} and $s$ controls spatial denoising strength.

\begin{figure}[!htb]
\captionsetup[subfigure]{justification=centering}
	\begin{center}
		\begin{subfigure}[t]{.24\linewidth}
			\begin{center}
				\includegraphics[width=\linewidth]{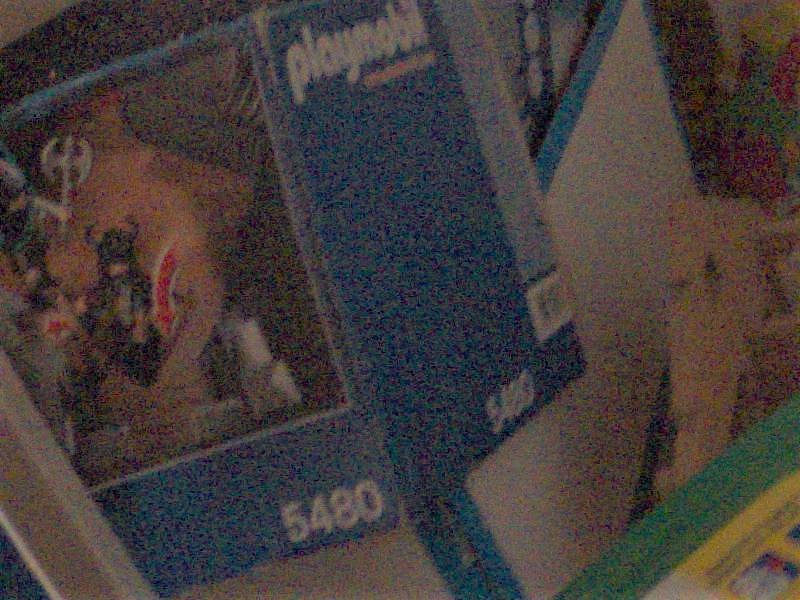}
				\caption{Reference image}
			\end{center}
		\end{subfigure}
		\begin{subfigure}[t]{.24\linewidth}
			\begin{center}
				\includegraphics[width=\linewidth]{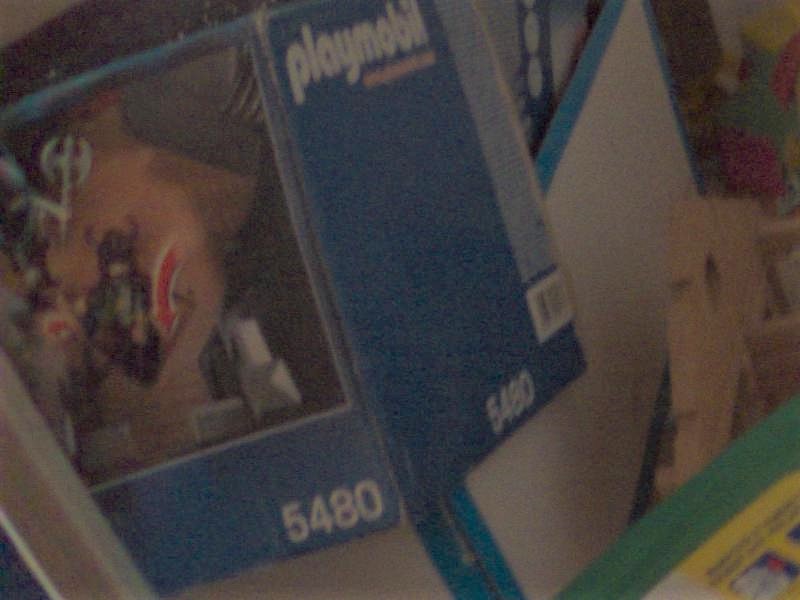}
				\caption{Temporal denoise \\($s=0$)}
			\end{center}
		\end{subfigure}
		\begin{subfigure}[t]{.24\linewidth}
			\begin{center}
				\includegraphics[width=\linewidth]{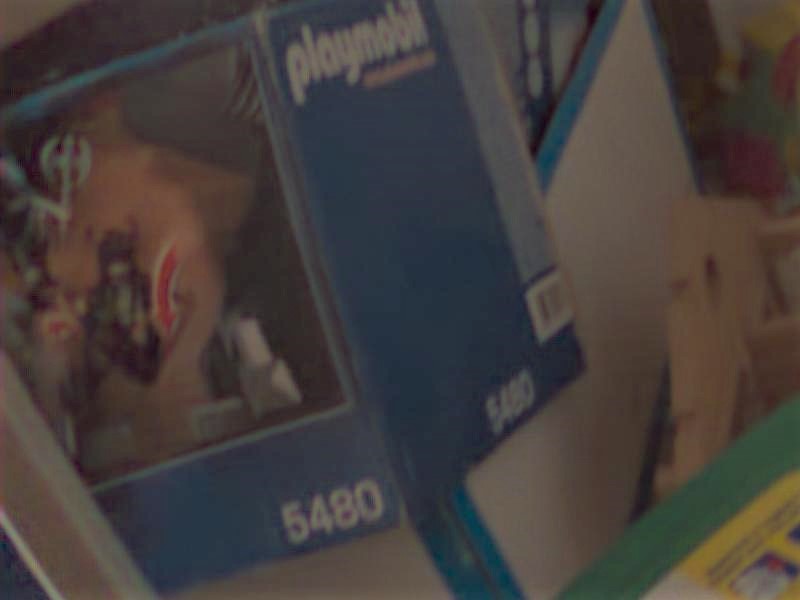}
				\caption{Temporal + spatial denoise ($s=10$)}
			\end{center}
		\end{subfigure}
		\begin{subfigure}[t]{.24\linewidth}
			\begin{center}
				\includegraphics[width=\linewidth]{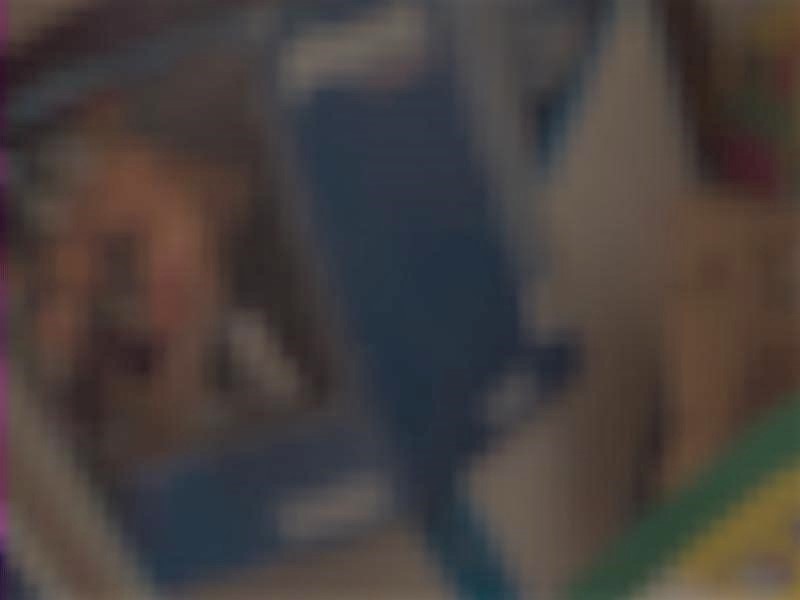}
				\caption{Temporal + spatial denoise ($s=10^{12}$)}
			\end{center}
		\end{subfigure}
		\caption{Impact of the tuning factor $s$ on the result of spatial denoising}
		\label{fig:spatialdenoise}
	\end{center}
\end{figure}

Mathematically, if $s=0$ we do not perform any spatial denoising. The higher the $s$, the more aggressively we filter higher spatial frequencies and obtain a lower-frequency final image (with little to no edges and textures within a tile in extreme cases). We typically set it to 0.1 to remove a bit of high frequency residual noise without loosing too much detail. The influence of $s$ on the spatial denoising can be observed in Figure~\ref{fig:spatialdenoise}.

This frequency-based denoising technique (and particularly our simplified implementation) is not expected to be as efficient as state-of-the-art spatial denoising algorithms~\cite{dabov2009bm3d, lebrun2013nonlocal, zhang2018ffdnet}. However, this step is used as a very computationally light complement to temporal denoising (since we are already in frequency space), which can be tuned to be fairly gentle and not remove image detail.

\subsection{Overlapped Tiles and Raised Cosine Window}

\begin{figure}[!htbp]
	\begin{center}
		\begin{subfigure}[t]{.49\linewidth}
			\begin{center}
				\includegraphics[width=\linewidth]{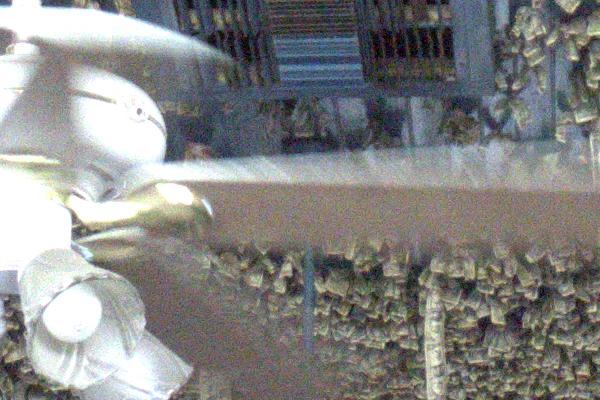}
				\caption{Noisy reference image crop}
			\end{center}
		\end{subfigure}
		\begin{subfigure}[t]{.49\linewidth}
			\begin{center}
				\includegraphics[width=\linewidth]{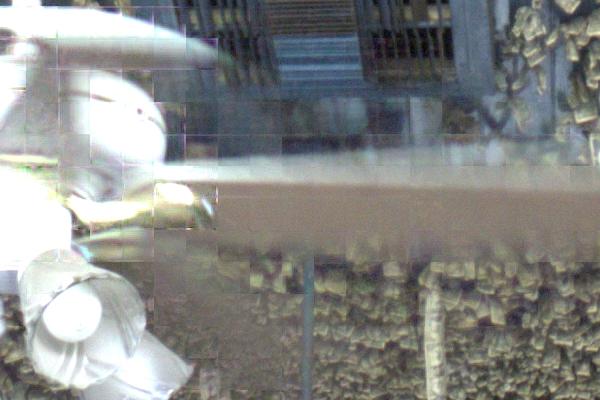}
				\caption{Merged image crop. Discontinuities between tiles and artifacts on tile edges due to DFT can be observed}
				\label{fig:nooverlapnocosine}
			\end{center}
		\end{subfigure}
		\caption{Results obtained with the steps described so far}
		\label{fig:artifactsnooverlap}		
	\end{center}
\end{figure}

The steps described so far are the following: we start by dividing the reference image in equally spaced tiles; we then associate each reference tile to a specific stack of alternate tiles (one for each other image of the burst according to the result of the alignment step); after that, each stack is merged using the strategy described in the previous steps of Section~\ref{merging}. Once all stacks are merged, we obtain a set of denoised tiles of the same number and disposition we had after dividing the reference frame, which we can call tiles of the \textit{merged} image. Unfortunately, these steps are not sufficient to create both a less noisy image and a realistic looking one. Because we merge each stack independently of others, and because different stacks have different image content, a different noise variance estimation, and can feature different alignment errors, continuity between adjacent tiles is not guaranteed. Additional artifacts can be seen on tile edges, because the merging operation is performed in the DFT domain (we compute the FFT of a tile stack, merge it, and then compute the inverse FFT to obtain a merged tile). These issues can be seen in Figure~\ref{fig:artifactsnooverlap}.

In order to create smoother images with less visually noticeable artifacts while retaining image detail, the solution of the authors of~\cite{hasinoff2016burst} is twofold:
\begin{itemize}
	\item \textbf{Tiles} (for both the alignment and merging steps) are \textbf{overlapped by half in each spatial dimension}. This means that the total number of tiles (and thus the number of operations for alignment and merging) is actually multiplied by a factor of 4.
	\item The window used for blending is a \textbf{modified raised cosine window}, defined in 1D as
	\begin{equation}\label{eq:cosine1D}
		w(x) = \frac{1}{2}-\frac{1}{2} \cos \left(2 \pi\left(x+\frac{1}{2}\right) / n\right), 0 \leq x \leq n-1.
	\end{equation}
	If $x$ is a vector of positions from $0$ to $n-1$, this window function is centered, and has nonzero values at $0$ and $n-1$ which means that content on edges is not fully discarded. As can be seen in Figure~\ref{fig:cosine1D}, another interesting property is that when they overlap by half, the sum of two such windows is always equal to 1. When dealing with tiles that overlap by half in two spatial dimensions, the actual window used for blending (visible in Figure~\ref{fig:cosine2D}) is the product of the 1D version in each dimension
	\begin{equation}\label{eq:cosine2D}
		w(i,j) = \left(\frac{1}{2}-\frac{1}{2} \cos \left(2 \pi\left(i+\frac{1}{2}\right) / n\right)\right) \left(\frac{1}{2}-\frac{1}{2} \cos \left(2 \pi\left(j+\frac{1}{2}\right) / n\right)\right), 0 \leq i,j \leq n-1.
	\end{equation}
	\begin{figure}[!htb]
		\begin{subfigure}[b]{.49\linewidth}
			\begin{center}
				\includegraphics[width=\linewidth]{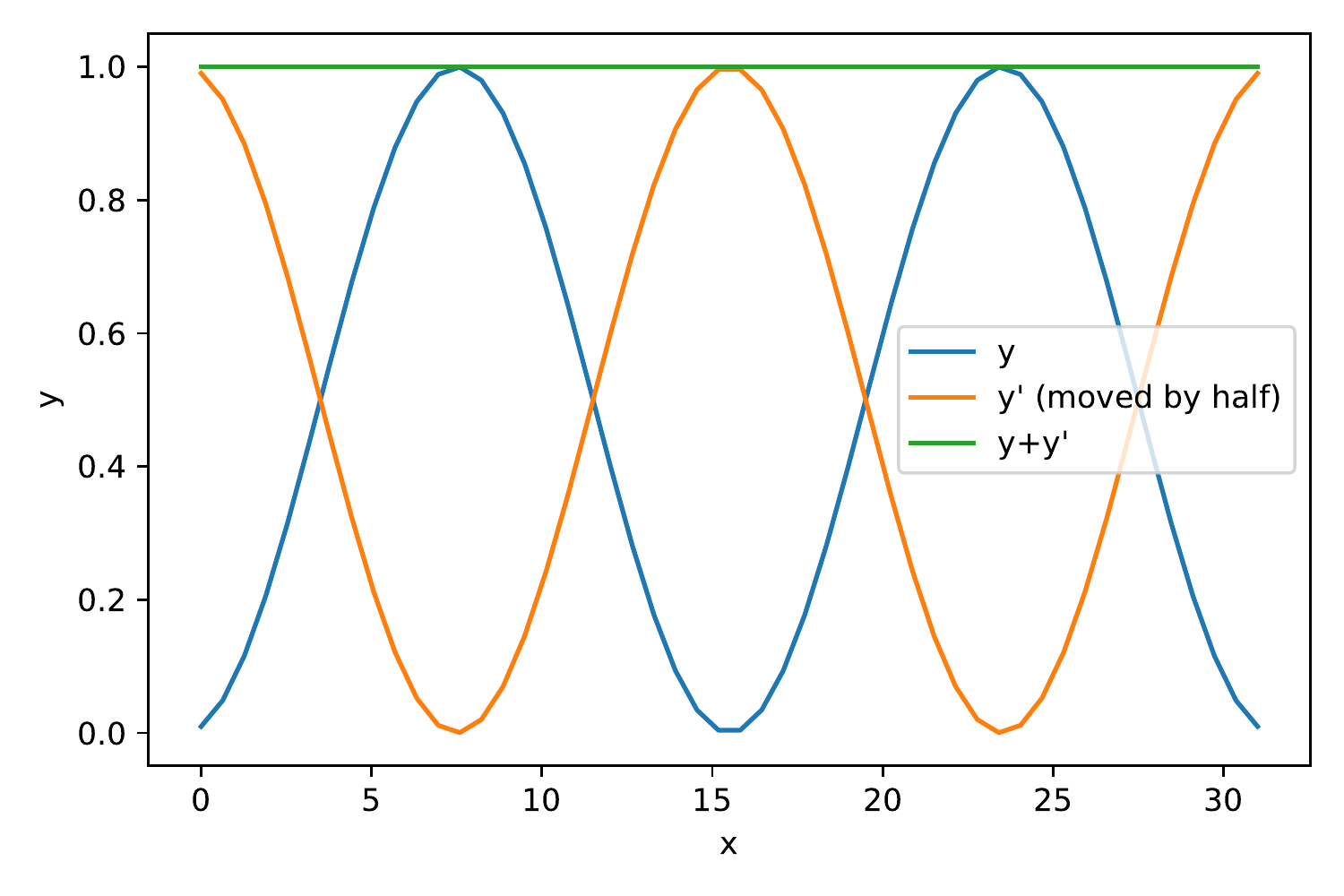}
				\caption{1D window ($n=16$)}
				\label{fig:cosine1D}
			\end{center}
		\end{subfigure}
		\begin{subfigure}[b]{.49\linewidth}
			\begin{center}
				\includegraphics[width=\linewidth]{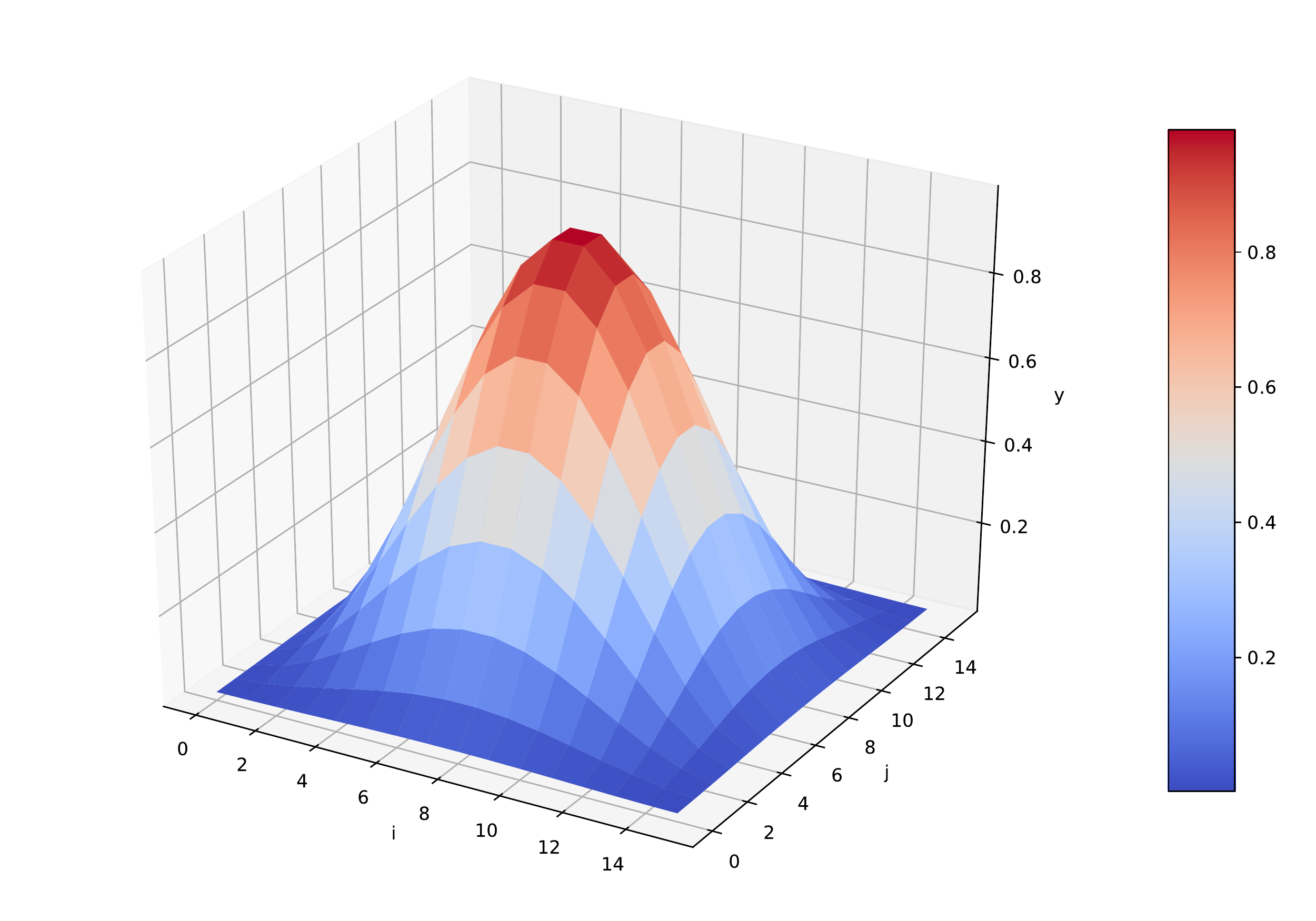}
				\caption{2D window}
				\label{fig:cosine2D}
			\end{center}
		\end{subfigure}
		\caption{Modified raised cosine window.}
		\label{fig:cosinewindow}
	\end{figure}
	This window allows the smooth blending of the overlapped tiles and has the added benefit of removing DFT artifacts on tile edges.
\end{itemize}

\begin{figure}[!htb]
	\begin{center}
		\begin{subfigure}[t]{.49\linewidth}
			\begin{center}
				\includegraphics[width=\linewidth]{./images/merged_no_overlap_no_cosine_cropped.jpg}
				\caption{Merged image crop without overlapped tiles}
				\label{fig:nooverlapnocosine2}
			\end{center}
		\end{subfigure}
		\begin{subfigure}[t]{.49\linewidth}
			\begin{center}
				\includegraphics[width=\linewidth]{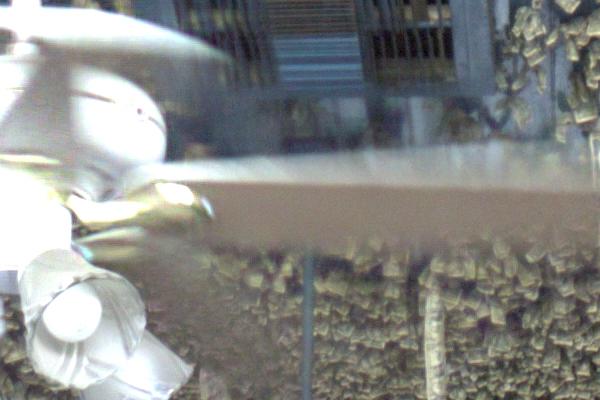}
				\caption{Merged image crop with overlapped tiles + cosine window}
				\label{fig:no_overlap}
			\end{center}
		\end{subfigure}
		\caption{Using overlapped tiles and the modified raised cosine window prevents observable discontinuities and edge artifacts while retaining image detail.}
		\label{fig:artifactsoverlap}
	\end{center}
\end{figure}

Going back to our real burst denoising example, Figure~\ref{fig:artifactsoverlap} showcases that using the aforementioned blending technique does create a more pleasing image without loosing image detail.

\section{Finishing}

After the alignment and merging steps, we obtain a single raw image based on the reference but with significantly less noise. The system described in~\cite{hasinoff2016burst} being a full camera pipeline, the subsequent steps that lead to a final \texttt{.jpg} image are also presented. Even though this article mainly focuses on the temporal denoising part of the HDR+ solution, this section discusses the remainder of the image processing pipeline. 

\subsection{Google's Pipeline} \label{googlepipeline}

The system described in~\cite{hasinoff2016burst} being a full camera pipeline, the subsequent steps that lead to the final image are also presented:

\begin{itemize}
	\item Black level subtraction: saved values of additional sensor pixels that are shielded from the light are subtracted so that image pixels that do not receive any light have a $0$ value.
	\item Lens shading correction: areas on the image corners that are darker are brightened. Low-resolution lens shading maps that are provided by the ISP are used for that compensation.
	\item White balance: colors are shifted so that objects human observers would perceive as gray actually appear as gray in the final image. Gains are applied to each one of the 4 RGGB channels individually, and those gains are provided by the ISP.
	\item Demosaicking: the raw Bayer image where each pixel corresponds to a single color channel is converted to an RGB image of the same size where each pixel has 3 color channels. The raw image being undersampled, an interpolation algorithm is used (the algorithm in the HDR+ pipeline is said to use a combination of techniques featured in~\cite{gunturk2005demosaicking} and is most likely proprietary).
	\item Chroma denoising: even after the temporal and spatial denoising steps, the image can still contain artifacts such as red and green blocks in dark areas, especially in the case of low-light images. To that end, the HDR+ authors apply a sparse $3\times 3$ tap non-linear kernel in two passes to the image in the YUV space, acting as an approximate bilateral filter.
	\item Color correction: a $3\times 3$ matrix supplied by the ISP is used to convert the image from sensor RGB to linear sRGB values.
	\item Dynamic range compression / local tone mapping: aside from burst denoising, the HDR+ pipeline aims at improving the dynamic range of the final image in specific scenarios (hence its name): in scenes that feature a high dynamic range, darker areas of the image must be brightened, while brighter content must remain unsaturated and local contrast must be preserved. To that end, a local tone mapping method derived from the exposure fusion algorithm~\cite{mertens2009exposure, IPOLexposure} is employed. However, since this method requires several images of different brightness (usually several captures at different exposure times), two gamma-corrected ``synthetic exposures'' are created from the intermediate result: one short (a gamma-corrected grayscale version of the current image, which features the exposure time used during the burst capture) and one long exposure (a gamma-corrected grayscale version of the current image with a gain applied according to the auto-exposure algorithm specific to that pipeline). From these two images, the exposure fusion algorithm uses image pyramids to create a smoothly blended image where each pixel is best-exposed. The gamma correction is then inverted and the image re-colorized by keeping the chroma ratios of the intermediate image.
	\item Dehazing: an extra global tone curve that pushes low pixel values even lower while preserving highlights and midtones is applied in order to mitigate the effect of veiling glare.
	\item Global tone adjustment / gamma curve: an S-shaped contrast enhancing tone curve is concatenated to the sRGB color component transfer function to transform the linear sRGB image to a nonlinear sRGB image where contrast is increased.
	\item Chromatic aberration correction: longitudinal and lateral chromatic aberration are corrected by replacing chroma values on high contrast edges by those of nearby pixels that are less likely to be affected by chromatic aberration.
	\item Sharpening: the authors implement unsharp masking\footnote{\url{https://micro.magnet.fsu.edu/primer/java/digitalimaging/processing/unsharpmask/index.html}} through a sum of Gaussian kernels constructed from a 3-level convolution pyramid.
	\item Hue-specific color adjustments: some custom transformations are applied to the image colors (shifting bluish
	cyans and purples towards light blue, and increasing the saturation of blues and greens generally), which is said to make vegetation and blue skies more visually appealing.
	\item Dithering: to mitigate quantization artifacts when reducing from 10/12 to 8 bits per pixel (sRGB images are typically destined for viewing on 8-bit displays), dithering is implemented by adding blue noise from a precomputed table.
	\item JPEG quantization and compression: the image is quantized and encoded to 8 bits where the lossy compression is set to a quality level of 95, resulting in the final \texttt{.jpg} file.
\end{itemize}

All these steps are performed via software (i.e.\ in a smartphone, these operations can run on the CPU/GPU and not necessarily on the ISP).

\subsection{Our Simplified Pipeline} \label{ourpipeline}

Since we decided to mainly focus on the denoising part of the HDR+ algorithm for this article, we implemented a simpler finishing pipeline. It still features parts of Google's own solution, and it is presented as a proof of concept, to showcase some of the decisions they made when designing their pipeline. It does not aim for parity with Google in terms of visual quality, as it would require a considerable amount of additional development time and a lot of trial and error, especially given the relatively sparse description of these steps in the original article.

Using rawpy, a Python wrapper of the LibRaw library\footnote{rawpy: RAW image processing for Python, a wrapper for LibRaw, \url{https://pypi.org/project/rawpy/}, 2014} we transform a temporally and spatially denoised 16 bit Bayer array (even though smartphone raw images typically have a 10 or 12 bit precision, they are often stored on 16 bit arrays) into a 16 bit linear RGB image of the same resolution with the following operations:
\begin{itemize}
	\item \textbf{Black level subtraction}
	\item \textbf{White balance}: rawpy uses the gains stored in the \texttt{As Shot Neutral} Exif metadata tag in the \texttt{.dng} file of the reference image to apply scales to the red and blue Bayer channels.
	\item \textbf{Demosaicking}: the raw Bayer image where each pixel corresponds to a single color channel is converted to a RGB image of the same size where each pixel has 3 color channels. We use the AHD algorithm~\cite{AHDdemosaic} since it was both used for burst fusion comparisons in the HDR+ supplement (where it was deemed ``representative of the algorithms used by mobile ISPs'') and available in rawpy. That algorithm is not the proprietary one used in Google's implementation.
	%DHT produces slightly better results but is extremely poorly documented
	\item \textbf{Color correction}: rawpy uses image metadata such as the \texttt{Color Matrix} Exif tag in the \texttt{.dng} file of the reference image to convert the image from sensor linear RGB to standard linear sRGB color space.
\end{itemize}
Once we're done with rawpy postprocessing, we apply additional custom operations:
\begin{itemize}
	\item \textbf{Hdr tone mapping}: we use a technique close to the one employed in~\cite{hasinoff2016burst}, where two synthetic exposures are created from a single image and combined with exposure fusion~\cite{mertens2009exposure} (a similar strategy is extended and demonstrated in~\cite{simulatedexposure}). From the result of all previous steps (which stems from an underexposed burst), we get a grayscale image by averaging the 3 color channels and then synthesize two exposures: we apply the standard sRGB gamma correction to get a short exposure, and we apply a gain followed by the same gamma correction to synthesize the long exposure. We then perform exposure fusion using the OpenCV \texttt{mergeMertens} implementation\footnote{\url{https://docs.opencv.org/master/d7/dd6/classcv_1_1MergeMertens.html}}. Given that we are blending grayscale images, and that in their implementation, the authors of~\cite{hasinoff2016burst} use ``a fixed weighting function of luma that favors moderately bright pixels'', we found that only taking the ``well-exposedness'' weights of exposure fusion into account for blending (we set the \texttt{exposure\_weight} parameter to 1 and \texttt{contrast\_weight} and \texttt{saturation\_weight} to 0 when calling the \texttt{createMergeMertens} function) produced convincing results (Figure~\ref{fig:hdrtonemap} shows an example of the applied exposure fusion).
	\begin{figure}[!htbp]
		\begin{center}
			\begin{subfigure}[t]{.3\linewidth}
				\begin{center}
					\includegraphics[width=\linewidth]{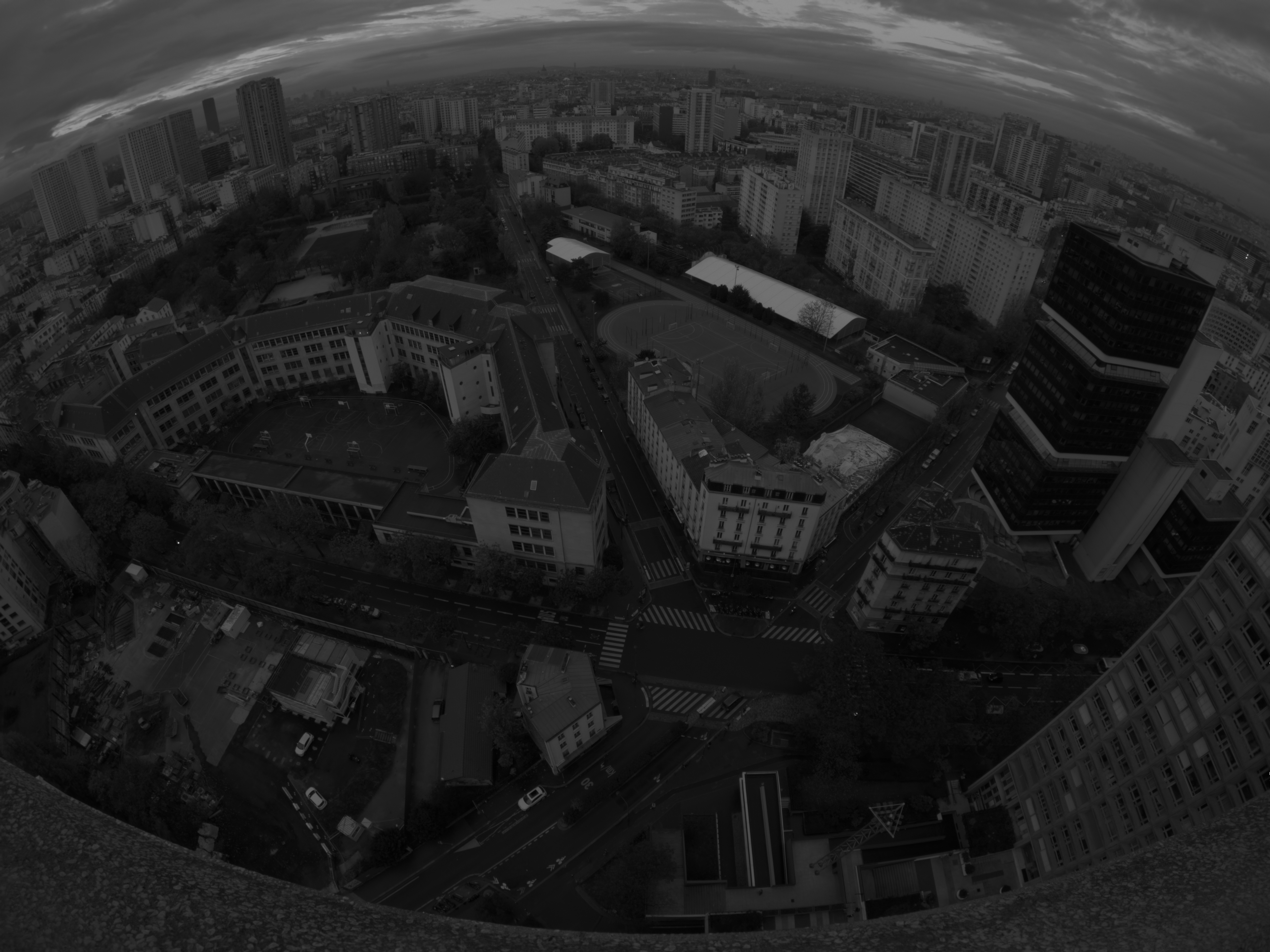}
					\caption{Synthetic short exposure}
					\label{fig:shortexposure}
				\end{center}
			\end{subfigure}
			\begin{subfigure}[t]{.3\linewidth}
				\begin{center}
					\includegraphics[width=\linewidth]{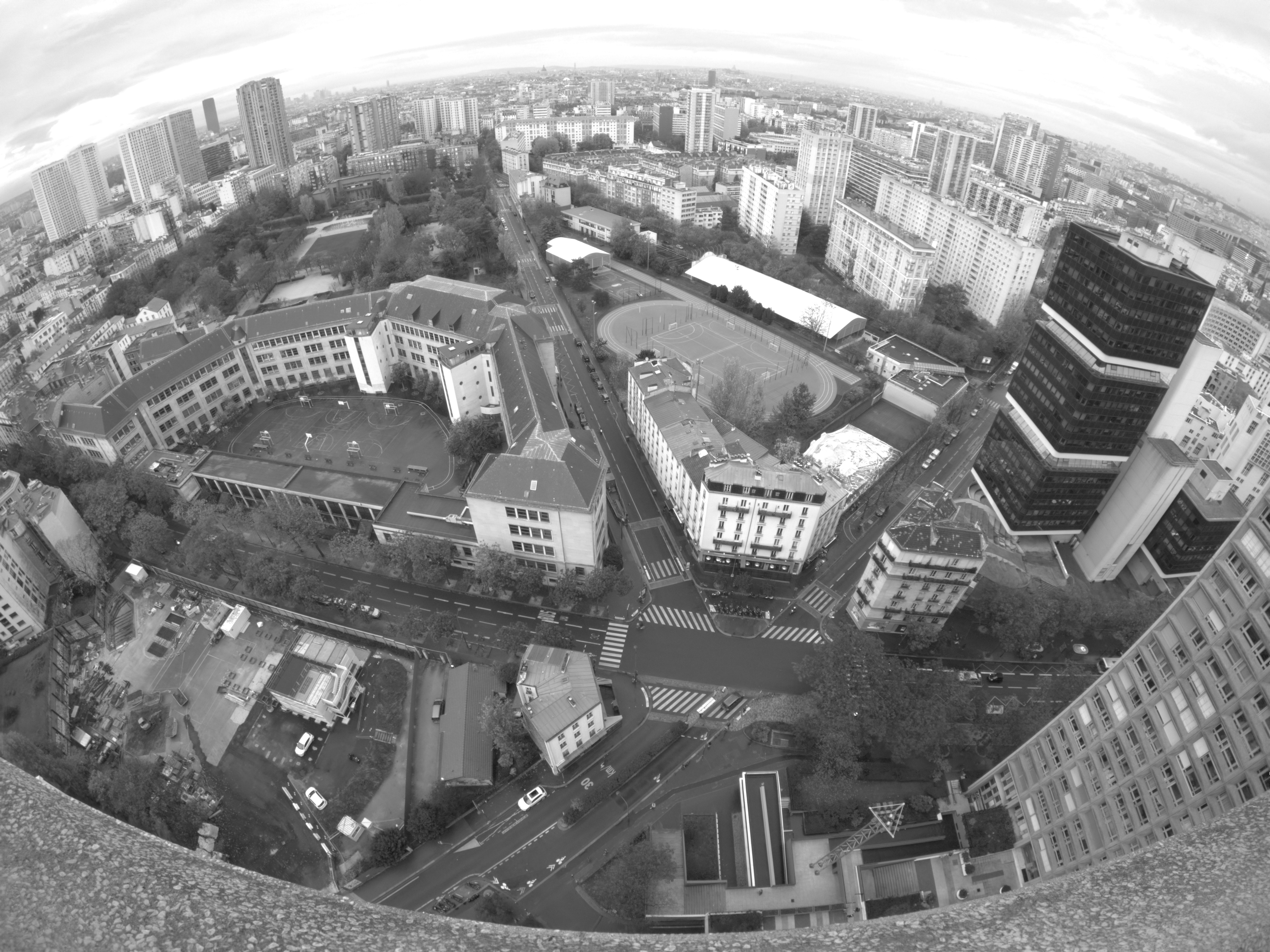}
					\caption{Long exposure (gain=15)}
					\label{fig:longexposure}
				\end{center}
			\end{subfigure}
			\begin{subfigure}[t]{.3\linewidth}
				\begin{center}
					\includegraphics[width=\linewidth]{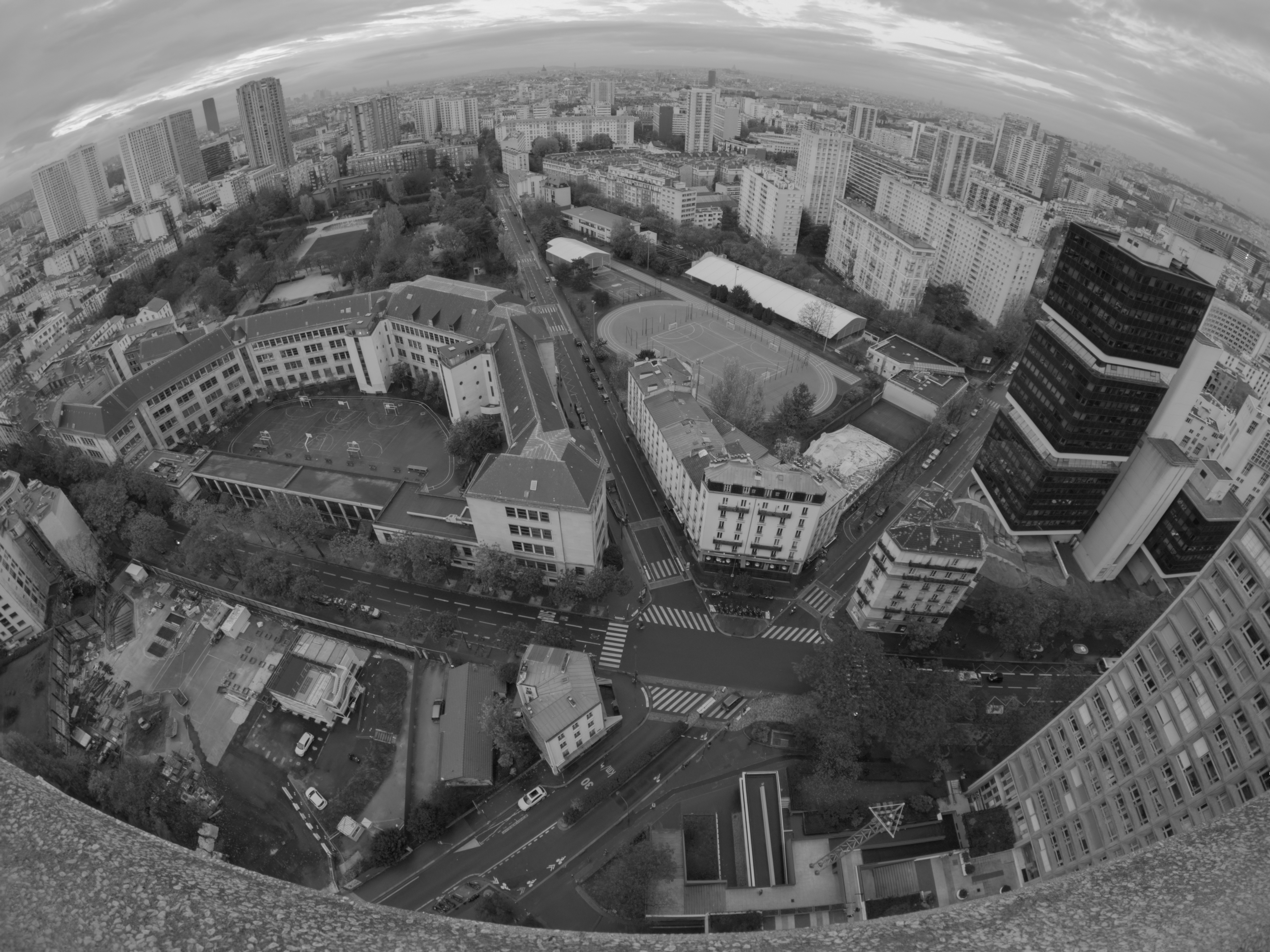}
					\caption{Exposure fusion result}
					\label{fig:fusedexposure}
				\end{center}
			\end{subfigure}
		\caption{Dynamic range compression / local tone mapping.}
		\label{fig:hdrtonemap}
		\end{center}
	\end{figure}
	We undo the gamma correction of the fused grayscale image, and compute a per-pixel scaling with the element-wise division of the result by the original grayscale image. We then apply that scaling to each channel of the underexposed RGB image to obtain the tone-mapped result.
	\item \textbf{Contrast enhancement} / \textbf{global tone mapping}: our next step is to increase image contrast by applying the following simple S-shaped function to each color channel
	\begin{equation}\label{eq:scontrast}
		y = \max(0, \min(x - \alpha \sin(2\pi x), 1)).
	\end{equation}
	\begin{figure}[!htbp]
	\begin{center}
		\includegraphics[width=.5\linewidth]{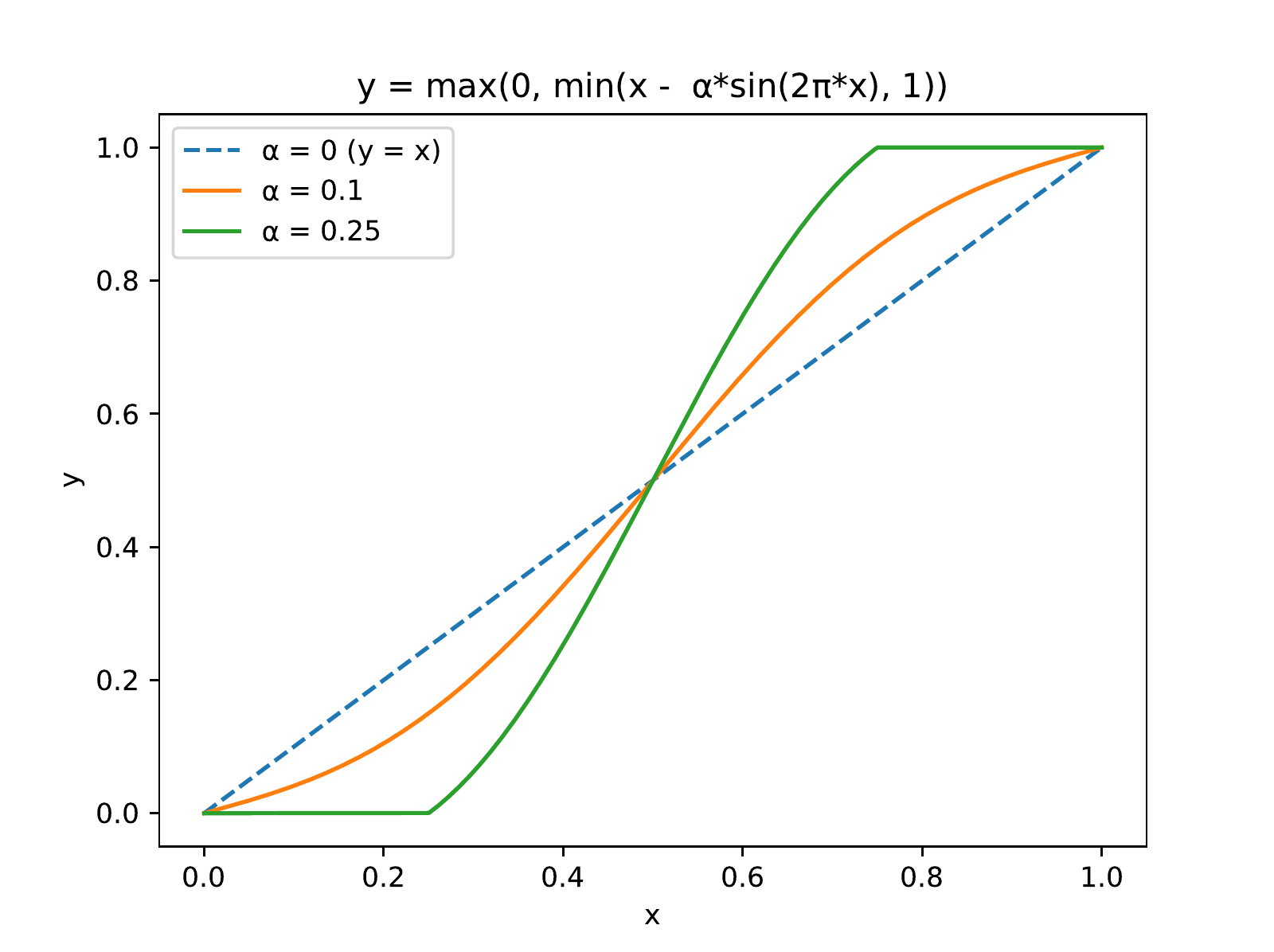}
		\caption{S-shaped contrast enhancement curve.}
		\label{fig:contrastcurve}
	\end{center}
	\end{figure}
	This makes dark areas of the image darker and bright areas brighter, thus enhancing contrast. As illustrated in Figure~\ref{fig:contrastcurve}, the $\alpha$ parameter must be carefully selected to avoid completely black or white regions while retaining a visually pleasing result. Although it is probably different from the one used in Google's own HDR+ implementation, this S-shaped contrast enhancing curve is actually very similar to the one referenced and used in another Google publication~\cite{brooks2019unprocessing}.
	\item \textbf{Gamma compression}: pixel values are shifted from linear to non-linear sRGB space in order to both reduce the required bandwidth and produce a more pleasing and identifiable look, taking advantage of the human visual sensitivity which is also non-linear (for example, humans are more sensitive to brightness differences in darker tones versus brightness differences in brighter tones). The sRGB transfer function is defined by the IEC 61966-2-1 international standard\footnote{\url{http://www.color.org/chardata/rgb/srgb.xalter}} as
	\begin{equation}\label{key}
		V_{out} = 
		\begin{cases}
			12.94 V_{in}, & \text{$0 \leq V_{in}\leq0.0031308$}, \\
			1.055 V_{in}^{\frac{1}{2.4}}-0.55, & 	\text{$0.0031308<V_{in}\leq1$}.\\
		\end{cases}	
	\end{equation}

	\item \textbf{Sharpening}: we get a final sharper image by computing the average of 3 images obtained via unsharp masking
	\begin{equation}\label{sharpentriple}
		I_{sharp} = \frac{1}{3} \sum_{m=1}^{3} I_{sharp_m},
	\end{equation}
	where
	\begin{equation}\label{sharpensingle}
		I_{sharp_m}(x,y) = 
		\begin{cases}
		I(x,y) + \alpha_m (I(x,y) - I^{\sigma_m}_{blur}(x,y)) & \text{If 	$\left| I(x,y) - I^{\sigma_m}_{blur}(x,y)\right| > \tau_m$}, \\
		I(x,y) & \text{otherwise},\\
		\end{cases}	
	\end{equation}
	where $I^{\sigma_m}_{blur}$ is the result of the convolution of $I$ with a Gaussian kernel of standard deviation $\sigma_m$, $\alpha_m$ controls the sharpening strength and $\tau_m$ is a threshold that controls which pixels will actually be sharpened. We empirically found that setting $\alpha_m \in \left\{ 1, 0.5, 0.5 \right\}$, $\sigma_m \in \left\{ 1, 2, 4 \right\}$ and $\tau_m \in \left\{ 2\%, 4\%, 6\% \right\}$, $m \in \left\{ 1, 2, 3 \right\}$ produced visually pleasing results (image that looks a bit sharper while remaining natural-looking, little to no noise reintroduced after sharpening homogeneous regions). This mask is certainly different from the one employed in~\cite{hasinoff2016burst} given that its parameters are unknown.
	\item \textbf{JPEG quantization and compression}: the image is quantized and encoded to 8 bits, resulting in the final \texttt{.jpg} file. We set the quality level of the lossy compression to 100 to minimize compression artifacts.
\end{itemize}
Although these steps produce less visually pleasing images than the full HDR+ finishing pipeline, they still showcase one of its most defining aspects: HDR tone mapping from a single underexposed merged image. A full or partial combination of these steps can be used for any raw .dng image, including the noisy reference image, or intermediate results of Google's own implementation. This will come in handy for visual comparisons, which we'll discuss in Sections~\ref{vsgooglemerged} and~\ref{vsgooglefinal}.

\section{Comparison to the Original HDR+ Implementation}

\subsection{The HDR+ Dataset}\label{hdrplusdata}

In 2018,~\cite{hasinoff2016burst} was completed by a very large dataset captured with several smartphones, simply named the HDR+ dataset\footnote{HDR+ Burst Photography Dataset, \url{http://www.hdrplusdata.org/}}. It is comprised of 3640 raw bursts (28461 images total), for a total of 765 GB\@. Each burst contains the following files:
\begin{itemize}
	\item a set of raw images stored in the open and commonly used Adobe Digital Negative (\texttt{.dng}) format, which can be read by rawpy and contains useful metadata such as exposure time, ISO, black level as well as white balance gains estimated by the camera.
	\item additional metadata stored in separate files, such as maps for lens shading correction and color correction matrices.
\end{itemize}
A curated subset of 153 bursts is also available. Since the HDR+ algorithm is not learning-based, we mostly focused on that subset for the purposes of this article.

Two sets of Google's own results are available on the dataset website: the 2016 results, said to correspond to the algorithm and parameters described in~\cite{hasinoff2016burst}, while the 2017 ones correspond to a more recent version of the pipeline with ``algorithm refinements and updated tuning''. For each raw burst, a set of results comes with the following data:
\begin{itemize}
	\item A raw \texttt{merged.dng} file which is said to correspond to the result of the alignment and temporal denoising steps (the dataset authors claim no additional processing of the raw image).
	\item A \texttt{final.jpg} file, which is obtained after the full finishing pipeline described in Section~\ref{googlepipeline} is applied to the merged result.
	\item A \texttt{reference.txt} file that contains the index of the picked reference frame.
\end{itemize}

\subsection{Comparison to Google's Merged Results}\label{googlemergedcomparison}

The inclusion of Google's raw alignment and temporal denoising results along with the chosen reference frame allows for an apples-to-apples comparison with our open-source Python implementation of the burst alignment and merging algorithm: we can measure similarity between the raw images and/or put the alignment and merging results through the same finishing pipeline for visual comparison.

\subsubsection{Statistical Comparison of Raw Merged Images}
\begin{figure}[!htb]
	\begin{center}
		\begin{subfigure}[t]{.49\linewidth}
			\begin{center}
				\includegraphics[width=.9\linewidth]{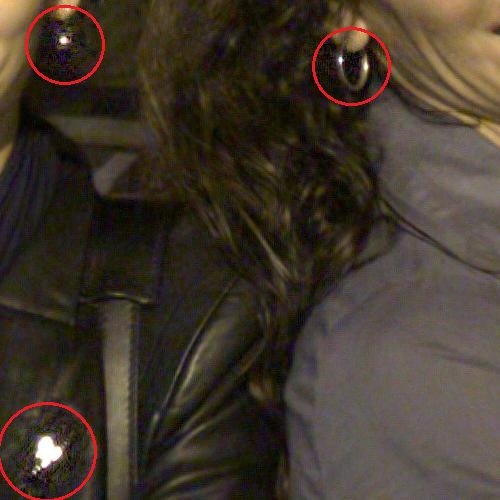}
				\caption{2016 ``Merged'' image (post-processed)}
			\end{center}
		\end{subfigure}
		\begin{subfigure}[t]{.49\linewidth}
			\begin{center}
				\includegraphics[width=.9\linewidth]{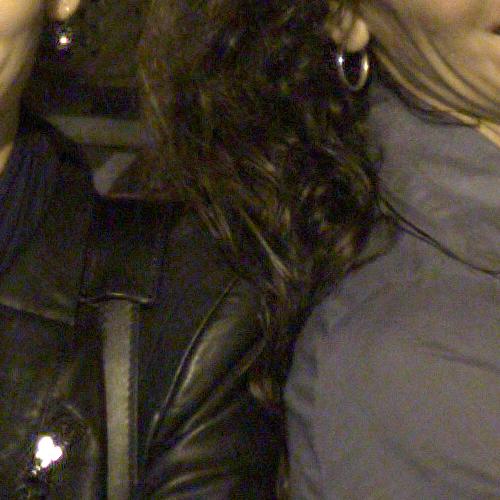}
				\caption{2017 ``Merged'' image (post-processed)}
			\end{center}
		\end{subfigure}
		\caption{The Google 2016 results seem to feature less residual noise except around strong high contrast features / specular highlights, which is typical of spatial denoising as stated in~\cite{hasinoff2016burst}.}
		\label{fig:tuning2016vs2017}
	\end{center}
\end{figure}

We first wanted to compare our raw merged results to the 2016 Google ones, because they are said to be the closest to the implementation presented in the article. However, by putting these files through a simplified finishing pipeline (black level subtraction, white balance, demosaicking, color correction, gamma curve), we actually found that it is very likely that the 2016 version of the \texttt{merged.dng} files also features spatial denoising, while the 2017 might not. Significant differences can be observed in Figure~\ref{fig:tuning2016vs2017}.

On the subject of spatial denoising, the article provides no indication of the slopes of the ``piecewise linear function'' of spatial frequency (or of any potential tuning of the spatial denoising strength). Therefore, it would be ill-advised to compare the 2016 results to our own. We instead decided to perform comparisons against the 2017 results. Even though they are said to feature different tunings and algorithm refinements (we do not know all the 2016 tuning factors either), the alignment and merging results seem more in line with what we were able to obtain with our own implementation.

Using the subset of 153 bursts provided in the HDR+ dataset, we thus took the bursts where the raw reference image and the 2017 Google ``\texttt{merged.dng}'' result were of the same size (for an unknown reason, some Google images are larger than the reference), and set out to compute two PSNRs:
\begin{itemize}
	\item the PSNR between Google's result and the noisier reference image it is based on.
	\item the PSNR between Google's result and our own result, tuned to match Google's result as much as possible:
		\begin{itemize}
			\item Coarse-to-fine upsampling factors of the 4-level Gaussian pyramid: 4, 4, 2 (as suggested in the HDR+ supplement).
			\item Coarse-to-fine tile sizes: 8, 16, 16, 16 (as suggested in the HDR+ supplement).
			\item Type of norm used for tile distance computation: L2 at all levels except the finest, L1 for the finest (as suggested in the HDR+ supplement).
			\item Temporal denoising factor $\tau$: 75 (it is claimed to be set to 8 in~\cite{hasinoff2016burst}, but we were not able to reproduce similar temporal denoising performance with our own way of computing the noise parameters $(\lambda_s, \lambda_r)$ and thus $\sigma^{2}$).
		\end{itemize}
\end{itemize}
On that topic, Google decided to increase the bit depth (and in turn change the black and white levels) of their raw merged result to put emphasis on the accuracy gained after alignment and temporal denoising, so we had to normalize the images before computing those PSNRs (our merged result has the same black and white levels as the reference image).
\begin{figure}
\begin{center}
	\includegraphics[width=.8\linewidth]{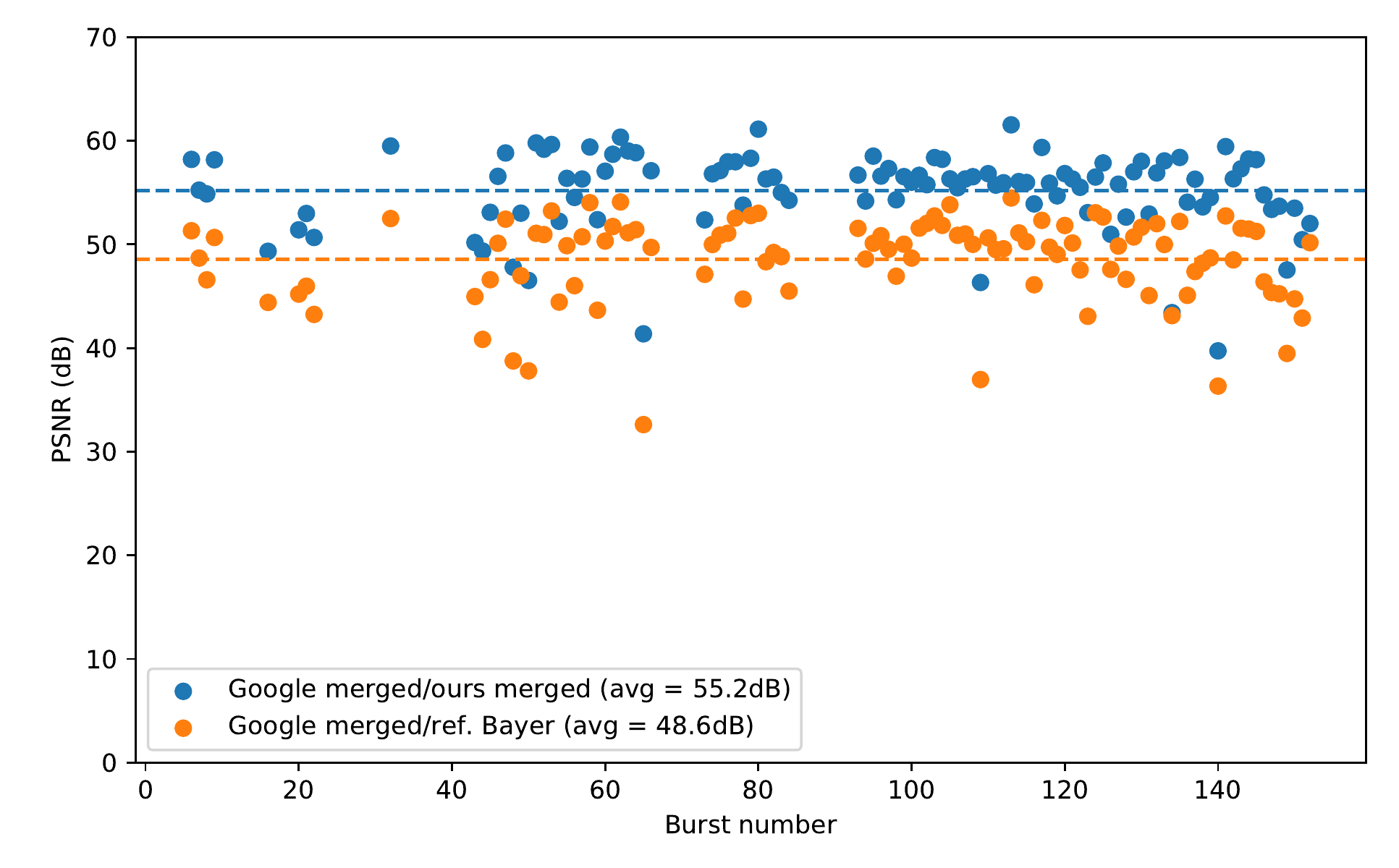}
	\caption{PSNR between Google's and our merged result on the applicable bursts of the ``\texttt{20171106\_subset}''.}
	\label{fig:psnrmerged}
\end{center}
\end{figure}
With these settings, we achieve an approximate +7dB PSNR improvement when compared to the reference on average, our results having higher PSNR for every burst. This suggests that our implementation does perform alignment and temporal denoising somewhat similarly to Google's implementation. Bear in mind that we're trying to be close to images whose tuning and algorithm refinements are unknown. These similarity measurements can be seen in Figure~\ref{fig:psnrmerged}.

\subsubsection{Visual Comparison of Minimally Processed Merged Images}\label{vsgooglemerged}
If we put the three raw images (the reference image, our merged result, and Google's) through the same minimal finishing pipeline (identical demosaicking + white balance + color matrix + gamma curve), we can visually assess the similarity between Google's implementation of the alignment and temporal denoising steps and our own.

\begin{figure}[!htbp]
	\begin{center}
		\begin{subfigure}[t]{.3\linewidth}
			\begin{center}
				\includegraphics[width=\linewidth]{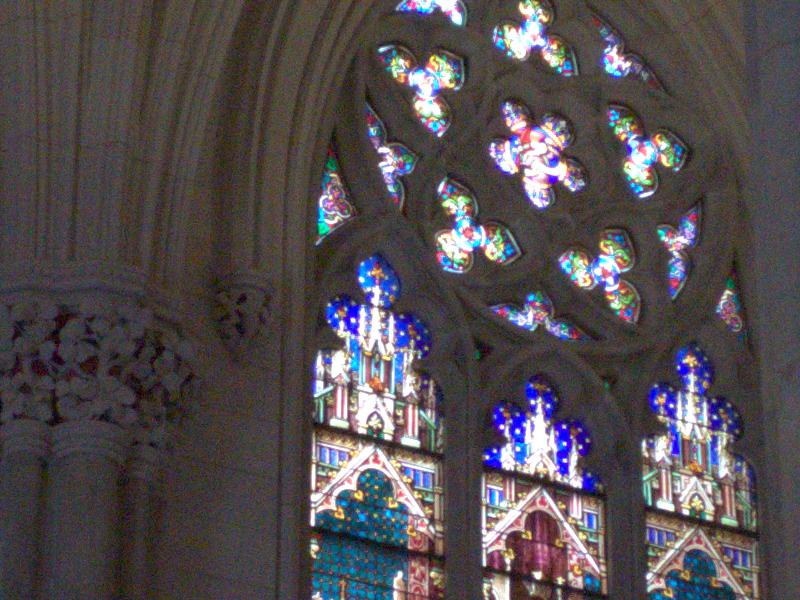}
			\end{center}
		\end{subfigure}
		\begin{subfigure}[t]{.3\linewidth}
			\begin{center}
				\includegraphics[width=\linewidth]{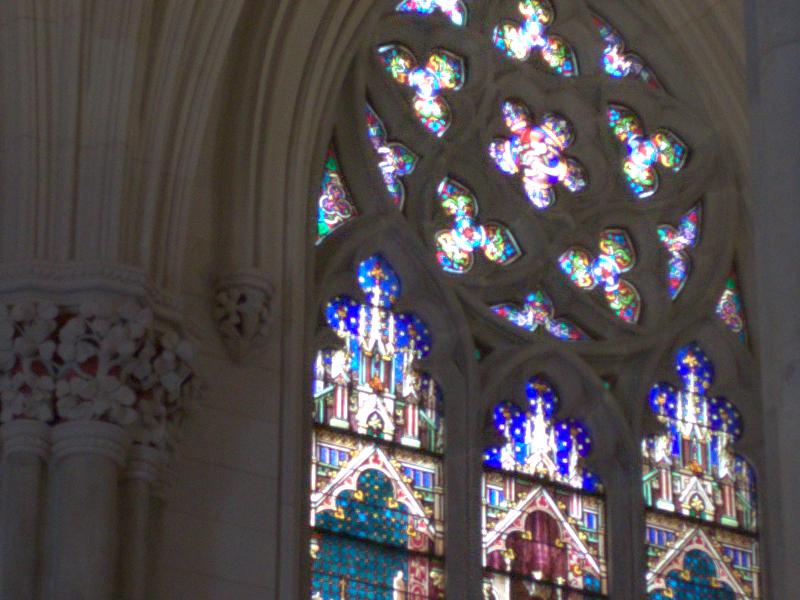}
			\end{center}
		\end{subfigure}
		\begin{subfigure}[t]{.3\linewidth}
			\begin{center}
				\includegraphics[width=\linewidth]{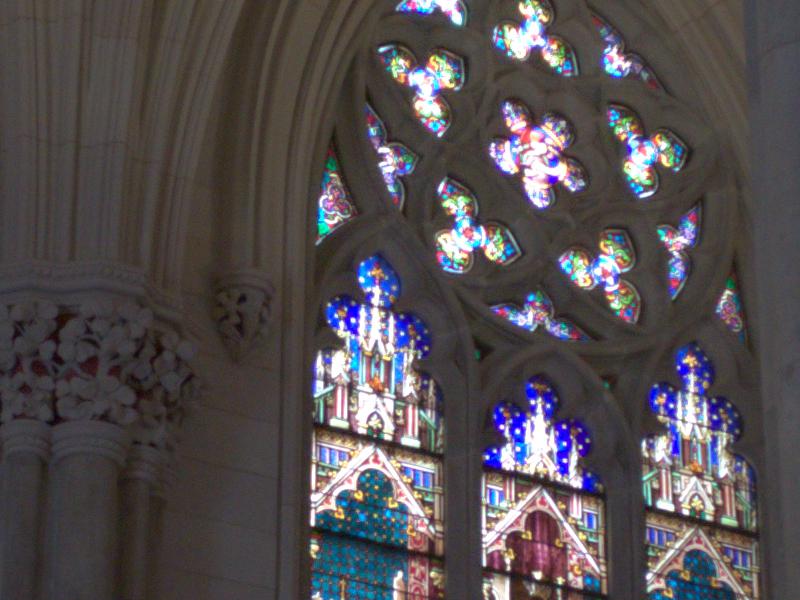}
			\end{center}
		\end{subfigure}
		\begin{subfigure}[t]{.3\linewidth}
			\begin{center}
				\includegraphics[width=\linewidth]{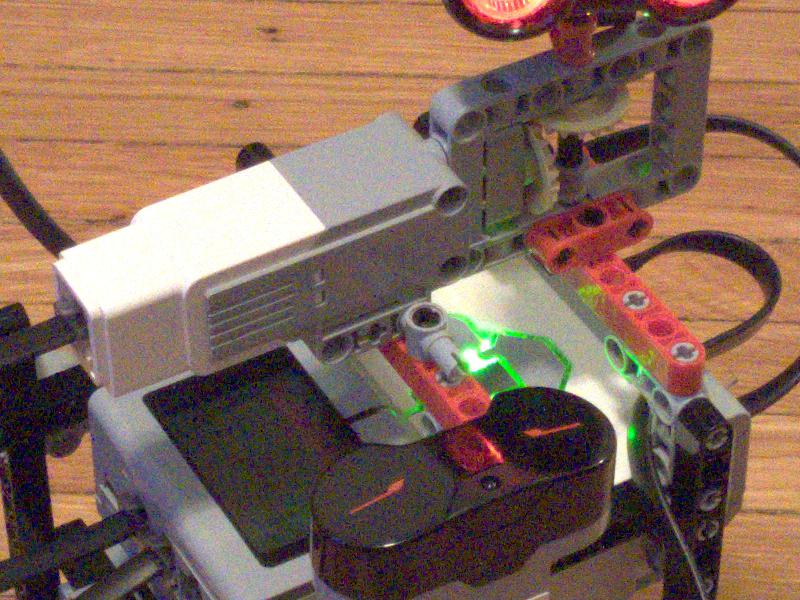}
			\end{center}
		\end{subfigure}
		\begin{subfigure}[t]{.3\linewidth}
			\begin{center}
				\includegraphics[width=\linewidth]{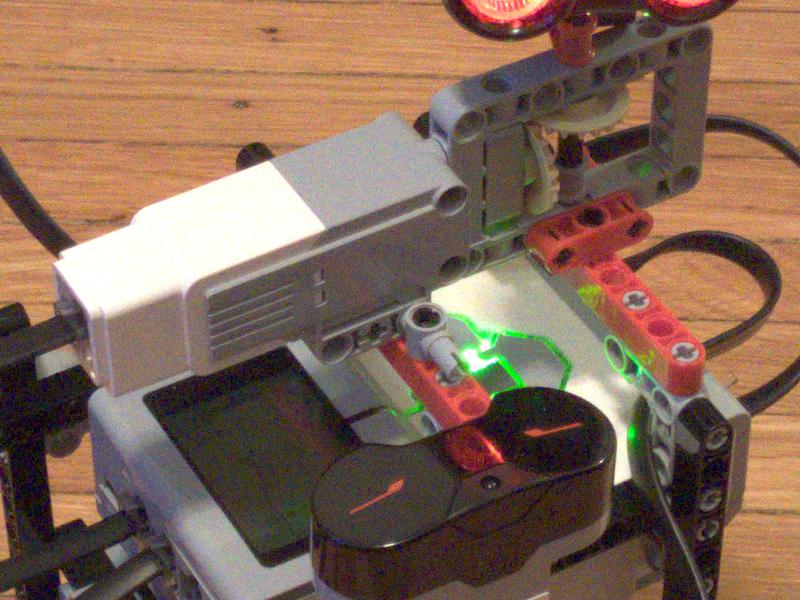}
			\end{center}
		\end{subfigure}
		\begin{subfigure}[t]{.3\linewidth}
			\begin{center}
				\includegraphics[width=\linewidth]{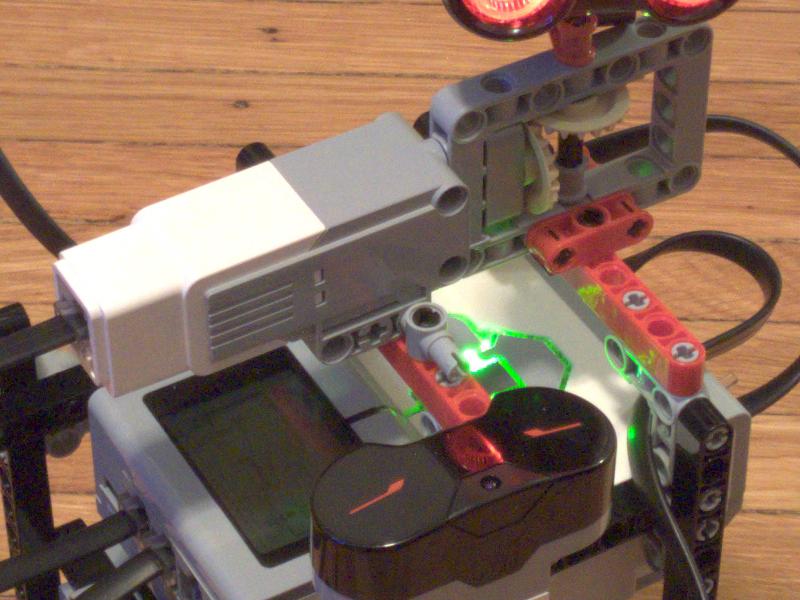}
			\end{center}
		\end{subfigure}
		\begin{subfigure}[t]{.3\linewidth}
			\begin{center}
				\includegraphics[width=\linewidth]{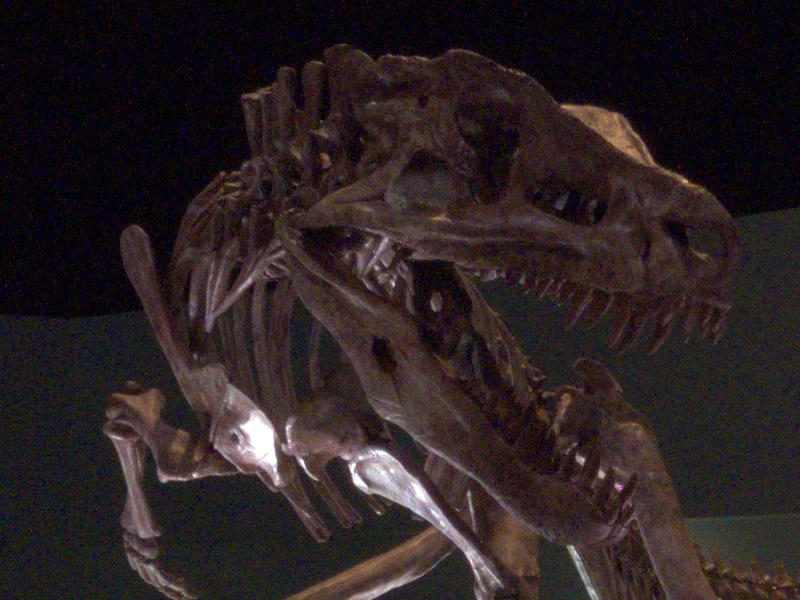}
			\end{center}
		\end{subfigure}
		\begin{subfigure}[t]{.3\linewidth}
			\begin{center}
				\includegraphics[width=\linewidth]{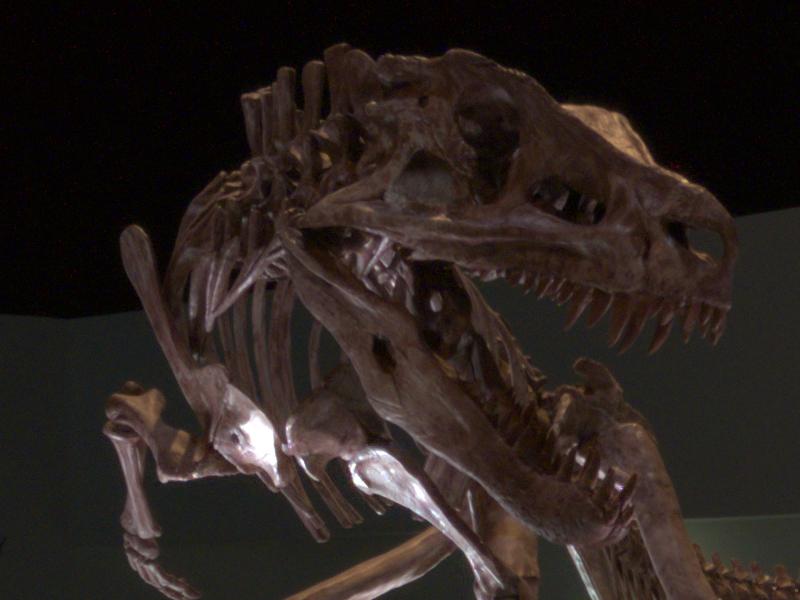}
			\end{center}
		\end{subfigure}
		\begin{subfigure}[t]{.3\linewidth}
			\begin{center}
				\includegraphics[width=\linewidth]{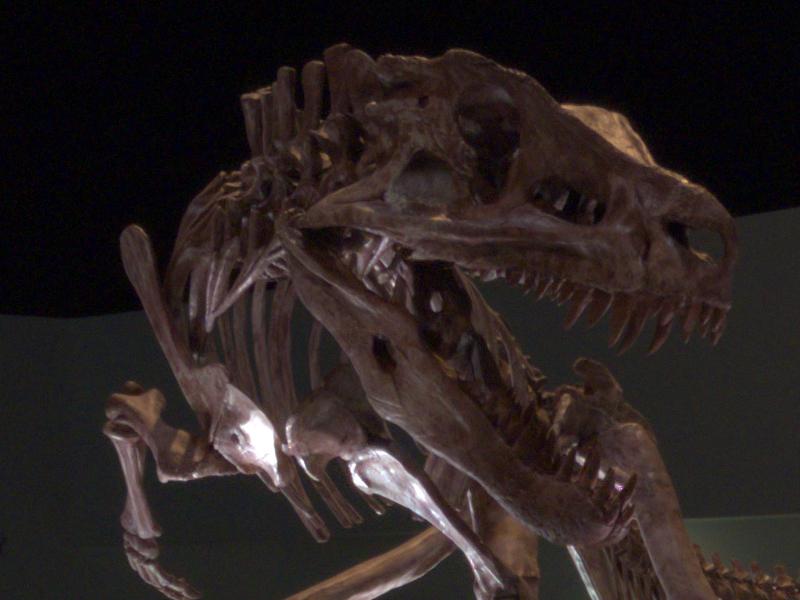}
			\end{center}
		\end{subfigure}
		\begin{subfigure}[t]{.3\linewidth}
			\begin{center}
				\includegraphics[width=\linewidth]{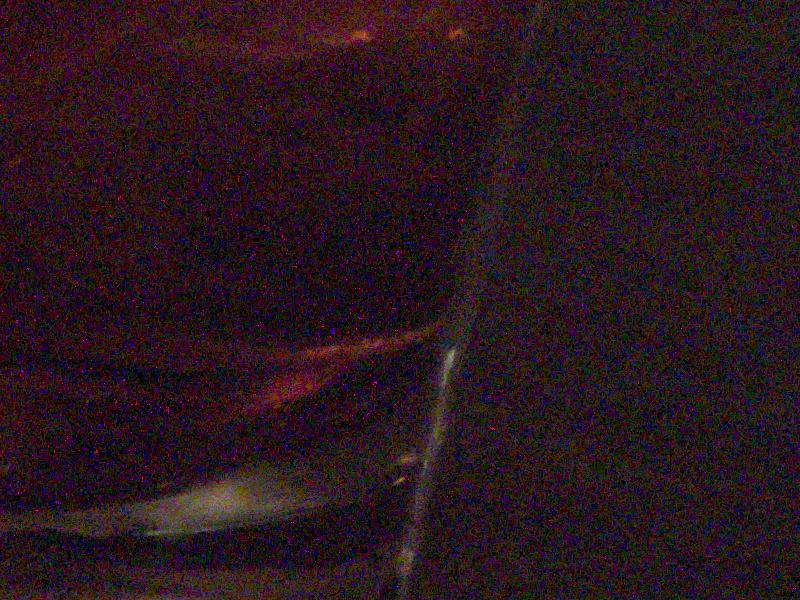}
			\end{center}
		\end{subfigure}
		\begin{subfigure}[t]{.3\linewidth}
			\begin{center}
				\includegraphics[width=\linewidth]{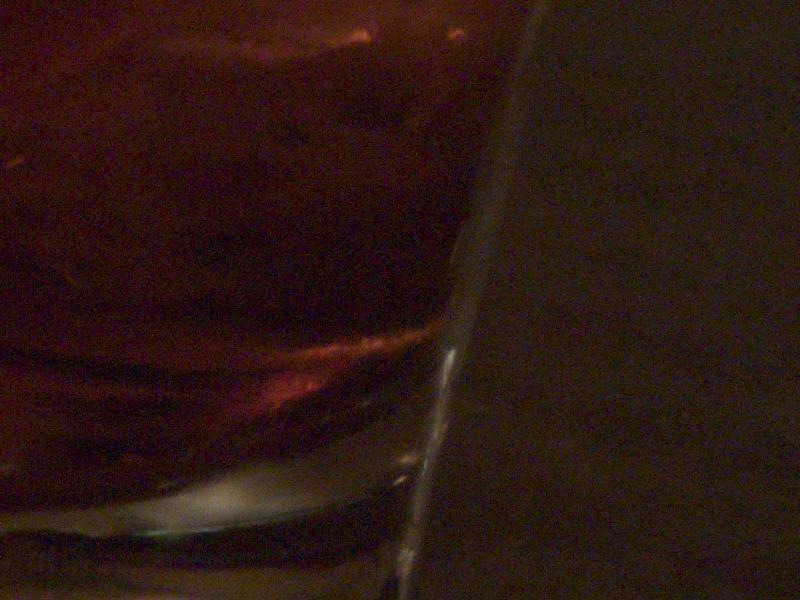}
			\end{center}
		\end{subfigure}
		\begin{subfigure}[t]{.3\linewidth}
			\begin{center}
				\includegraphics[width=\linewidth]{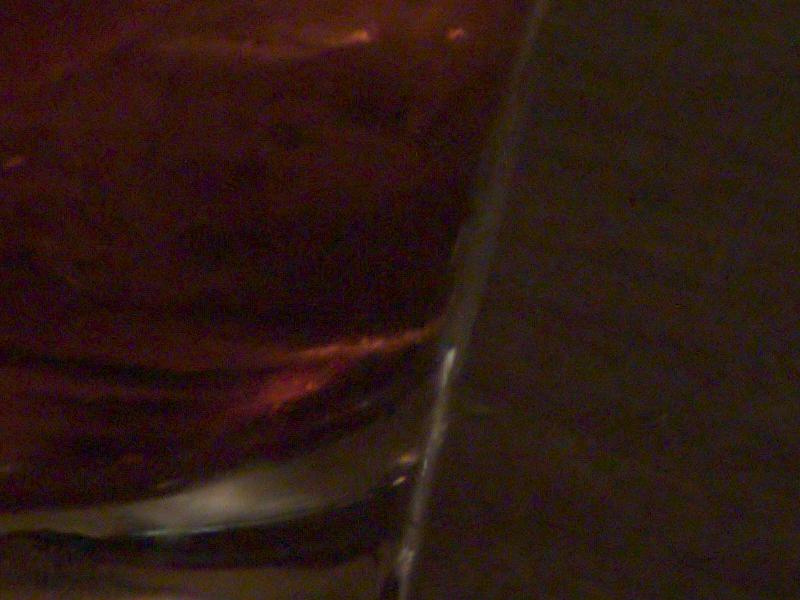}
			\end{center}
		\end{subfigure}
		\begin{subfigure}[t]{.3\linewidth}
			\begin{center}
				\includegraphics[width=\linewidth]{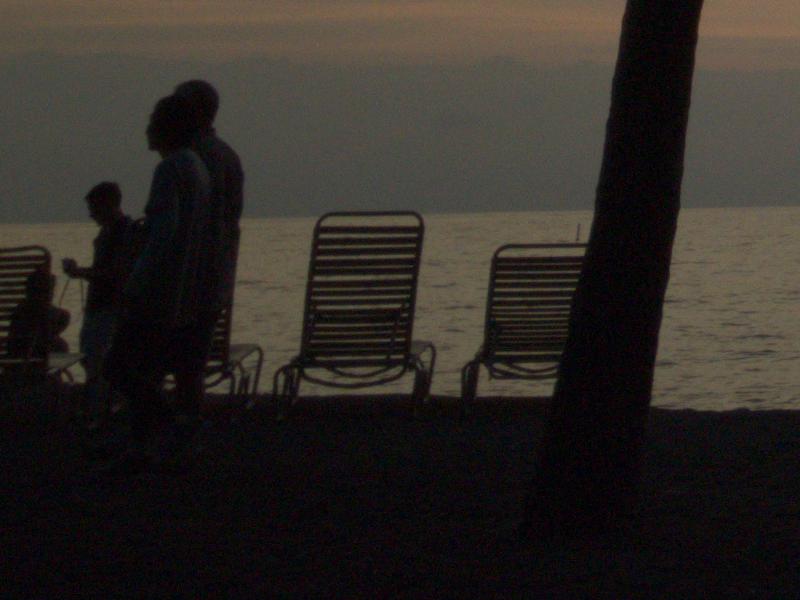}
				\caption*{\textbf{Reference}}
			\end{center}
		\end{subfigure}
		\begin{subfigure}[t]{.3\linewidth}
			\begin{center}
				\includegraphics[width=\linewidth]{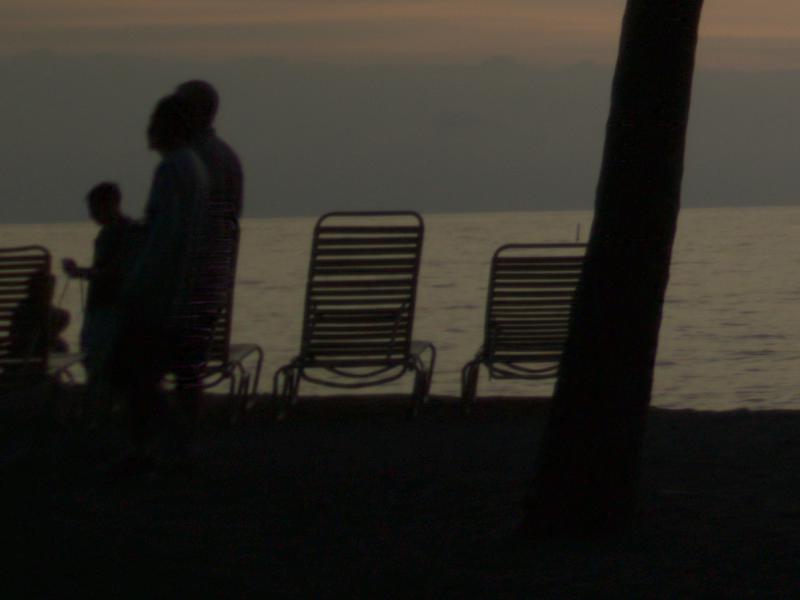}
				\caption*{\textbf{Ours merged}}			
			\end{center}
		\end{subfigure}
		\begin{subfigure}[t]{.3\linewidth}
			\begin{center}
				\includegraphics[width=\linewidth]{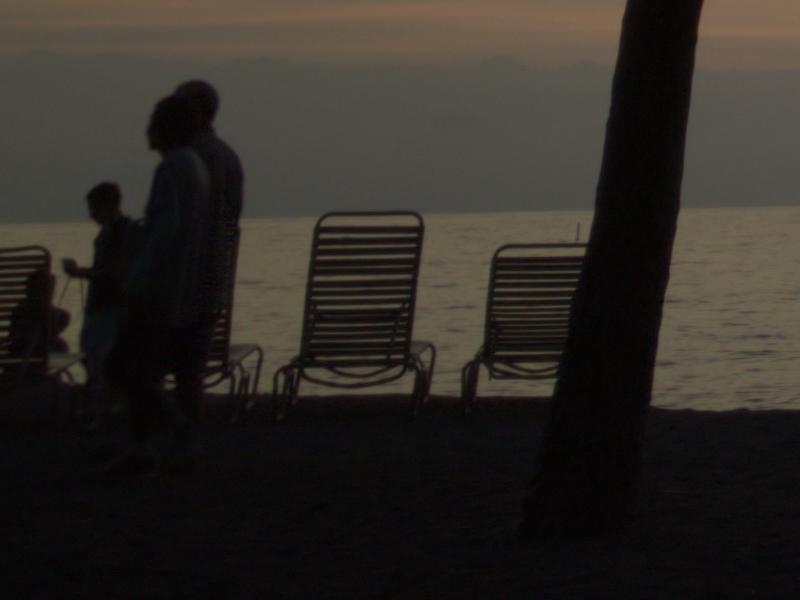}
				\caption*{\textbf{Google merged}}
			\end{center}
		\end{subfigure}
	\caption{Visual comparison of crops of minimally processed versions of the reference image, our ``merged'' result and Google's ``merged'' result (viewers are invited to zoom in). Both alignment and merging procedures significantly reduce noise. \textit{Top row}: both merged results are virtually indistinguishable. \textit{Second row}: Google seems to further reduce noise. \textit{Third row}: our result seems more denoised. \textit{Fourth row}: Google's residual noise seems to be of lower spatial frequency compared to ours (they might be using a larger tile size for this low-light image). \textit{Bottom row}: for this image, ghosting is slightly stronger in our result.}	
	\label{fig:merged_vs_google}
	\end{center}
\end{figure}

Looking at the subset of 153 bursts, some images seem to have less residual noise on our end, while others look more noisy than Google's result. The most likely candidates for noticeable differences between the two are:
\begin{itemize}
	\item The way we compute the $(\lambda_s, \lambda_r)$ noise curve parameters: we either extract them from \texttt{.dng} metadata when possible, or compute them as a function of baseline ISO 100 values and actual ISO (as explained in Equation~\eqref{eq:noiseparams} of Section~\ref{noiseestimation}), while Google can derive these parameters from the analog and digital gain set at capture time.
	\item In~\cite{hasinoff2016burst}, the authors claim to typically use $16\times 16$ tiles for alignment (at the finest pyramid level) and merging, except for very dark scenes where they use $32\times 32$ tiles. We do not make that distinction and always use $16\times 16$ tiles.
	\item Google could use different tuning factors from ours (this interacts with the two previous potential differences anyways).
	\item The so called ``algorithm refinements'' of the 2017 results we're unaware of.
\end{itemize}
Selected crops can be observed in Figure~\ref{fig:merged_vs_google}. Full-size results for the whole subset of 153 bursts can be consulted and downloaded from a Google Photos album\footnote{\url{https://photos.app.goo.gl/mEhyNrqKc2x7rbqj9}}.

\subsection{Visual Comparison to Google's Final Results}\label{vsgooglefinal}

The presence of their final results also enables visual comparison of their consumer-grade raw to JPEG finishing pipeline against our own. That said, given the simplified nature of our own finishing step, if we compare our final images to Google's cleverly chroma denoised, tone-mapped, sharpened, dehazed, color corrected (and so on and so forth) final HDR+ results, such comparison is definitely not in our favor in the vast majority of cases. It would certainly make sense to fully implement these additional finishing steps, but that would require a lot of additional development, testing and fine-tuning time to reach a similar level of quality. The added complexity would also likely imply longer processing times in our Python implementation, and would stir this article away from our main subject of focus, the raw burst alignment and merging algorithm. Still, if we compare our images to a simple post-processing of the reference image (demosaicking + white balance + color matrix + gamma curve), we can see that our pipeline does provide significant visual improvement.
\begin{figure}[!htbp]
	\begin{center}
		\begin{subfigure}[t]{.3\linewidth}
			\begin{center}
				\includegraphics[width=\linewidth]{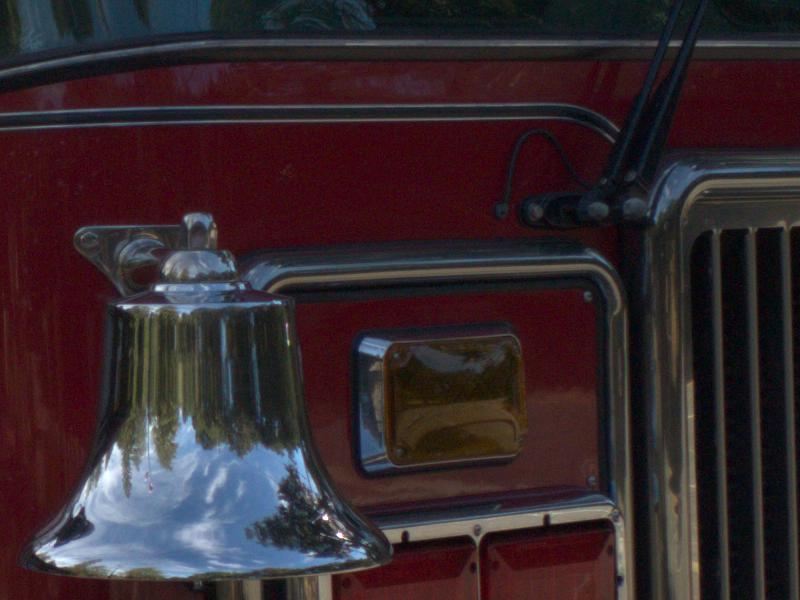}
			\end{center}
		\end{subfigure}
		\begin{subfigure}[t]{.3\linewidth}
			\begin{center}
				\includegraphics[width=\linewidth]{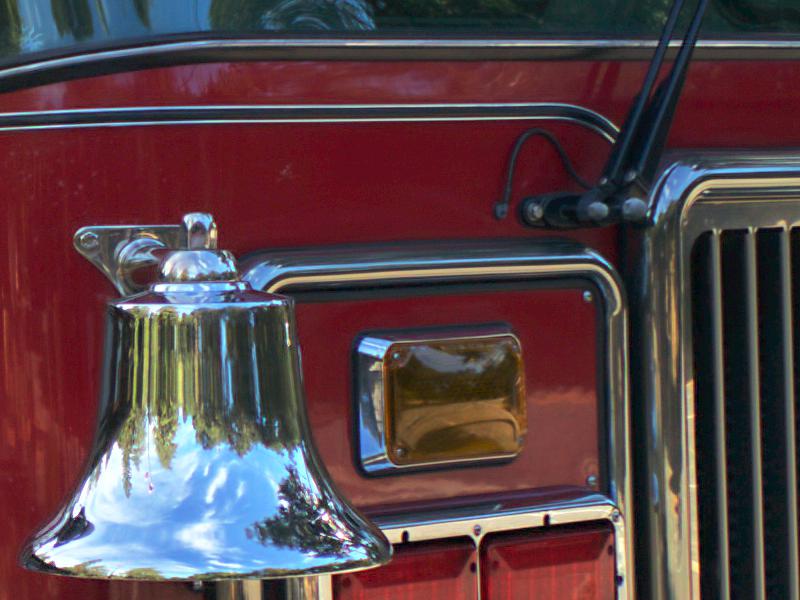}
			\end{center}
		\end{subfigure}
		\begin{subfigure}[t]{.3\linewidth}
			\begin{center}
				\includegraphics[width=\linewidth]{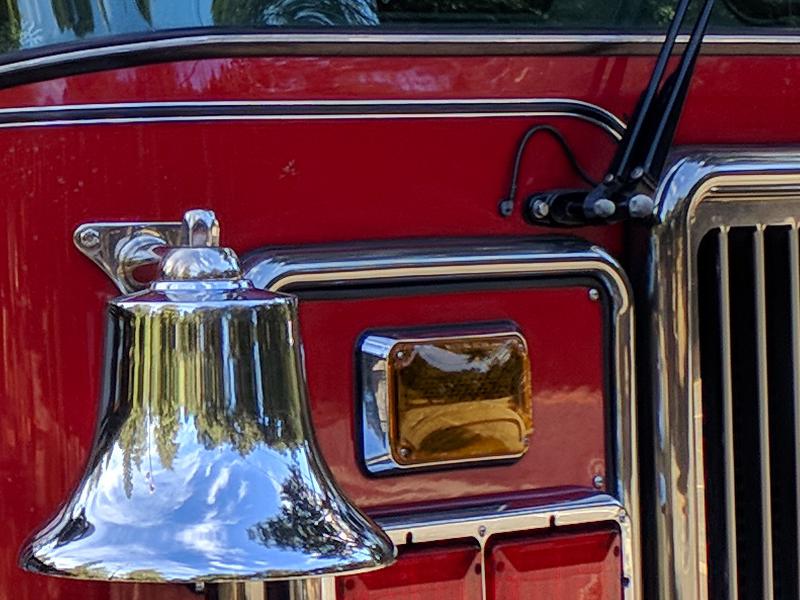}
			\end{center}
		\end{subfigure}
		\begin{subfigure}[t]{.3\linewidth}
			\begin{center}
				\includegraphics[width=\linewidth]{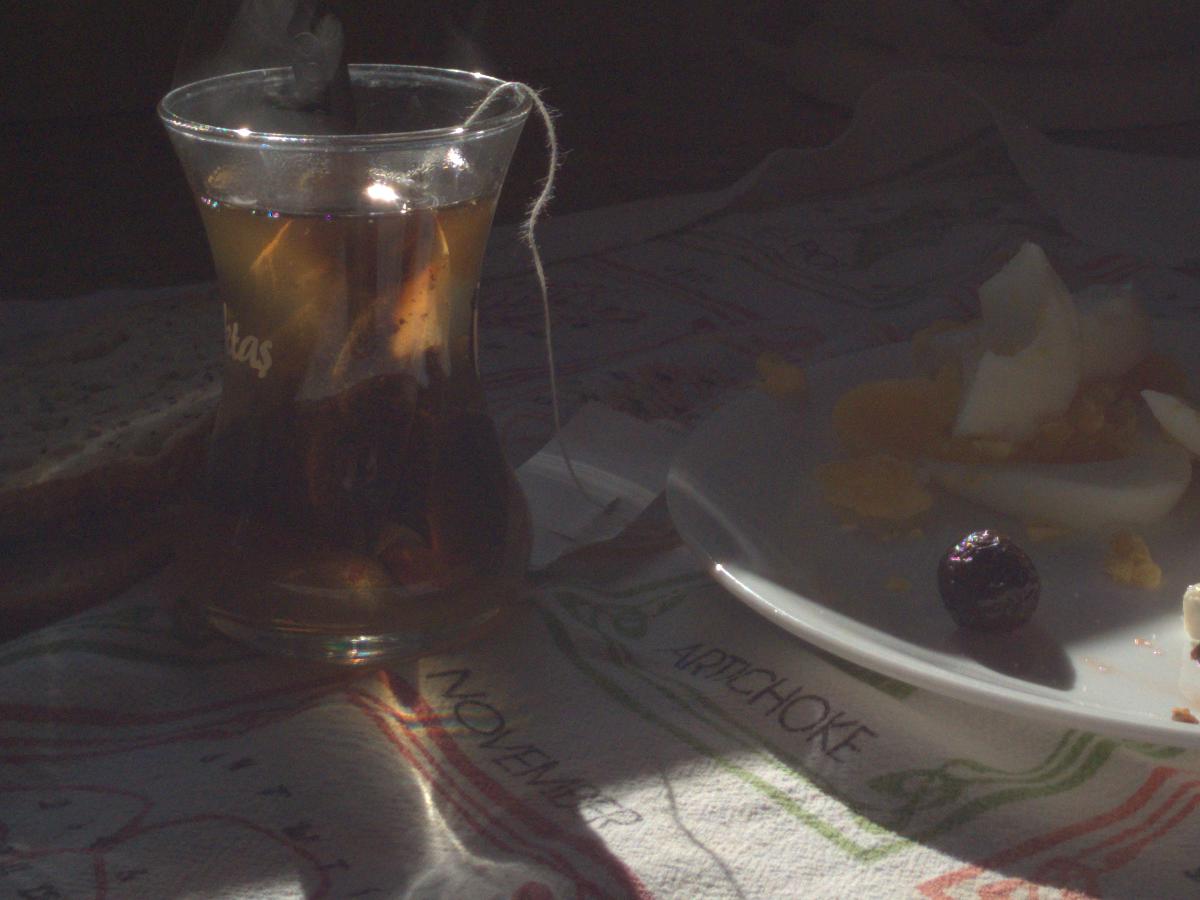}
			\end{center}
		\end{subfigure}
		\begin{subfigure}[t]{.3\linewidth}
			\begin{center}
				\includegraphics[width=\linewidth]{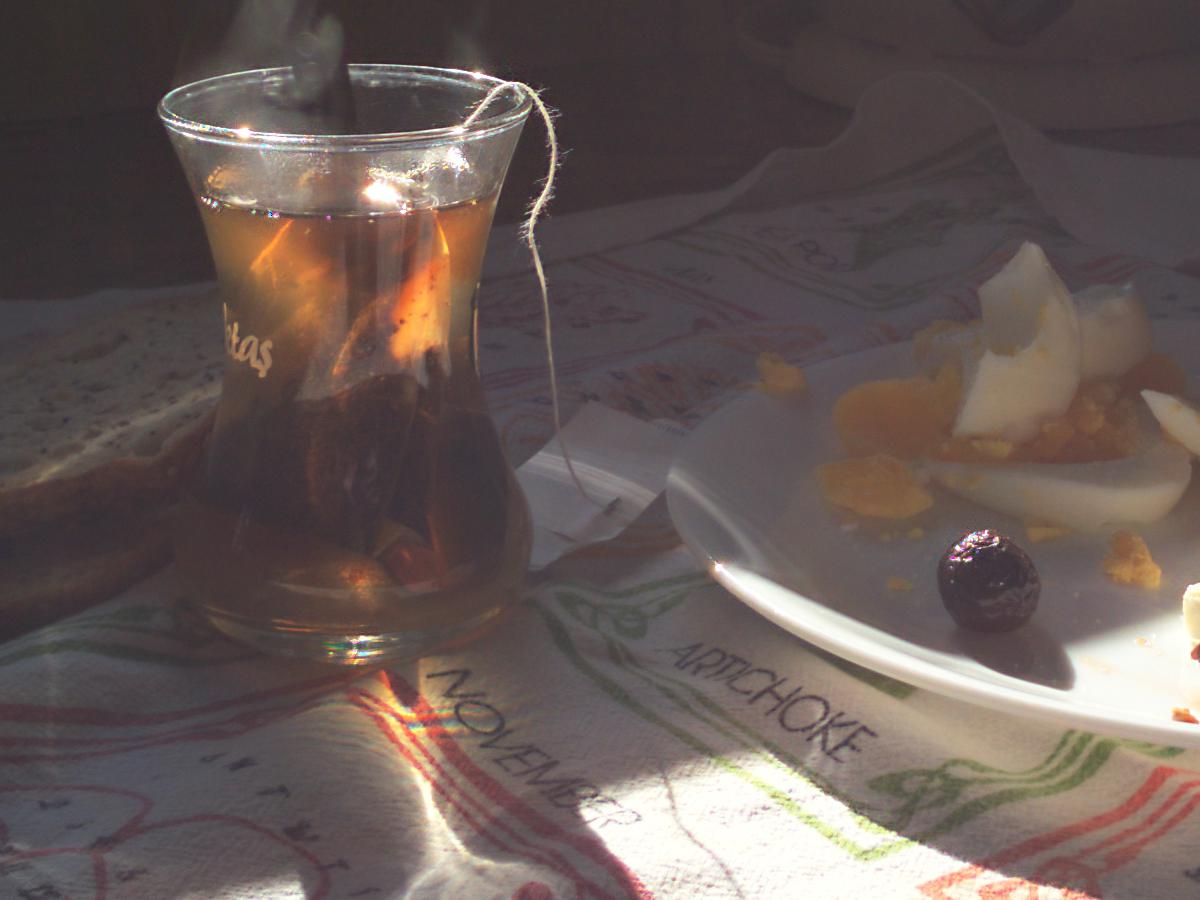}
			\end{center}
		\end{subfigure}
		\begin{subfigure}[t]{.3\linewidth}
			\begin{center}
				\includegraphics[width=\linewidth]{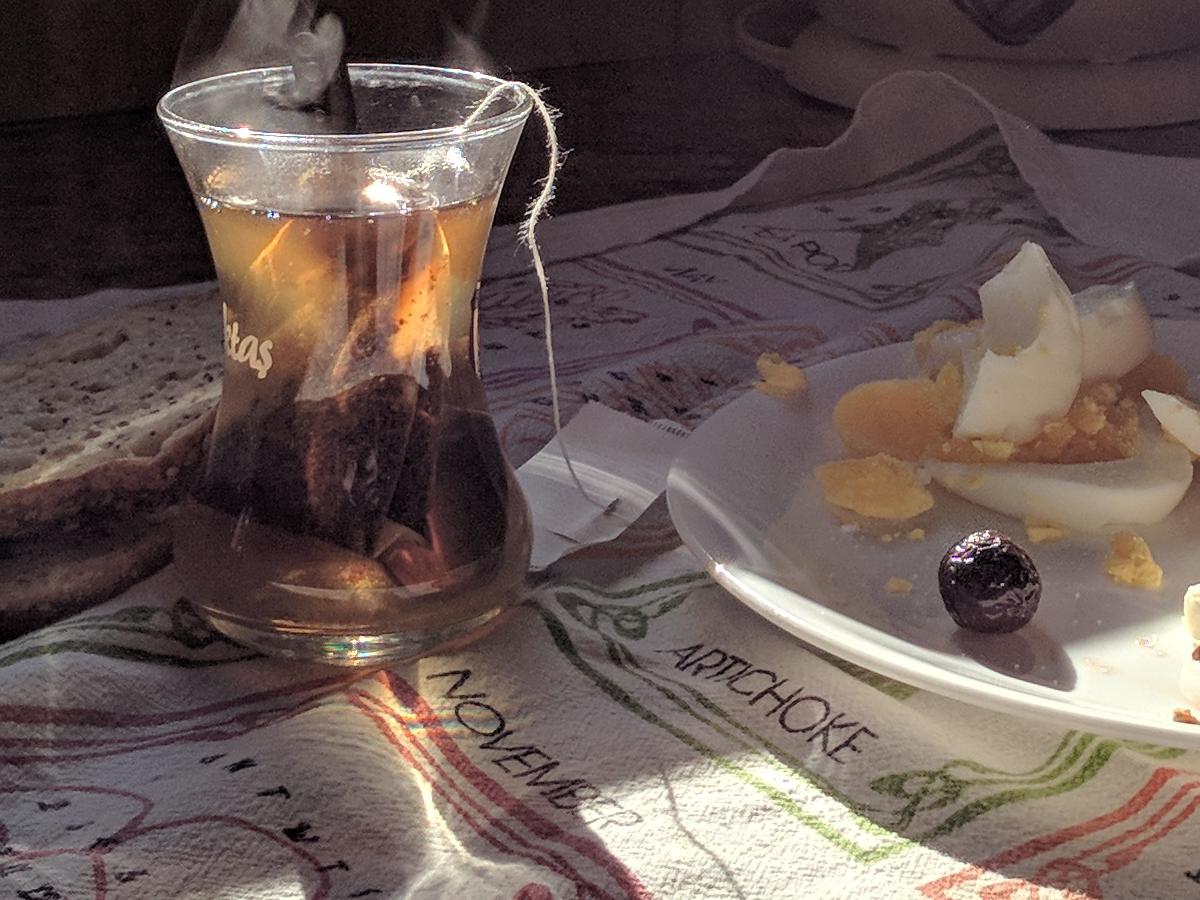}
			\end{center}
		\end{subfigure}
		\begin{subfigure}[t]{.3\linewidth}
			\begin{center}
				\includegraphics[width=\linewidth]{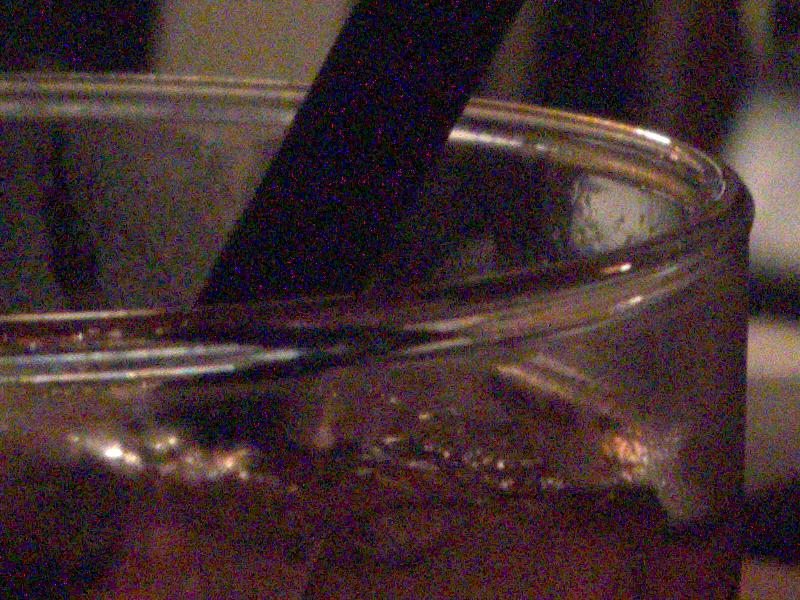}
			\end{center}
		\end{subfigure}
		\begin{subfigure}[t]{.3\linewidth}
			\begin{center}
				\includegraphics[width=\linewidth]{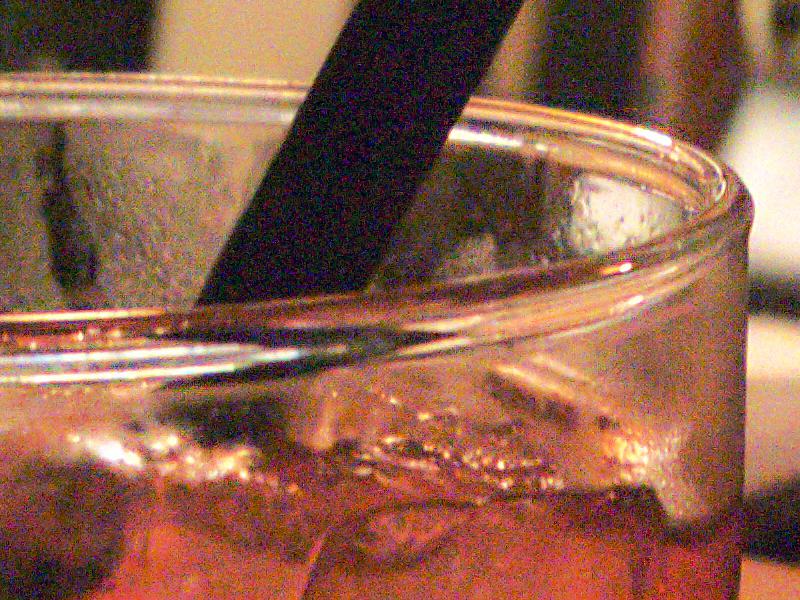}
			\end{center}
		\end{subfigure}
		\begin{subfigure}[t]{.3\linewidth}
			\begin{center}
				\includegraphics[width=\linewidth]{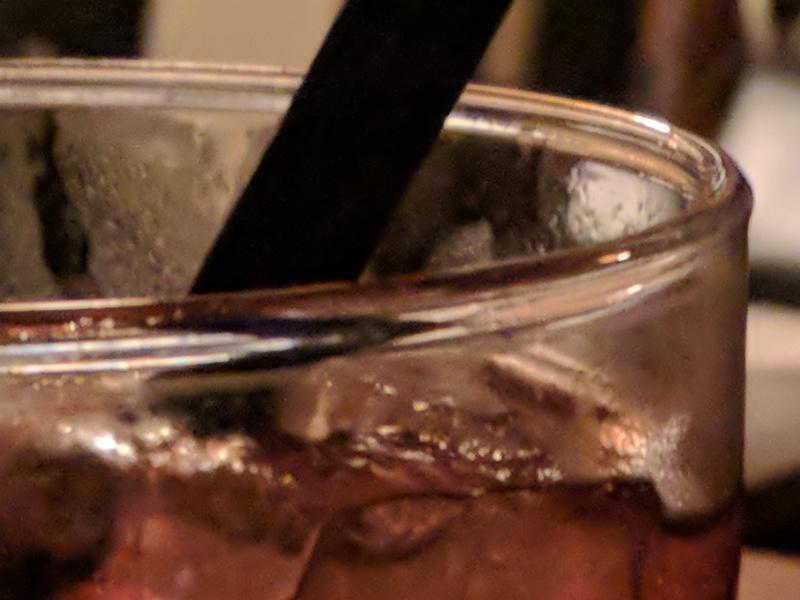}
			\end{center}
		\end{subfigure}
		\begin{subfigure}[t]{.3\linewidth}
			\begin{center}
				\includegraphics[width=\linewidth]{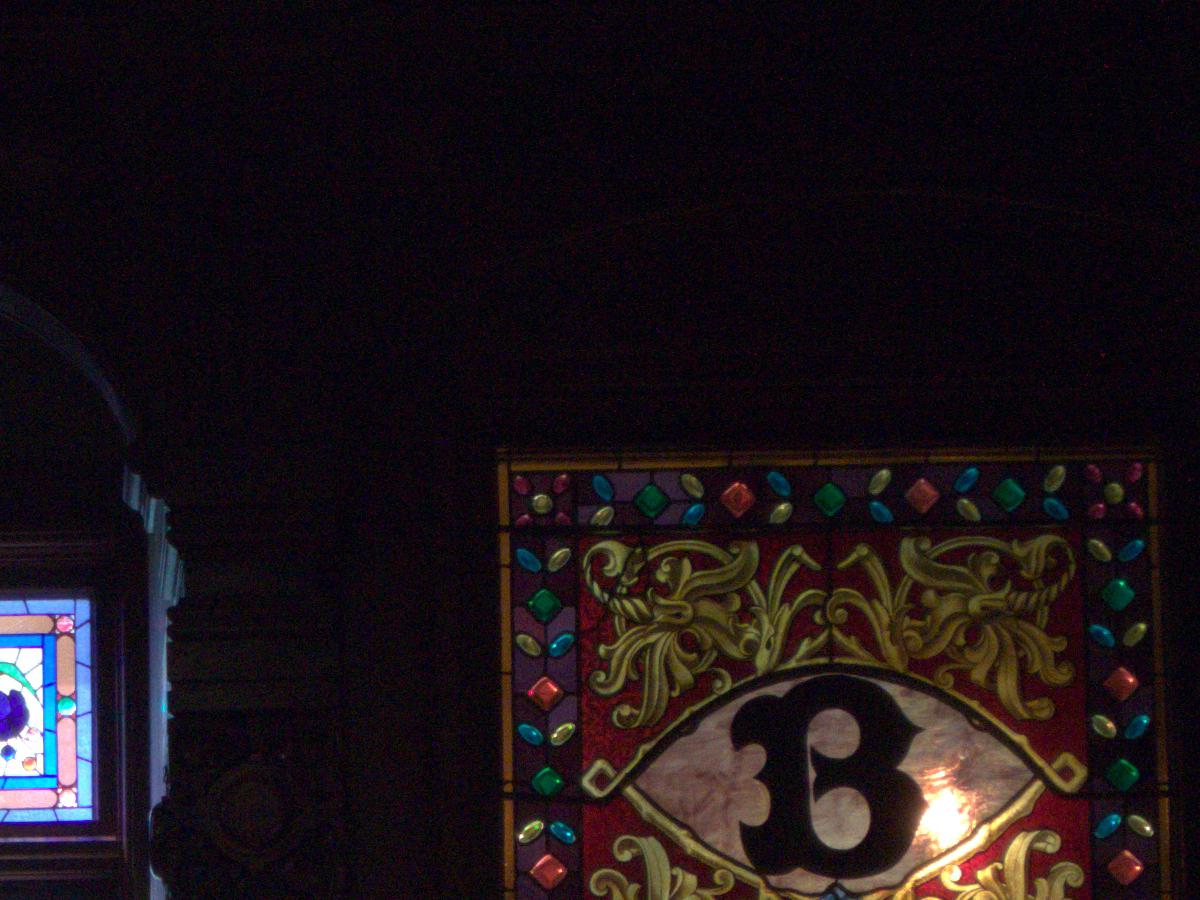}
				\caption*{\textbf{Reference}}
			\end{center}
		\end{subfigure}
		\begin{subfigure}[t]{.3\linewidth}
			\begin{center}
				\includegraphics[width=\linewidth]{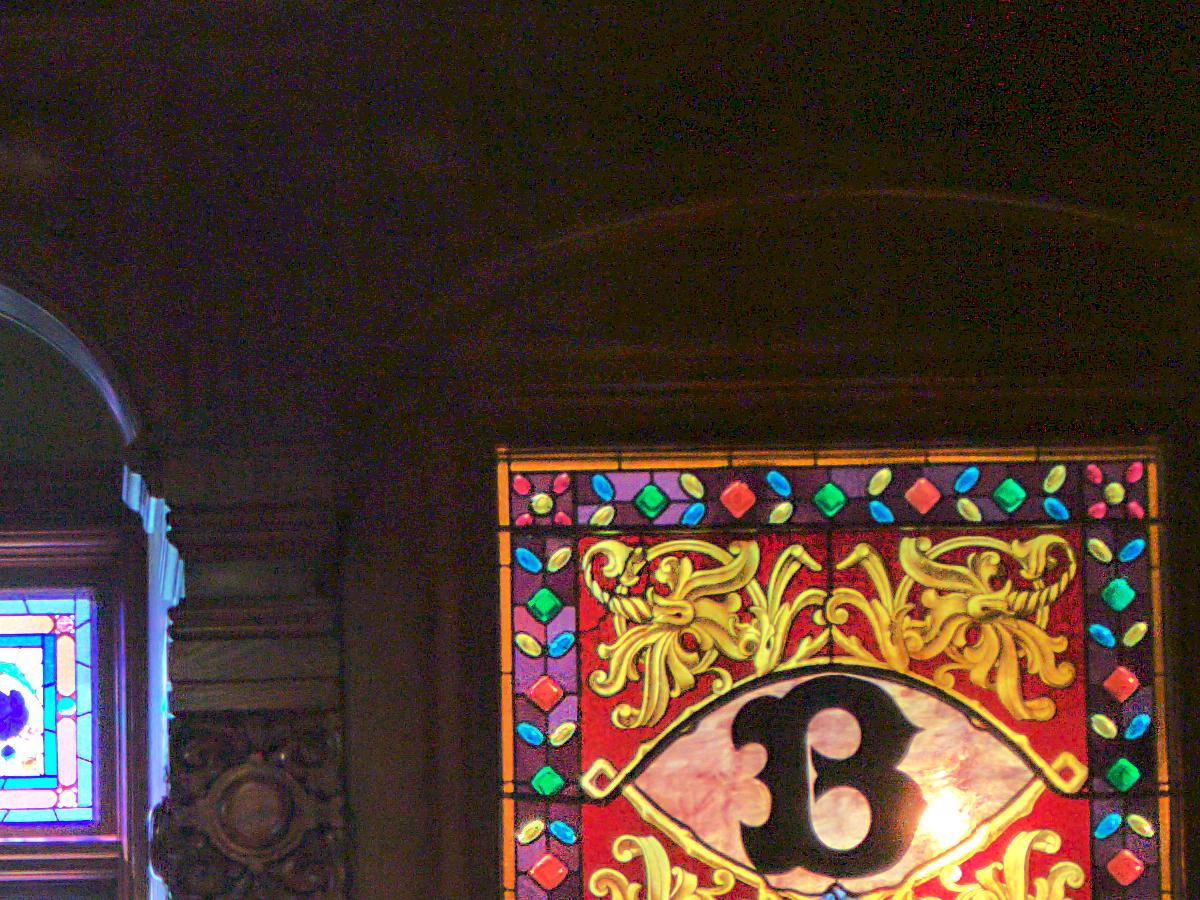}
				\caption*{\textbf{Ours final}}
			\end{center}
		\end{subfigure}
		\begin{subfigure}[t]{.3\linewidth}
			\begin{center}
				\includegraphics[width=\linewidth]{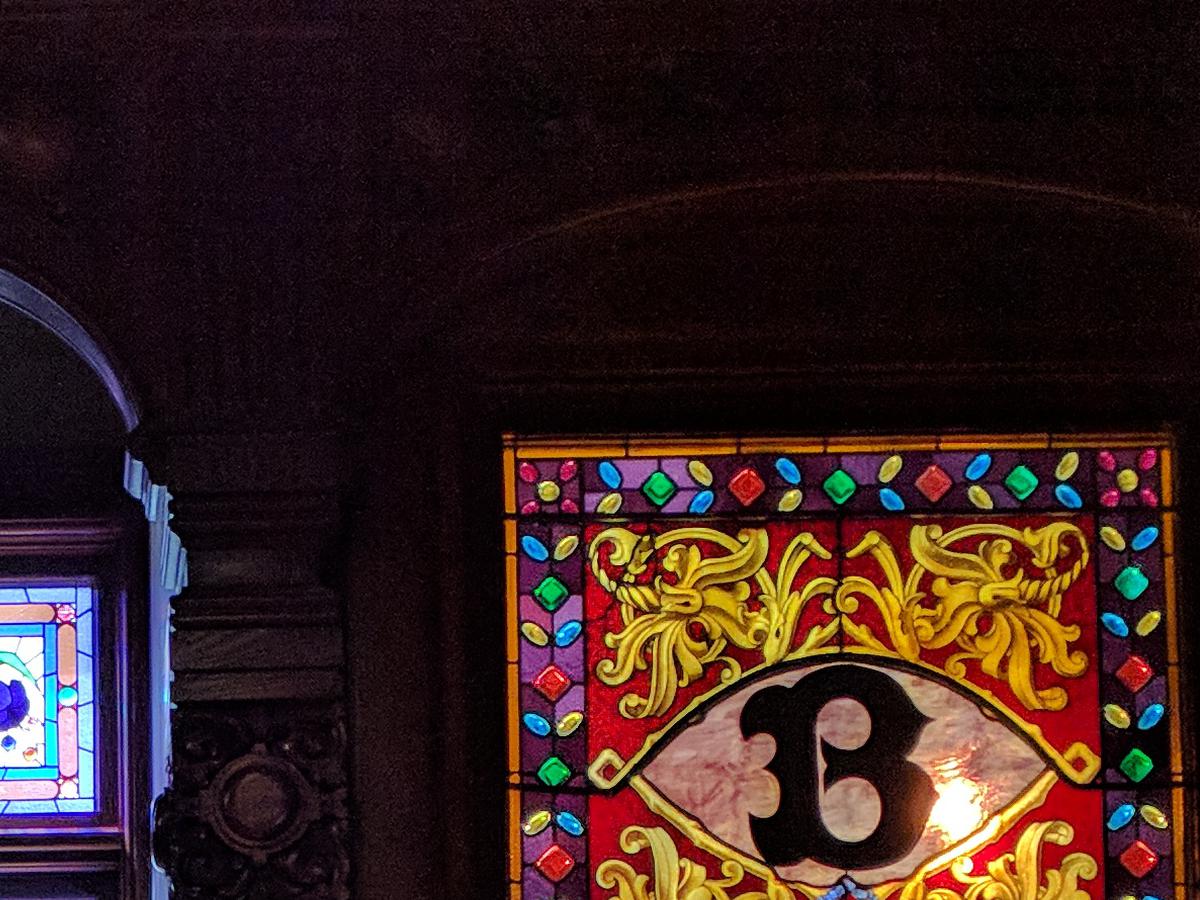}
				\caption*{\textbf{Google final}}
			\end{center}
		\end{subfigure}
		\caption{Visual comparison of crops of the minimally processed reference image, our final result and Google's final result (viewers are invited to zoom in). Google's different spatial denoising, tone mapping, sharpening and additional operations (chroma denoising, lens shading correction, dehazing, color adjustment, dithering, etc) ensures better results in almost any scenario. \textit{First Row}: our results can be similar to Google's in some static, well lit environments. \textit{Second row}: Google's more aggressive tone mapping produces an image with better dynamic range at the expense of more residual noise. \textit{Third row}: since we don't apply any chroma denoising, residual noise can be amplified by our tone mapping and sharpening operations. \textit{Bottom row}: Google also seems to perform hot pixel removal, which we do not (look at the top right corner of the crops).}	
		\label{fig:final_vs_google}
	\end{center}
\end{figure}
Some examples are featured in Figure~\ref{fig:final_vs_google}. Final full-size results for the whole subset of 153 bursts can be consulted and downloaded from another Google Photos album\footnote{\url{https://photos.app.goo.gl/QurTFcvUMc8i1DM7A}}.
\section{Results on GoPro Bursts}

In order to further verify the appeal of the HDR+ raw burst denoising algorithm and the improvements provided by parts of its finishing pipeline, we set out to test our own implementation on bursts that were not part of the HDR+ dataset. We used GoPro HERO 8 Black and Hero 9 Black action cameras as an alternative to the Android devices used by Google. They feature similar sensors (Sony CMOS sensor with relatively small pixels) but much wider lenses when compared to the main camera typically found in modern smartphones.

We captured sequences of 10 raw images using the dedicated burst mode of the HERO 8 and 9, all while trying to follow the base principles described in~\cite{hasinoff2016burst} (underexposed bursts, with identical focal length and short exposure time for all images, and scene and camera motion that is manageable with respect to exposure time). Here are the HERO9 parameters we had access to in order to respect these principles as much as possible:
\begin{itemize}
	\item We did not have direct control of the exposure time, as it is handled by the camera's auto-exposure algorithm. Instead, we could specify the burst rate (how many images we want in a specified amount of time). We set it to its fastest value (10 images in 1s when capturing raw images), ensuring a maximum exposure time of 0.1 s.
	\item We could also change the EV compensation parameter (which controls how bright the burst images are). Since we wanted the bursts to be underexposed, we typically set it between 0 and -2 depending on scene brightness, selected ISO and the expected behavior of the auto-exposure algorithm.
	\item Finally, we could set minimum and maximum ISO values. We actually had to set both parameters to the same value if we wanted all images of the burst to feature the same ISO (if we didn't, the AE algorithm might have selected different values for different images of the same burst). This means that we had to manually pick ISO before capture. We usually set ISO to 100 in most scenarios (in order to avoid clipped highlights) except in low light where we might set it between 200 and 800.
\end{itemize}
We deliberately shot most bursts in very low light / high dynamic range scenarios to ensure these would have a lot of noise. Since GoPro cameras output raw images in their proprietary \texttt{.gpr} format, we converted them to Adobe Digital Negative (\texttt{.dng}) files using Adobe DNG Converter\footnote{\url{https://helpx.adobe.com/photoshop/using/adobe-dng-converter.html}}, which is the same format used for the HDR+ dataset. We then simply executed the exact same code we ran on the images provided by Google, without changing any tuning parameter.

\begin{figure}[!htbp]
	\begin{center}
		\begin{subfigure}[t]{.3\linewidth}
			\begin{center}
				\includegraphics[width=\linewidth]{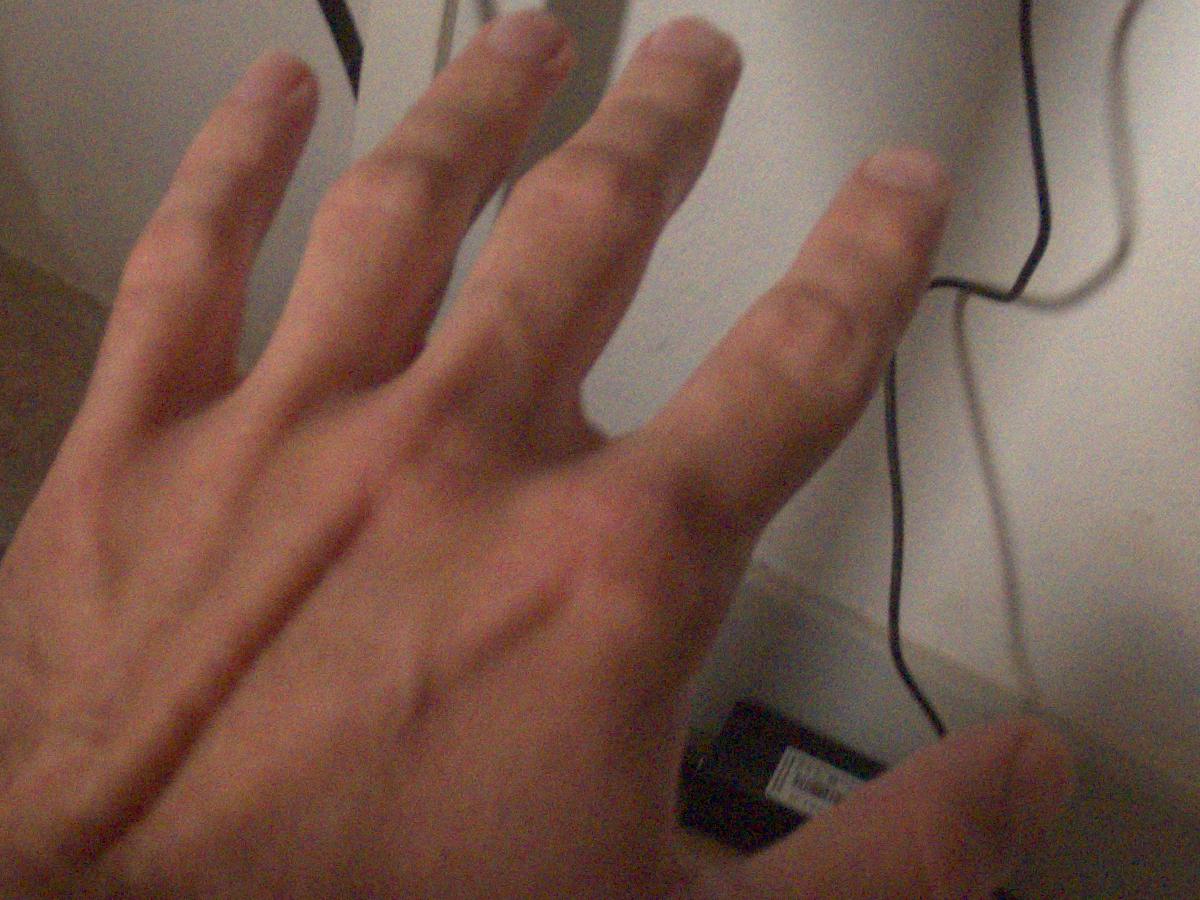}
			\end{center}
		\end{subfigure}
		\begin{subfigure}[t]{.3\linewidth}
			\begin{center}
				\includegraphics[width=\linewidth]{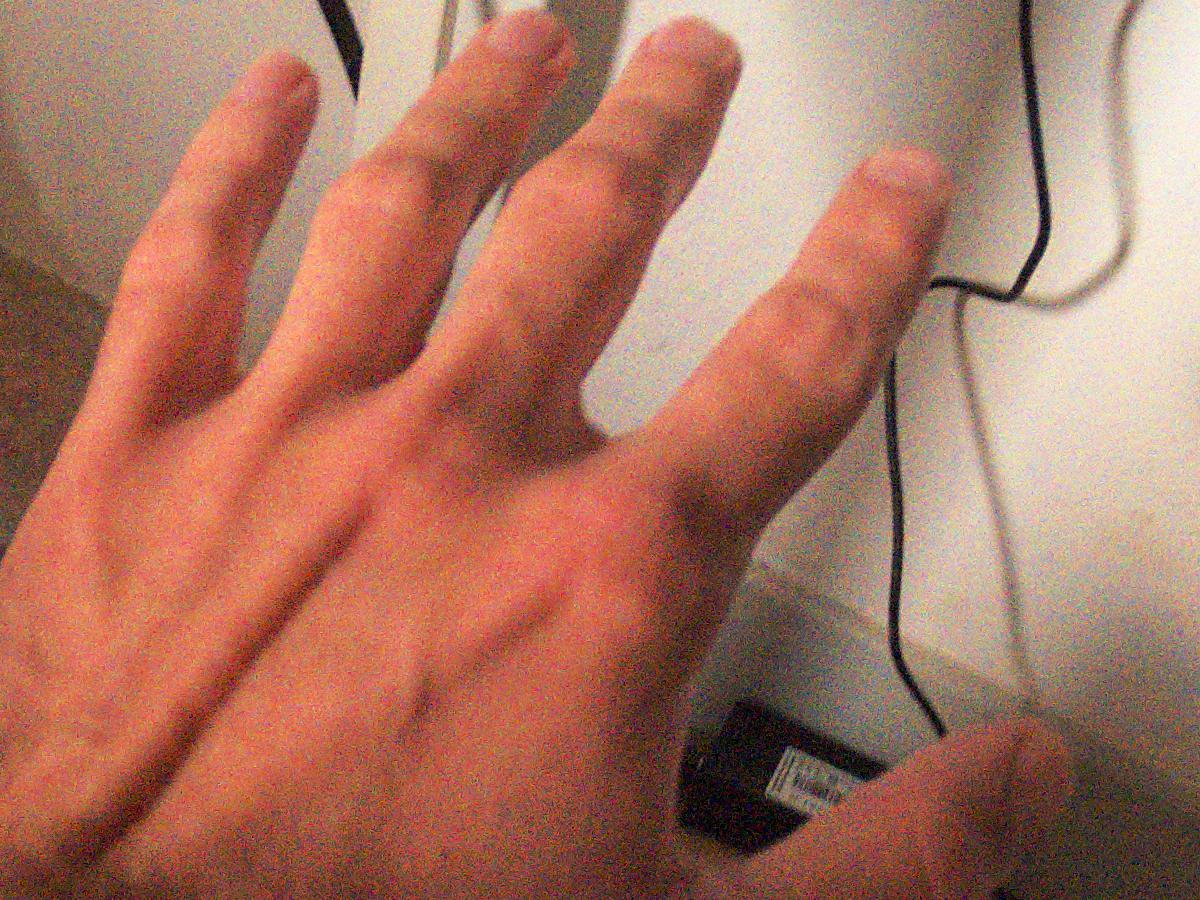}
			\end{center}
		\end{subfigure}
		\begin{subfigure}[t]{.3\linewidth}
			\begin{center}
				\includegraphics[width=\linewidth]{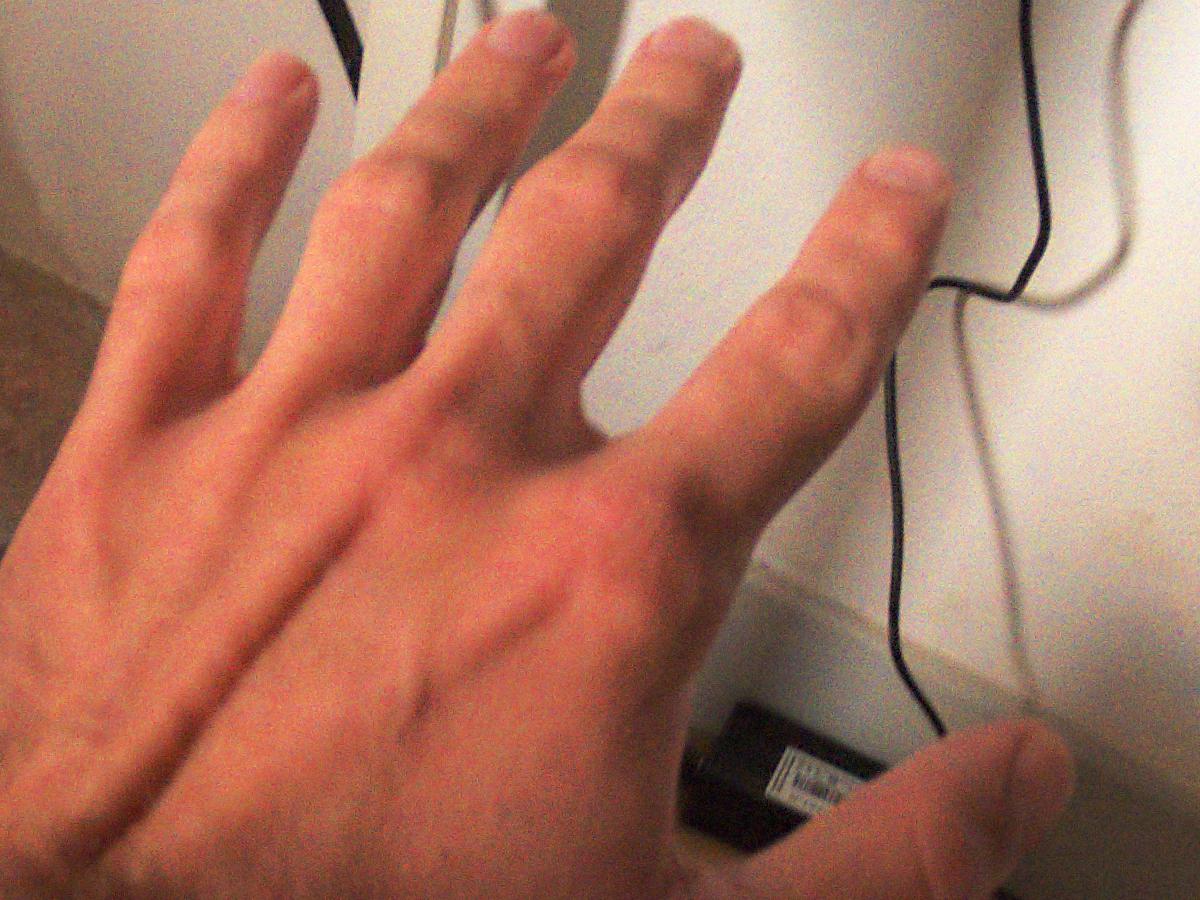}
			\end{center}
		\end{subfigure}
		\begin{subfigure}[t]{.3\linewidth}
			\begin{center}
				\includegraphics[width=\linewidth]{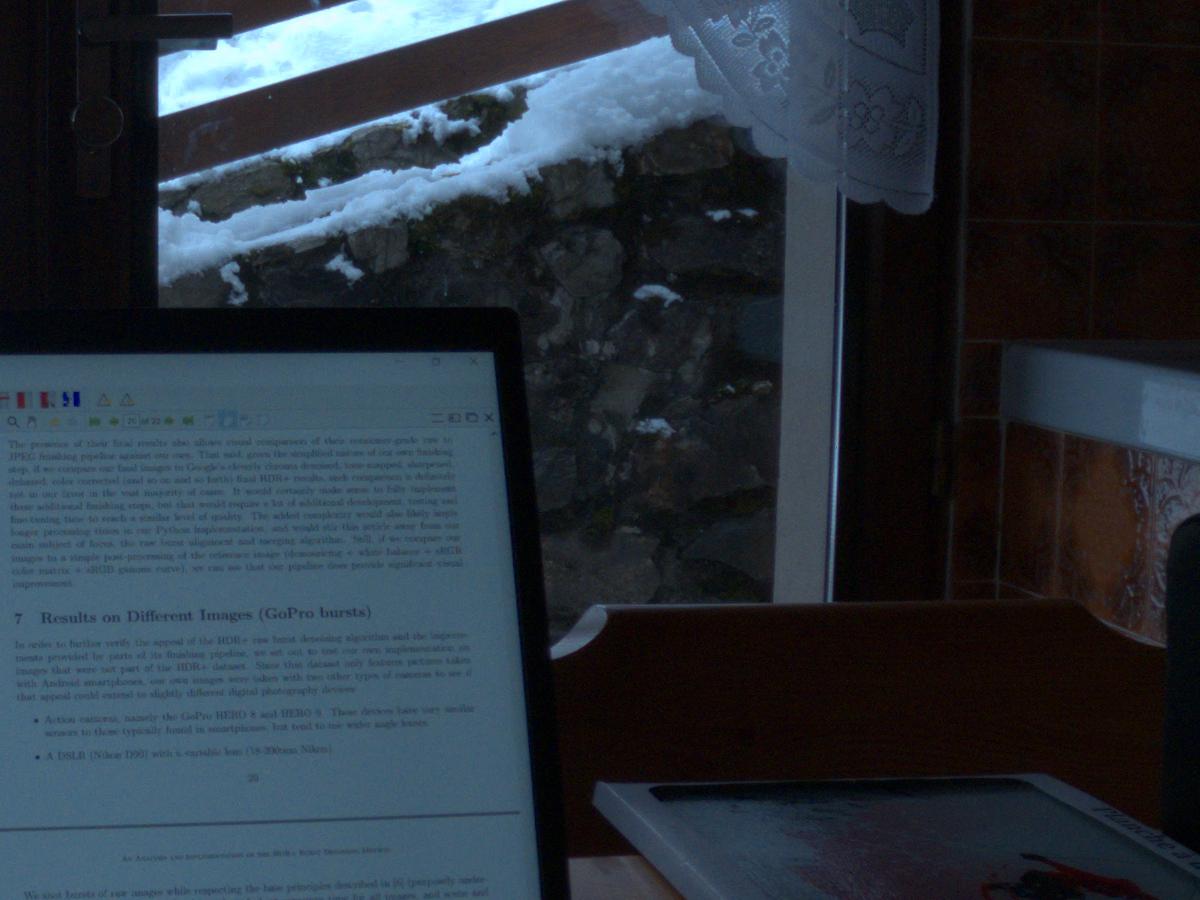}
			\end{center}
		\end{subfigure}
		\begin{subfigure}[t]{.3\linewidth}
			\begin{center}
				\includegraphics[width=\linewidth]{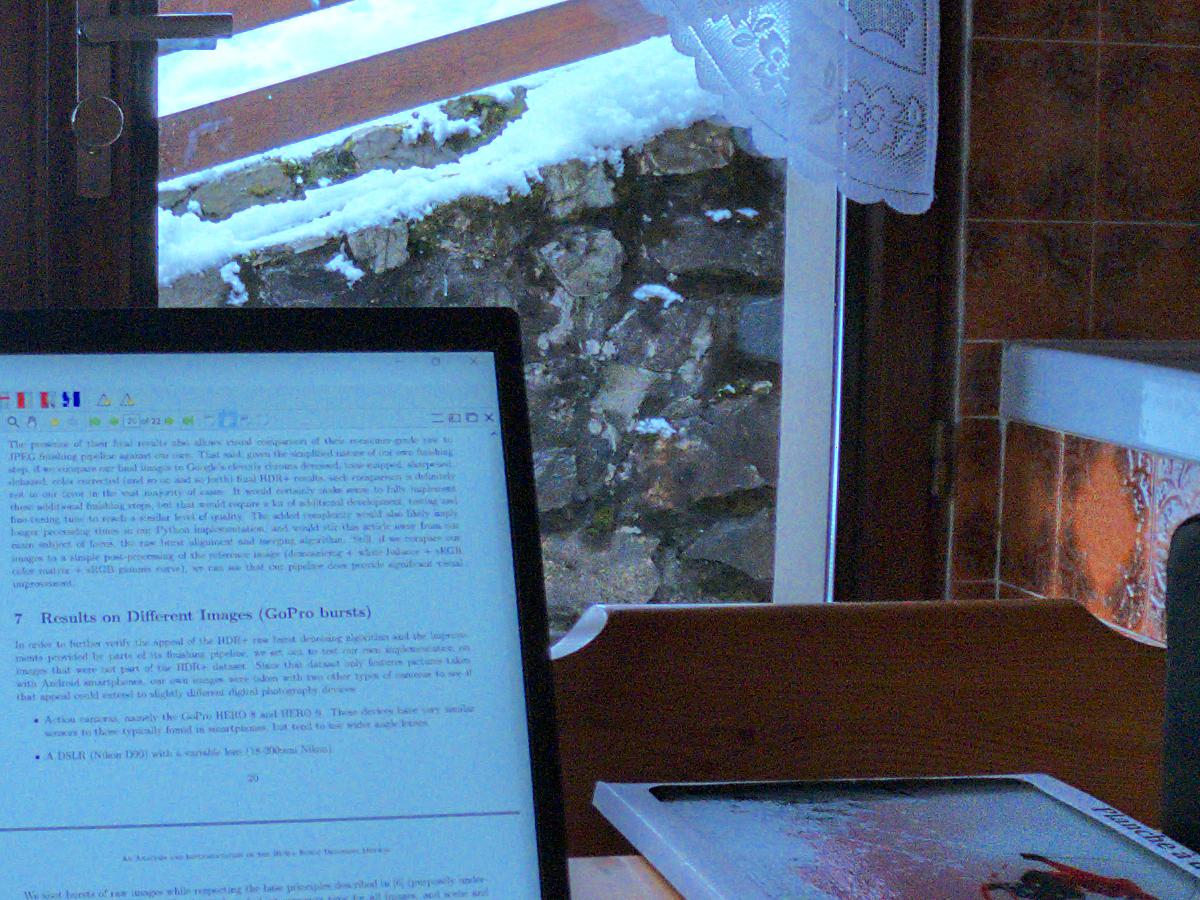}
			\end{center}
		\end{subfigure}
		\begin{subfigure}[t]{.3\linewidth}
			\begin{center}
				\includegraphics[width=\linewidth]{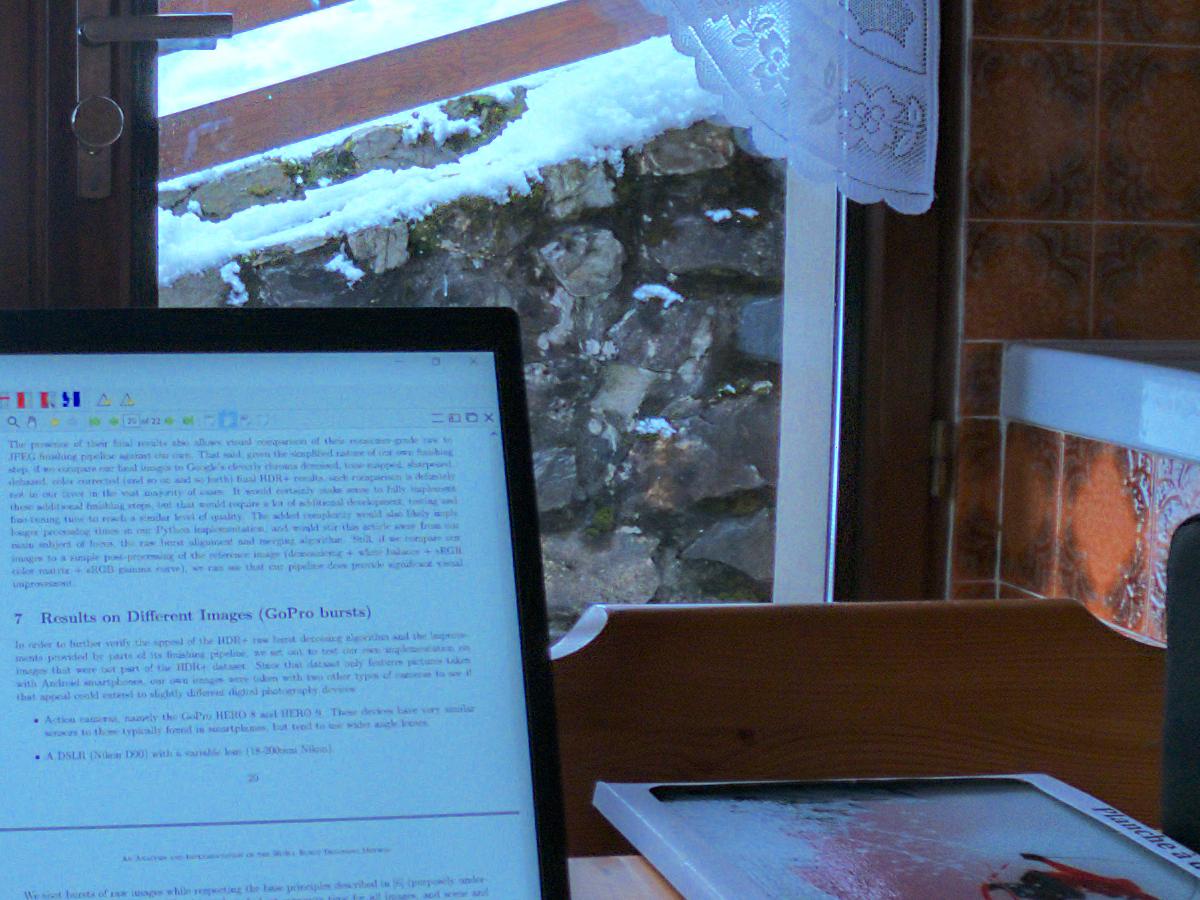}
			\end{center}
		\end{subfigure}
		\begin{subfigure}[t]{.3\linewidth}
			\begin{center}
				\includegraphics[width=\linewidth]{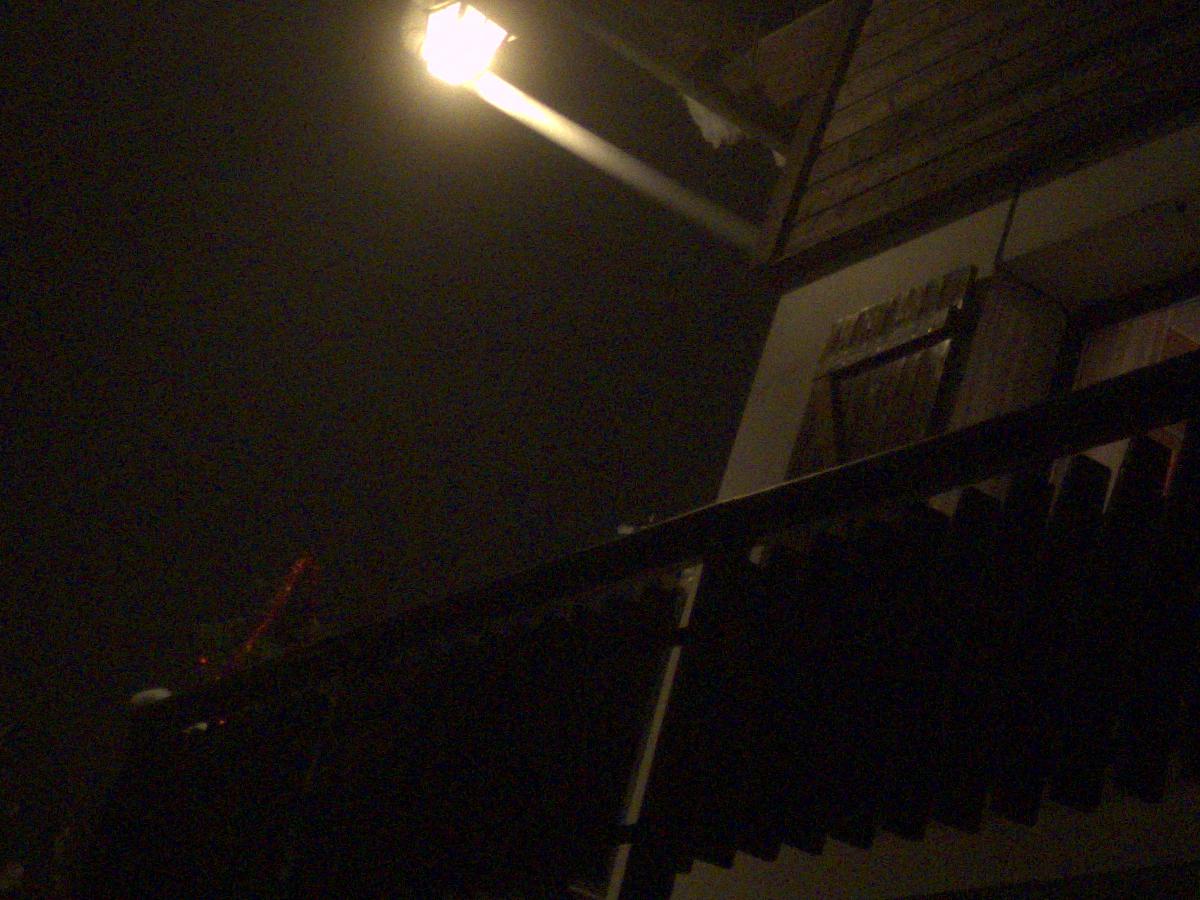}
			\end{center}
		\end{subfigure}
		\begin{subfigure}[t]{.3\linewidth}
			\begin{center}
				\includegraphics[width=\linewidth]{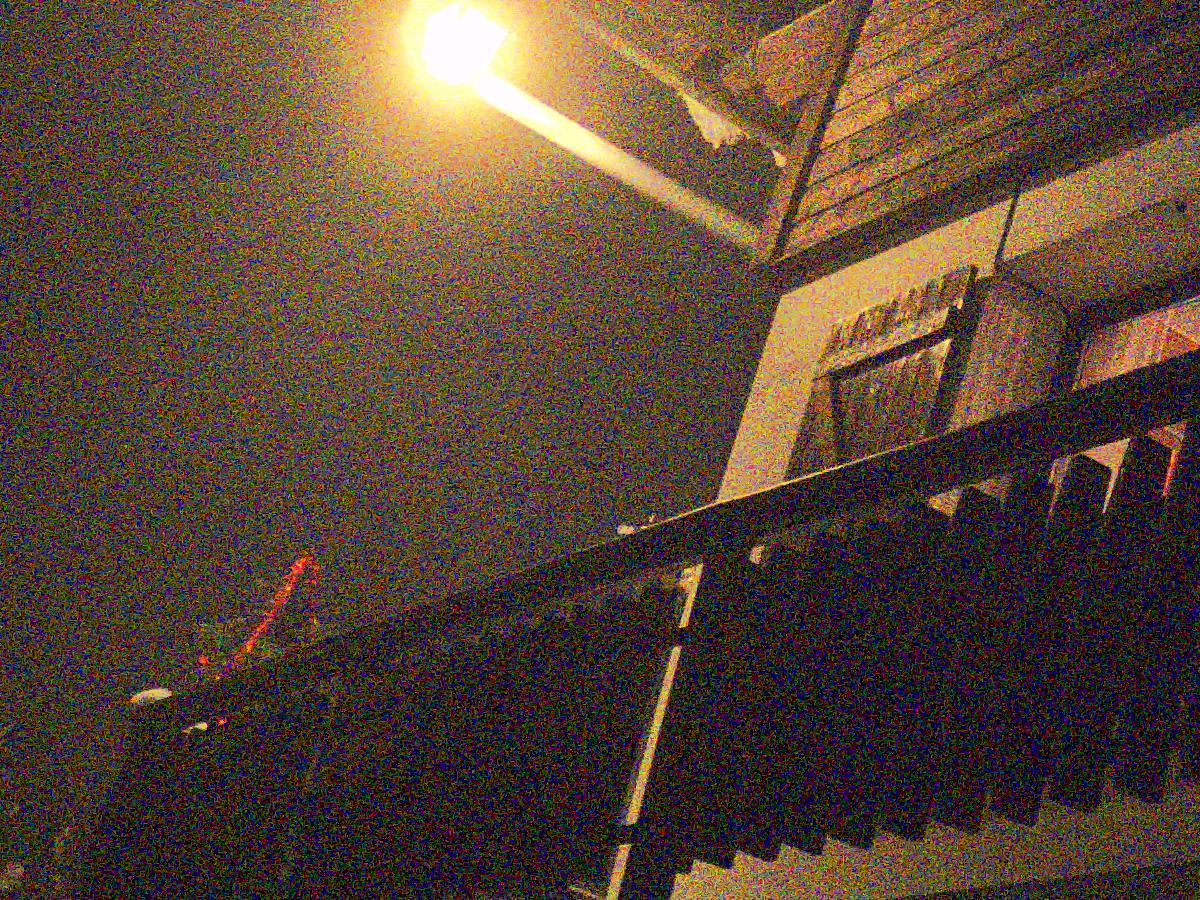}
			\end{center}
		\end{subfigure}
		\begin{subfigure}[t]{.3\linewidth}
			\begin{center}
				\includegraphics[width=\linewidth]{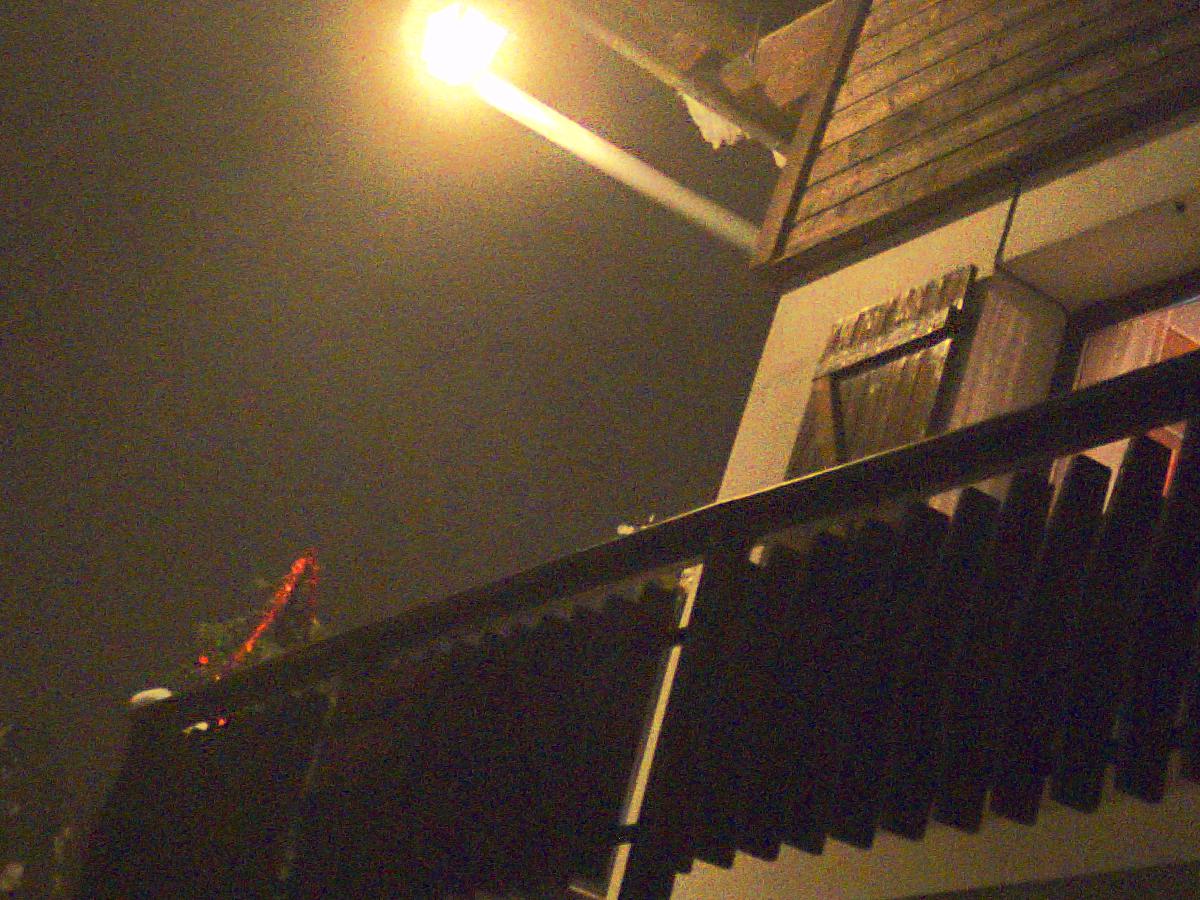}
			\end{center}
		\end{subfigure}
		\begin{subfigure}[t]{.3\linewidth}
			\begin{center}
				\includegraphics[width=\linewidth]{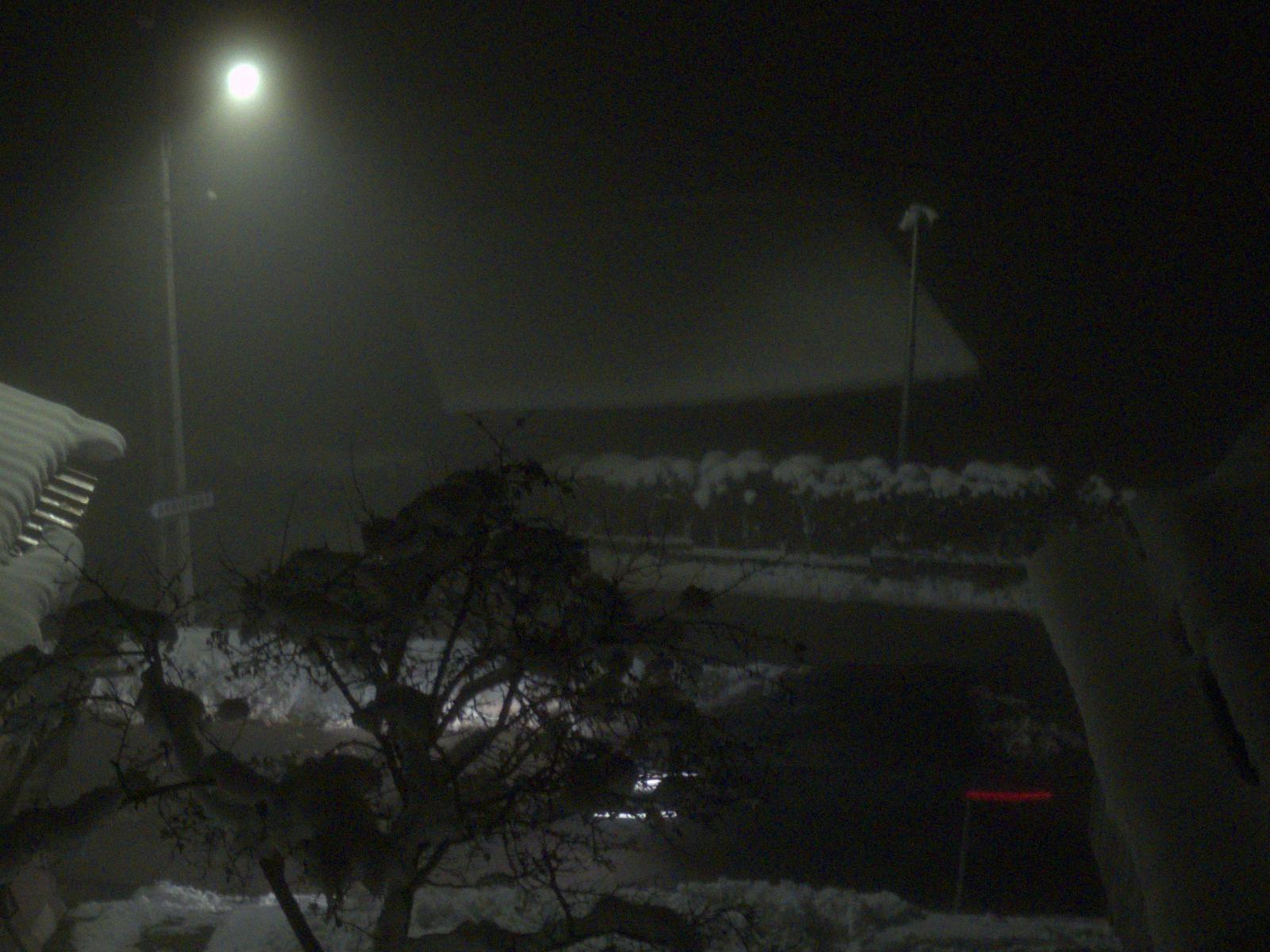}
			\end{center}
		\end{subfigure}
		\begin{subfigure}[t]{.3\linewidth}
			\begin{center}
				\includegraphics[width=\linewidth]{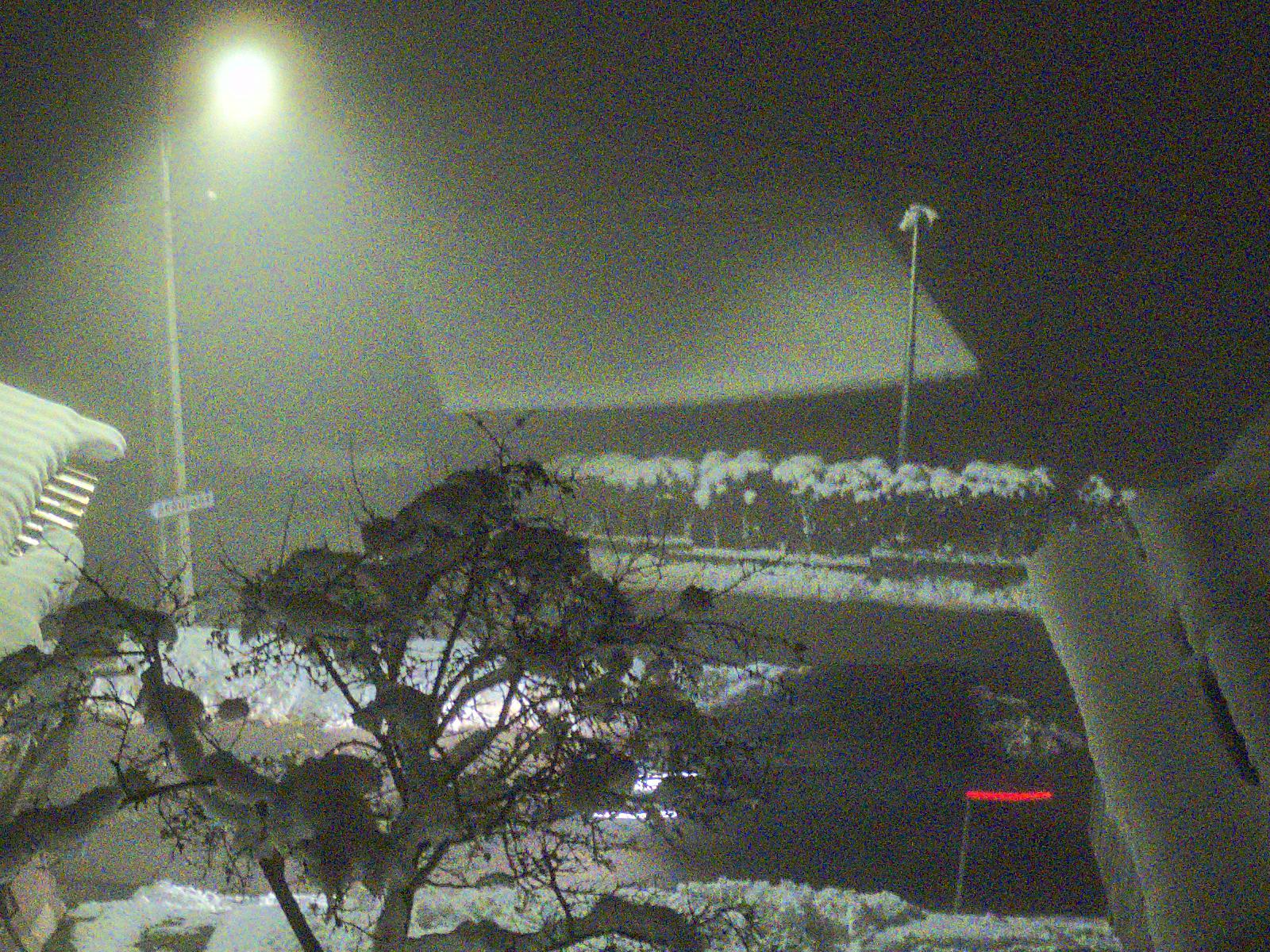}
			\end{center}
		\end{subfigure}
		\begin{subfigure}[t]{.3\linewidth}
			\begin{center}
				\includegraphics[width=\linewidth]{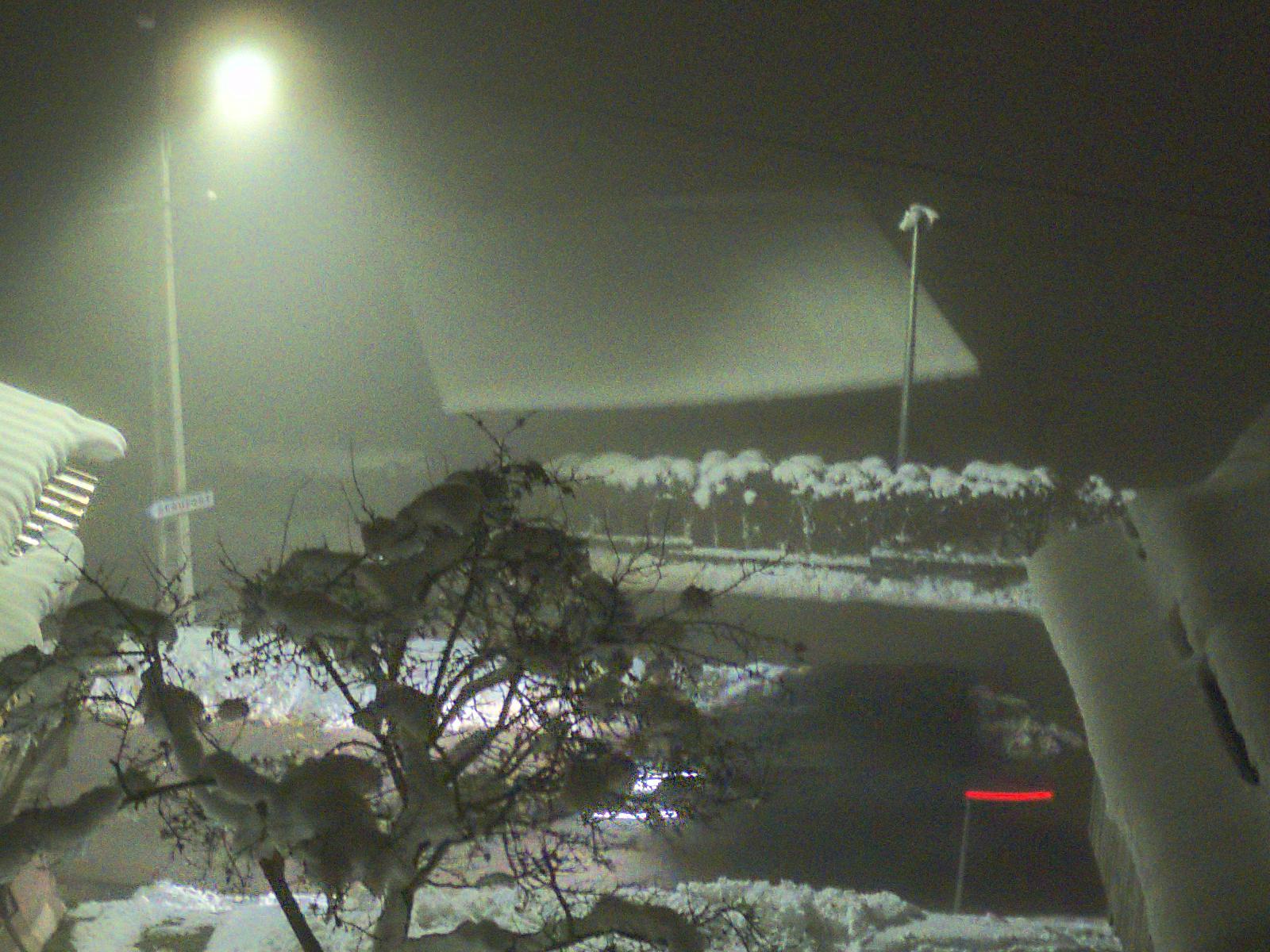}
			\end{center}
		\end{subfigure}
		\begin{subfigure}[t]{.3\linewidth}
			\begin{center}
				\includegraphics[width=\linewidth]{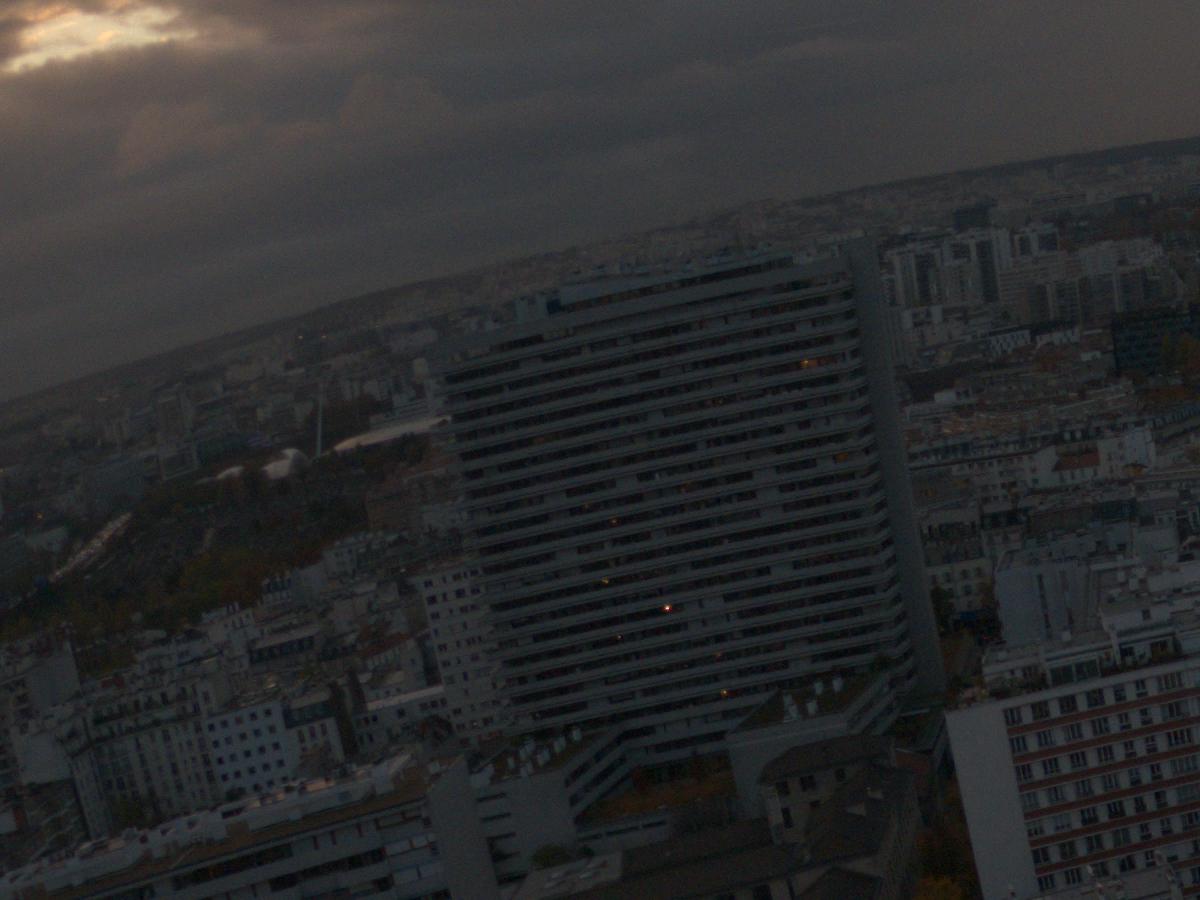}
				\caption*{\textbf{Reference} (minimal processing)}
			\end{center}
		\end{subfigure}
		\begin{subfigure}[t]{.3\linewidth}
			\begin{center}
				\includegraphics[width=\linewidth]{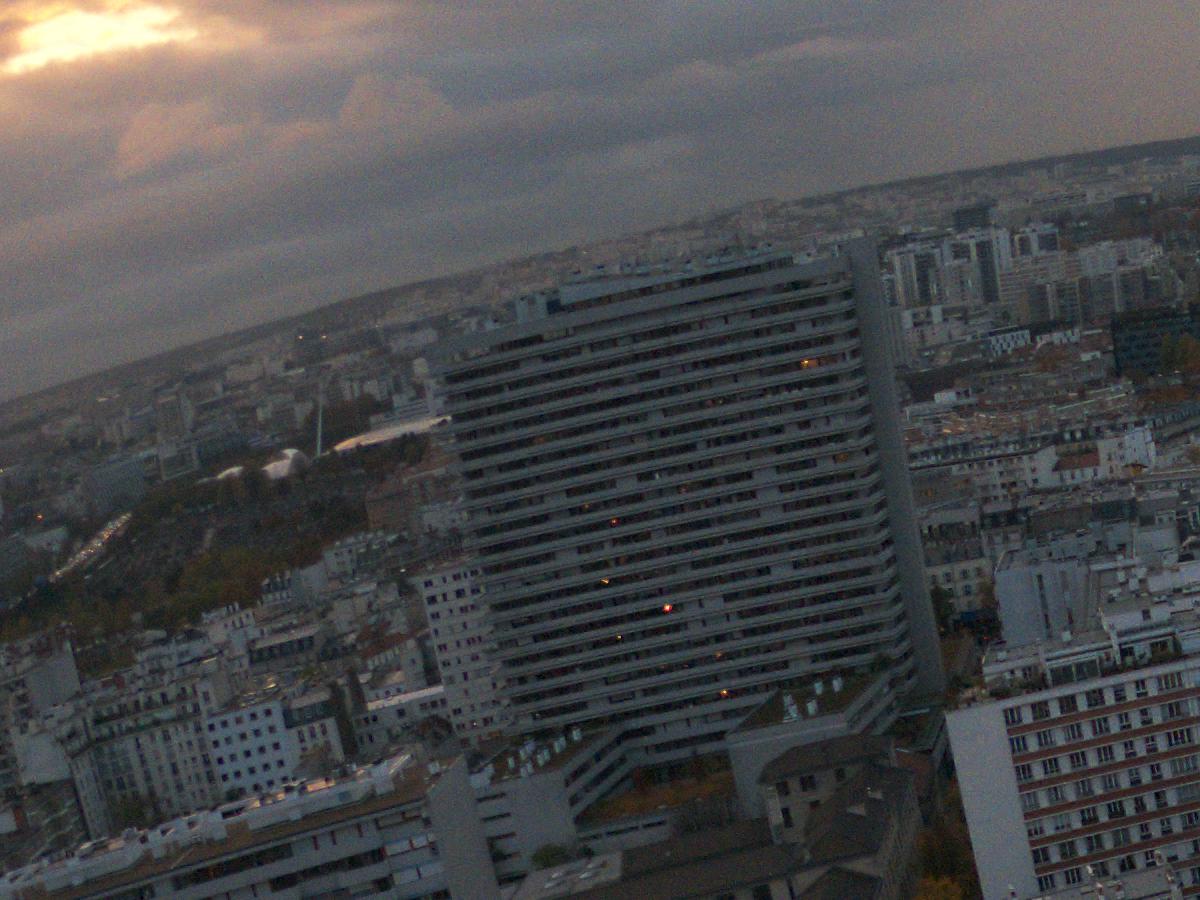}
				\caption*{\textbf{Reference} + \textbf{finishing} pipeline}
			\end{center}
		\end{subfigure}
		\begin{subfigure}[t]{.3\linewidth}
			\begin{center}
				\includegraphics[width=\linewidth]{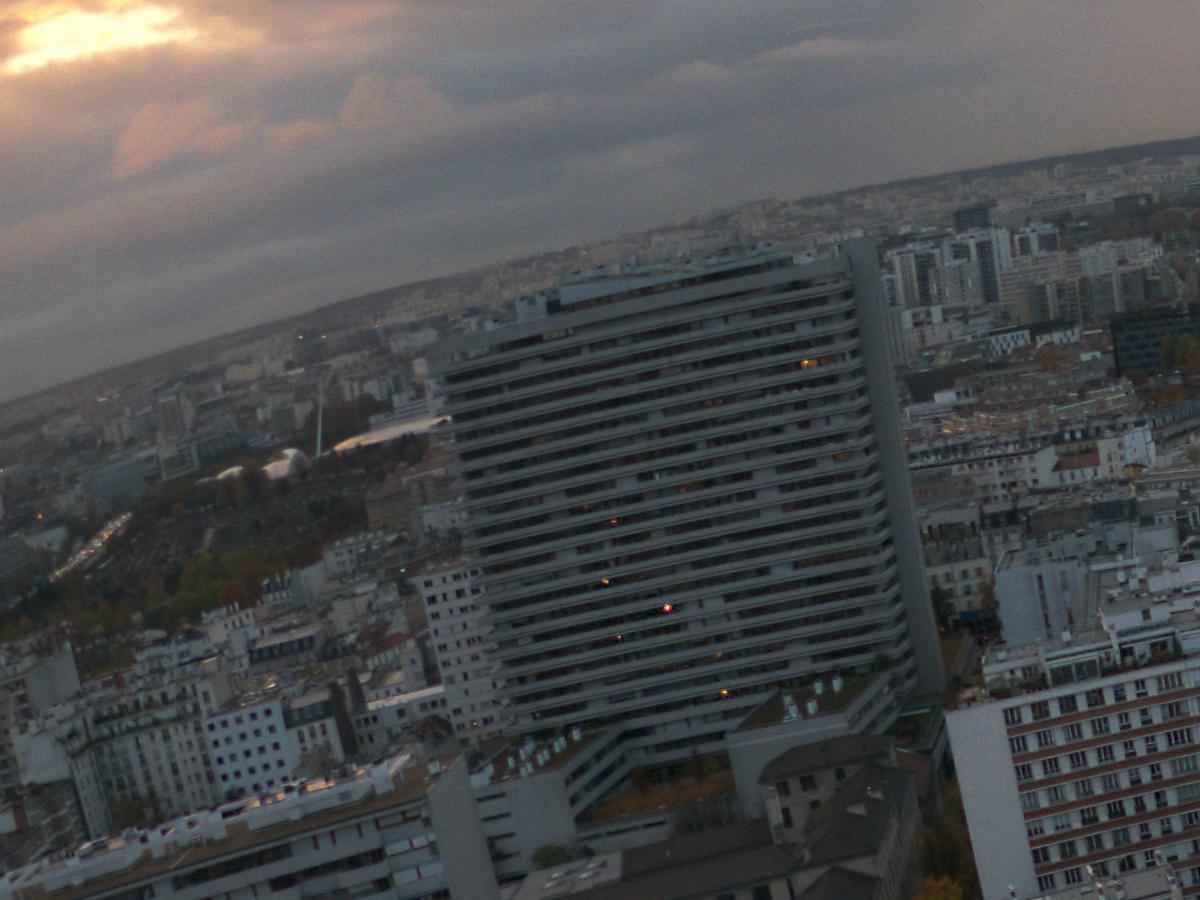}
				\caption*{\textbf{Merged} + \textbf{finishing} pipeline}
			\end{center}
		\end{subfigure}
		\caption{Visual comparison of crops of the minimally processed reference image, the reference processed by our finishing pipeline, and our alignment and merging result processed by our finishing pipeline (viewers are invited to zoom in).}
		\label{fig:gopro_bursts}
	\end{center}
\end{figure}

We applied our own finishing pipeline (described in Section~\ref{ourpipeline}) to both the reference image and the result of the alignment and merging steps in order to assess their impact on noise reduction. We also added a minimally processed version of the reference image (with no tone-mapping) to show how bright the original raw image actually is. We can see that noise is significantly reduced by the alignment and merging procedure (even on image edges, where these steps might perform worse because of image distortion induced by the wide lens). Residual noise can however still be quite present in some images (particularly in very low light), and some of it can be amplified by the tone mapping and sharpening steps of our pipeline. Figure~\ref{fig:gopro_bursts} showcases some of these results, while the full-size results on 17 GoPro bursts can be consulted and downloaded from Google Photos\footnote{\url{https://photos.app.goo.gl/XKiiSSNmqUEMYpYS8}}. The 17 raw DNG bursts can also be downloaded from Google Drive\footnote{\url{https://drive.google.com/drive/folders/1j2NIEPSnrdjS0sjL1kzl3VEmYKBkChJD?usp=sharing}}.

\newpage
\section{Conclusion}

Throughout this paper, we presented the core principle at the heart of the HDR+ digital photography pipeline: a raw burst alignment and merging algorithm. It uses a Wiener filter variant to combine the information of aligned tiles stacks in the 2D DFT space, to produce individual tiles with significantly less noise. Little priors and image metadata are required to produce convincing results, the most important one being a crude estimation of the noise level. When tuned properly, a good compromise between residual noise and ghosting artifacts can be found, resulting in natural, visually pleasing images. Many additional steps are required after alignment and merging to obtain a final RGB image acceptable for consumer-grade products, especially when trying to stand out from competitors. When compared to more recent state-of-the art approaches, the constrained computational complexity of the whole pipeline makes it well suited for mobile photography, but the strategies employed are versatile enough to improve image quality on many different types of cameras.

\section*{Acknowledgment}

The authors would like to thank Marc Lebrun for his support optimizing the Python code, which led to great reductions in total processing time. They would also like to thank Matias Tassano for his input over the course of this project.

%------------------------------------------------------------------------------
\section*{Image Credits}

{\small
	\includegraphics[height=2em]{./images/fused_exposure.jpg}
	\includegraphics[height=2em]{./images/spatial_ref_bright.jpg}
	\includegraphics[height=2em]{./images/gopro/hand_iso1600_final_cropped.jpg}
	\includegraphics[height=2em]{./images/gopro/kitchen_laptop_window_final_cropped.jpg}
	\includegraphics[height=2em]{./images/gopro/night_house_lamp_final_cropped.jpg}
	\includegraphics[height=2em]{./images/gopro/night_lamp_car_final_cropped.jpg}
	\includegraphics[height=2em]{./images/gopro/paris_north_west_final_cropped.jpg}
	Captured by the authors using GoPro Hero 8 and Hero 9 cameras\\\\
	\includegraphics[height=2em]{./images/pyramid_motion/imRef_scale3.png}
	\includegraphics[height=2em]{./images/reference_image_cropped.jpg}
	\includegraphics[height=2em]{./images/tf_ref_cropped.jpg}
	\includegraphics[height=2em]{./images/merged_2016_vs_2017/merged_2017_cropped.jpg}
	\includegraphics[height=2em]{./images/ours_google_merged/33TJ_20150626_104554_024_google_merged_cropped.jpg}
	\includegraphics[height=2em]{./images/ours_google_merged/c1b1_20141007_235657_307_google_merged_cropped.jpg}
	\includegraphics[height=2em]{./images/ours_google_merged/c1b1_20150424_192545_763_google_merged_cropped.jpg}
	\includegraphics[height=2em]{./images/ours_google_merged/33TJ_20150606_224837_294_google_merged_cropped.jpg}
	\includegraphics[height=2em]{./images/ours_google_merged/33TJ_20150822_185205_562_google_merged_cropped.jpg}
	\includegraphics[height=2em]{./images/ours_google_final/4KK2_20150829_162922_083_google_cropped.jpg}
	\includegraphics[height=2em]{./images/ours_google_final/5a9e_20150403_093136_401_google_cropped.jpg}
	\includegraphics[height=2em]{./images/ours_google_final/33TJ_20150606_224837_294_google_cropped.jpg}
	\includegraphics[height=2em]{./images/ours_google_final/6G7M_20150404_132319_709_google_cropped.jpg}
	Extracted from the HDR+ Burst Photography Dataset
}

\bibliographystyle{siam}
\bibliography{references}

\end{document}